\documentclass[journal,onecolumn,11pt]{IEEEtran}
\usepackage[T1]{fontenc}
\usepackage{amsmath,amssymb,amsthm,bm}
\usepackage{mathtools}
\usepackage{mleftright}
\mleftright
\usepackage{cite}
\usepackage{microtype}
\usepackage{booktabs}
\usepackage{tikz}
\usetikzlibrary{arrows.meta,positioning,calc,patterns}
\usepackage{pgfplots}
\pgfplotsset{compat=1.18}
\usepgfplotslibrary{fillbetween}
\definecolor{converse}{HTML}{2A78D6}
\definecolor{lowerbound}{HTML}{EB6834}
\definecolor{kbound}{HTML}{1BAF7A}
\tikzset{conv/.style={converse, solid, line width=1pt},
 lowb/.style={lowerbound, dash pattern=on 6pt off 3pt, line width=1pt},
 kdepb/.style={kbound, dash pattern=on 6pt off 2pt on 1.5pt off 2pt, line width=1pt}}
\definecolor{ballone}{HTML}{EDA100}
\definecolor{balltwo}{HTML}{E87BA4}
\definecolor{ballthree}{HTML}{008300}
\usepackage[hidelinks]{hyperref}
\hypersetup{pdftitle={The Asymptotic Capacity of the Grabbag Channel With and Without Feedback},pdfsubject={The order of the capacities of a multicolor finite-state channel},pdfkeywords={biological information theory, channel capacity, feedback, finite-state channels, grabbag channel, martingales, molecular communication, potential functions, random coding, trapdoor channel}}
\usepackage[capitalise]{cleveref}
\crefname{figure}{Fig.}{Figs.}
\Crefname{figure}{Figure}{Figures}
\crefname{table}{Table}{Tables}
\crefname{section}{Section}{Sections}
\crefname{subsection}{Section}{Sections}
\crefname{subsubsection}{Section}{Sections}
\crefname{appendix}{Appendix}{Appendices}
\crefname{theorem}{Theorem}{Theorems}
\crefname{lemma}{Lemma}{Lemmas}
\crefname{proposition}{Proposition}{Propositions}
\crefname{corollary}{Corollary}{Corollaries}
\crefname{definition}{Definition}{Definitions}
\crefname{remark}{Remark}{Remarks}
\crefname{example}{Example}{Examples}
\crefformat{equation}{\textup{(#2#1#3)}}
\Crefformat{equation}{Equation~\textup{(#2#1#3)}}
\crefrangeformat{equation}{\textup{(#3#1#4)}--\textup{(#5#2#6)}}
\crefmultiformat{equation}{\textup{(#2#1#3)}}{ and~\textup{(#2#1#3)}}{, \textup{(#2#1#3)}}{, and~\textup{(#2#1#3)}}
\allowdisplaybreaks[2]
\newtheorem{theorem}{Theorem}
\newtheorem{definition}{Definition}
\newtheorem{lemma}{Lemma}
\newtheorem{proposition}{Proposition}
\newtheorem{corollary}{Corollary}
\theoremstyle{remark}
\newtheorem{remark}{Remark}
\newtheorem{example}{Example}

\newcommand{\E}{\mathbb E}
\newcommand{\PP}{\mathbb P}
\newcommand{\ind}{\mathbf 1}
\newcommand{\ee}{\mathbf e}
\newcommand{\CC}{C_{M,K}}
\newcommand{\CF}{C^{\mathrm{fb}}_{M,K}}
\newcommand{\Ber}{\operatorname{Ber}}
\title{The Asymptotic Capacity of the Grabbag Channel
With and Without Feedback}

\author{Jun Chen and Yuta Sakai%
\thanks{Authors are listed alphabetically.}
  \\[1.5ex]
  {\normalsize\itshape Dedicated to the memory of Toby Berger}
}

\begin{document}
\maketitle
\begin{abstract}
The grabbag channel models communication through a randomly mixed store. At each use, an input ball of one of $K$ colors is inserted into a bag containing $M$ balls, and one of the resulting $M+1$ balls is selected uniformly, observed, and removed. We show that, when $2\leq K\leq M+1$, both the capacity without feedback and the capacity with noiseless strictly causal output feedback are at most $9.77(K/(M+1))^{2/3}$, and that, when moreover $M+1\geq2^{22}$, both are at least an absolute constant times $(K/(M+1))^{2/3}$. Hence, both capacities are of order $(K/M)^{2/3}$, and of order $M^{-2/3}$ for every fixed $K$. The converse is a single bound on the information that a feedback code conveys in $n$ uses, valid for every initial state. For each initial state, it bounds the information in each output by the change of a potential of the posterior law of the state given the earlier outputs, defined through noisy cube roots of the color counts. A budget on the true state bounds the total cost of the uses in which the inserted color is rare. The lower bound uses a statistic of two output histograms whose expectation does not depend on the unknown initial composition, an outer code, and words with prescribed compositions. We also show that the channel is indecomposable and relate its capacities to the finite-state-channel information formulas. 
\end{abstract}

\section{Introduction}\label{sec:intro}
Communication through a well-mixed pool has a simple physical interpretation: the transmitter adds a distinguishable object, while the receiver observes an object sampled from the pool. The object observed need not be the one just added. The pool therefore retains information about earlier inputs, and its composition creates channel memory. In biochemical communication, different molecular species can play the role of distinguishable colors, while a mixed compartment can play the role of the pool. The model studied here is an idealization of this mechanism: it assumes a fixed population, instantaneous mixing, one insertion and one removal per use, and exact observation of the removed object's type. It does not model diffusion times, reaction kinetics, or imperfect detection.

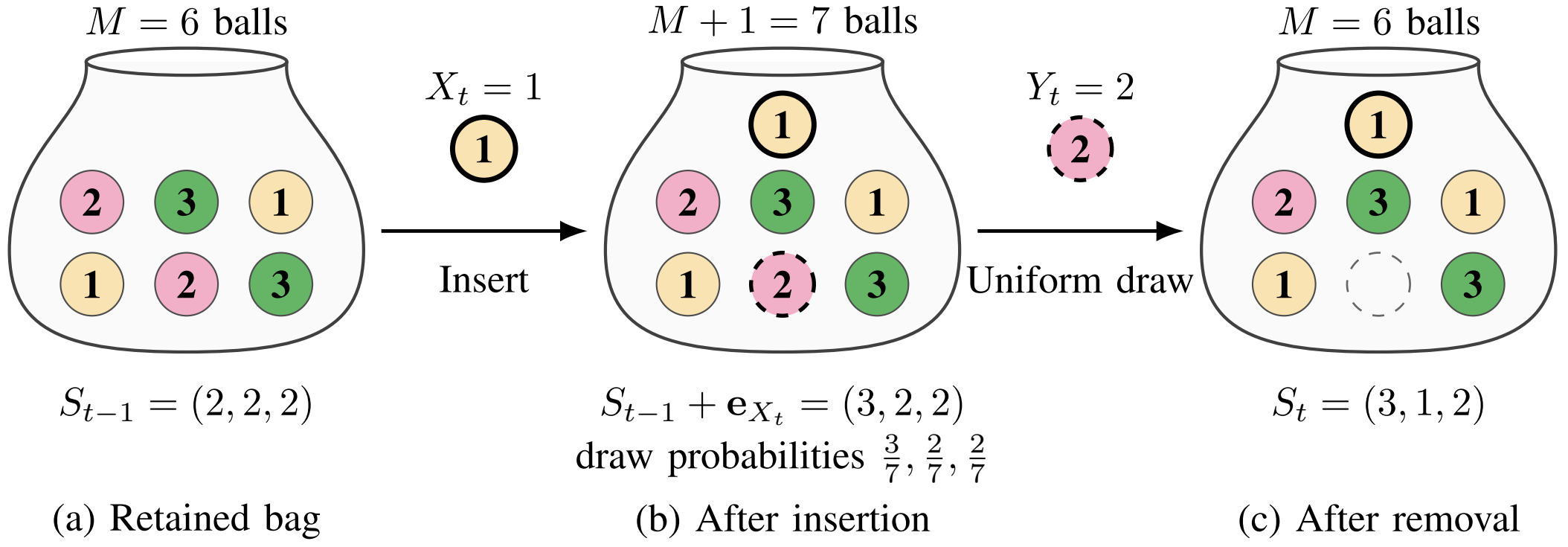
\begin{figure}[t]
\centering
\begin{tikzpicture}[x=1.05cm,y=0.86cm,>=Latex,
 ball/.style={circle,draw=black!70,line width=0.4pt,minimum size=5.6mm,inner sep=0pt,font=\small\bfseries},
 typeone/.style={ball,fill=ballone!30},typetwo/.style={ball,fill=balltwo!60},typethree/.style={ball,fill=ballthree!60},
 new/.style={draw=black,line width=1.3pt},
 drawn/.style={draw=black,dashed,line width=1pt},
 gone/.style={circle,draw=black!60,dashed,line width=0.5pt,minimum size=5.6mm,inner sep=0pt},
 stage/.style={font=\small,align=center},flow/.style={->,line width=0.9pt}]
\foreach \shift in {0,5.3,10.6}{
 \begin{scope}[xshift=\shift cm]
 \path[draw=black!75,fill=black!2,line width=0.8pt]
 (-.85,1.7) .. controls (-.8,1.15) and (-1.5,.8) .. (-1.5,-.25)
 .. controls (-1.5,-1.1) and (-1,-1.3) .. (0,-1.3)
 .. controls (1,-1.3) and (1.5,-1.1) .. (1.5,-.25)
 .. controls (1.5,.8) and (.8,1.15) .. (.85,1.7);
 \draw[black!75,line width=0.8pt] (0,1.7) ellipse [x radius=.85,y radius=.14];
 \end{scope}
}
\node[typeone] at (-.8,-.6) {1}; \node[typetwo] at (0,-.6) {2}; \node[typethree] at (.8,-.6) {3};
\node[typetwo] at (-.8,.25) {2}; \node[typethree] at (0,.25) {3}; \node[typeone] at (.8,.25) {1};
\begin{scope}[xshift=5.3cm]
\node[typeone] at (-.8,-.6) {1}; \node[typetwo,drawn] at (0,-.6) {2}; \node[typethree] at (.8,-.6) {3};
\node[typetwo] at (-.8,.25) {2}; \node[typethree] at (0,.25) {3}; \node[typeone] at (.8,.25) {1};
\node[typeone,new] at (0,1.05) {1};
\end{scope}
\begin{scope}[xshift=10.6cm]
\node[typeone] at (-.8,-.6) {1}; \node[gone] at (0,-.6) {}; \node[typethree] at (.8,-.6) {3};
\node[typetwo] at (-.8,.25) {2}; \node[typethree] at (0,.25) {3}; \node[typeone] at (.8,.25) {1};
\node[typeone,new] at (0,1.05) {1};
\end{scope}
\node[stage] at (0,-3.05) {(a) Retained bag};
\node[stage] at (5.05,-3.05) {(b) After insertion};
\node[stage] at (10.1,-3.05) {(c) After removal};
\node[stage] at (0,2.12) {$M=6$ balls};
\node[stage] at (5.05,2.12) {$M+1=7$ balls};
\node[stage] at (10.1,2.12) {$M=6$ balls};
\node[stage] at (0,-1.85) {$S_{t-1}=(2,2,2)$};
\node[stage] at (5.05,-1.85) {$S_{t-1}+\ee_{X_t}=(3,2,2)$};
\node[stage] at (5.05,-2.4) {draw probabilities $\frac37,\frac27,\frac27$};
\node[stage] at (10.1,-1.85) {$S_t=(3,1,2)$};
\draw[flow] (1.65,-.05) -- (3.4,-.05);
\node[stage] at (2.52,-.55) {Insert};
\node[typeone,new] at (2.52,.8) {1};
\node[stage] at (2.52,1.42) {$X_t=1$};
\draw[flow] (6.7,-.05) -- (8.45,-.05);
\node[stage] at (7.57,-.55) {Uniform draw};
\node[typetwo,drawn] at (7.57,.8) {2};
\node[stage] at (7.57,1.42) {$Y_t=2$};
\end{tikzpicture}
\caption{One use of the grabbag channel, illustrated with $K=3$ colors and $M=6$ retained balls. First insert the input ball, drawn with a heavy outline. Then sample one of all $M+1$ balls uniformly and remove it. Here, the input has color 1. The drawn color is 1, 2 or 3 with probabilities $3/7$, $2/7$ and $2/7$, and the output is color 2: the dashed ball in (b), whose place is empty in (c). In the drawing, every other ball keeps its place. The newly inserted ball is also eligible to be drawn.}
\label{fig:grabbag}
\end{figure}

We call this model the \emph{grabbag channel}. A bag contains $M$ balls, each having one of $K$ colors. The transmitter inserts one ball. The receiver then draws a uniformly selected ball from the $M+1$ balls in the bag, observes its color, and removes it. The $M$ balls left in the bag form the next state. Thus, the input and output alphabets are both the set of colors, and the state is the vector of retained color counts. The order ``insert, then draw'' is part of the definition. In particular, when $K=2$ and $M=1$, the channel is the classical binary trapdoor channel: the output is selected with equal probability from the retained ball and the newly inserted ball. A convention that draws before inserting has a different storage parameter and is not used in this paper.
\Cref{fig:grabbag} illustrates one channel use and the change in the retained composition.

The trapdoor channel was introduced by Blackwell~\cite{blackwell1961} and is discussed in early information-theory texts, including Ash~\cite{ash}. Even this seemingly simple finite-state channel has resisted a complete nonfeedback capacity characterization. The analytical bounds are $0.572\log2\leq C\leq\log(3/2)$ nats per channel use: the lower bound is recalled in~\cite[Eq.~(18)]{computable}, and the upper bound is proved in~\cite[Theorem~7]{computable}, where finite-state-channel duality and computational methods give further capacity upper bounds. A numerical evaluation of an upper bound on a delayed-feedback capacity gives $C\leq0.5765\log2$ nats per channel use~\cite[Eqs.~(16)--(17)]{delayed}. Permuter, Cuff, Van Roy, and Weissman~\cite{trapdoorfb} determined the feedback capacity of the trapdoor channel, which is $\log\bigl((1+\sqrt5)/2\bigr)$ nats per channel use. Their work also identifies larger stores and more ball types as natural generalizations. When the feedback is delayed by two uses, the capacity of the trapdoor channel is $\log(3/2)$ nats per channel use~\cite{delayed}. Ahlswede and Kaspi~\cite{ahlswede} studied its zero-error coding problem in the broader setting of permuting channels. In the permuting relay channel, the message sender chooses the removed balls, while a helper supplies an input sequence that does not depend on the message. Kobayashi~\cite{kobayashi} determined the capacity of this channel. These results concern fixed finite-state channels, whereas our question concerns a sequence of channels whose storage size tends to infinity.

Large stores have a history of their own. Berger studied generalizations of the trapdoor channel with many balls as part of his program of biological information theory. An archived abstract of his 2000 seminar~\cite{bergerseminar} announces that the capacity of a large-storage channel of this kind is of order $b^{-2/3}$, where $b$ denotes the storage size, together with coding schemes of that order. Giles and Hajek~\cite{gileshajek} describe the insert-then-draw channel with two colors and $B$ retained balls. Citing a 2002 draft manuscript of Berger and Zhang entitled ``On the capacity of the grabbag channel,'' they state that its capacity is at least a constant times $B^{-2/3}$ for large $B$. Gastpar~\cite{gastpar} describes a version with $b$ symbols of several types, in which a symbol is drawn before a new one is injected, and states the same behavior as a lower bound. Both 2002 sources credit the lower bound to unpublished work of Zhang and Berger. The present paper proves upper and lower bounds of the same order for the multicolor model, with and without feedback, and identifies the dependence on the number of colors.

Finite-state-channel coding theory provides the broader setting. Blackwell, Breiman, and Thomasian~\cite{bbt} established coding results for indecomposable finite-state channels. Gallager~\cite{gallager} developed the corresponding capacity theory and conditions under which the effect of the initial state disappears. With feedback, directed-information and stochastic-control formulations replace the ordinary input-distribution optimization (see, for example, Chen and Berger~\cite{chenberger}, Tatikonda and Mitter~\cite{tatikonda}, and Permuter, Weissman, and Goldsmith~\cite{fscfb}). Yang, Kav\v{c}i\'c, and Tatikonda~\cite{yangkavcictatikonda} developed a stochastic-control formulation for finite-state machine channels whose next state is determined by the preceding state and the input. For the grabbag channel, \cref{app:fsc} verifies the hypotheses of the unifilar-channel feedback capacity characterization in~\cite[Theorem~3]{trapdoorfb}. The direct proofs of our main bounds do not use these capacity formulas. Equality of feedback and nonfeedback capacities can hold for particular channel families. A recent example is the approximately memoryless surjective POST family studied in~\cite{post}. Our results concern the order of the capacities, which is a weaker assertion.

The biochemical motivation belongs to a broader program of biological information theory. Berger's Shannon lecture~\cite{bergerlecture} discusses information-theoretic questions arising in living systems. Molecular communication studies how chemical species and their transport or reactions can support information transmission (see~\cite{molecular} for a systematic treatment). Thomas and Eckford~\cite{bind}, for example, analyze a ligand-receptor signal-transduction channel with Markov memory and characterize the role of feedback in that model. Pierobon and Akyildiz~\cite{pierobonakyildiz} study diffusion memory and molecular noise in a model with Brownian motion and concentration measurements. Nakano, Eckford, and Haraguchi~\cite{molecular} refer to Blackwell's model as the chemical channel and regard a bag of many balls as a coarse model of diffusion.
Sampling from a pool is also the read-out model of DNA-based storage~\cite{shomoronyheckel}, of channels that encode information in molecular concentrations~\cite{gerzon}, and of noisy permutation channels~\cite{makur,tangpolyanskiy}. Weinberger and Merhav~\cite{weinbergermerhav} study a DNA storage channel in which the molecules of a codeword are stored without order, sampled with replacement, and sequenced over a discrete memoryless channel. Tamir, Weinberger, and Guill\'en i F\`abregas~\cite{tamirshort} study coding for the noiseless shuffling-sampling channel in the short-molecule regime at a fixed positive coverage depth. In those models the pool is written once and then read, whereas the grabbag channel is written while it is read. The grabbag channel isolates a different source of memory: random sampling from a population whose composition is changed by the inputs and outputs.
\Cref{tab:prior} summarizes what is known about the capacities of the trapdoor channel and of large stores, and \cref{tab:results} lists the results of this paper. Three conventions must be kept in mind when the rows are compared. First, the order of insertion and drawing matters. The description in~\cite{gileshajek} inserts before drawing, as we do, whereas the description in~\cite{gastpar} draws before inserting. For nonfeedback codes, the two conventions correspond after a shift of the input by one use, with $b$ symbols in the second corresponding to $M=b-1$ retained balls in the first (\cref{subsec:model-variants}). Second, the general feedback-capacity formula in~\cite{trapdoorfb} assumes an initial state known to both terminals, whereas \cref{def:codes} uses the worst initial state, known to neither terminal. Upper bounds transfer from the known-state setting to ours, but lower bounds need their own argument. Third, the rows on large stores concern the order of the capacity in the storage size, not its value for a fixed store.

\begin{table}[t]
\centering
\caption{Known capacity results for the trapdoor channel ($M=1$, $K=2$) and for large stores. All rates are in nats per channel use.}
\label{tab:prior}
\renewcommand{\arraystretch}{1.2}
\begin{tabular}{@{}lll@{}}
\toprule
Setting & Result & Source\\
\midrule
Trapdoor, feedback$^{\rm a}$ & $\log\frac{1+\sqrt5}2\approx0.4812$ & \cite{trapdoorfb}\\
Trapdoor, no feedback, analytical & $0.572\log2\leq C\leq\log(3/2)$, that is, $0.3964<C<0.4055$ & \cite{computable}\\
Trapdoor, no feedback, numerical & $C\leq0.5765\log2\approx0.3996$ & \cite[Eqs.~(16)--(17)]{delayed}\\
Trapdoor, feedback delayed by two uses$^{\rm a}$ & $\log(3/2)\approx0.4055$ & \cite{delayed}\\
\midrule
Two colors, store of size $b$, feedback not stated & order $b^{-2/3}$, announced & \cite{bergerseminar}\\
Two colors, $B$ retained balls, no feedback & at least a constant times $B^{-2/3}$, proof unpublished & \cite{gileshajek}\\
Several colors, $b$ balls, draw then insert, no feedback & at least of order $b^{-2/3}$, proof unpublished & \cite{gastpar}\\
\bottomrule
\multicolumn{3}{@{}l@{}}{\footnotesize $^{\rm a}$With the initial state known to both terminals.}
\end{tabular}
\end{table}

\begin{table}[t]
\centering
\caption{The results of this paper, with the constants $\gamma\eta/72>9\times10^{-4}$ and $c_0>2.2\times10^{-8}$ of~\cref{eq:gamma}, and rates in nats per channel use.}
\label{tab:results}
\renewcommand{\arraystretch}{1.2}
\begin{tabular}{@{}lp{0.6\textwidth}l@{}}
\toprule
Range & Statement & Where\\
\midrule
$2\leq K\leq M+1$ & $\CC\leq\CF<9.77\,(K/(M+1))^{2/3}$; for $K=2$, $<15.51\,(M+1)^{-2/3}$ & \cref{thm:converse}\\
$K\geq2$, $M\geq2^{16}$ (sufficient) & $\CC\geq{}$ the right side of~\cref{eq:lower-exact}; for $K=2$, $\CC\geq0.0253\,(M+1)^{-2/3}$ & \cref{thm:lower}\\
$2\leq k\leq K$ & $C_{M,k}\leq\CC$ and $C^{\mathrm{fb}}_{M,k}\leq\CF$ & \cref{prop:merging}\\
$K\geq2$, $M+1\geq2^{22}\lfloor K/2\rfloor$ & $\CC\geq\frac{\gamma\eta}{72}\bigl(\lfloor K/2\rfloor/(M+1)\bigr)^{2/3}$ & \cref{prop:kdependence}\\
$2\leq K\leq M+1$, $M+1\geq2^{22}$ & $c_0(K/(M+1))^{2/3}\leq\CC\leq\CF<9.77\,(K/(M+1))^{2/3}$ & \cref{cor:order}\\
\midrule
$K=2$, $M\to\infty$ & a bound $I(Q\wedge Y^n)\leq c_1n(M+1)^{-2/3}+R(M)$ for all $n$ and all codes needs $R(M)\geq cM^{1/3}$ & \cref{prop:accumulation}\\
\bottomrule
\end{tabular}
\end{table}

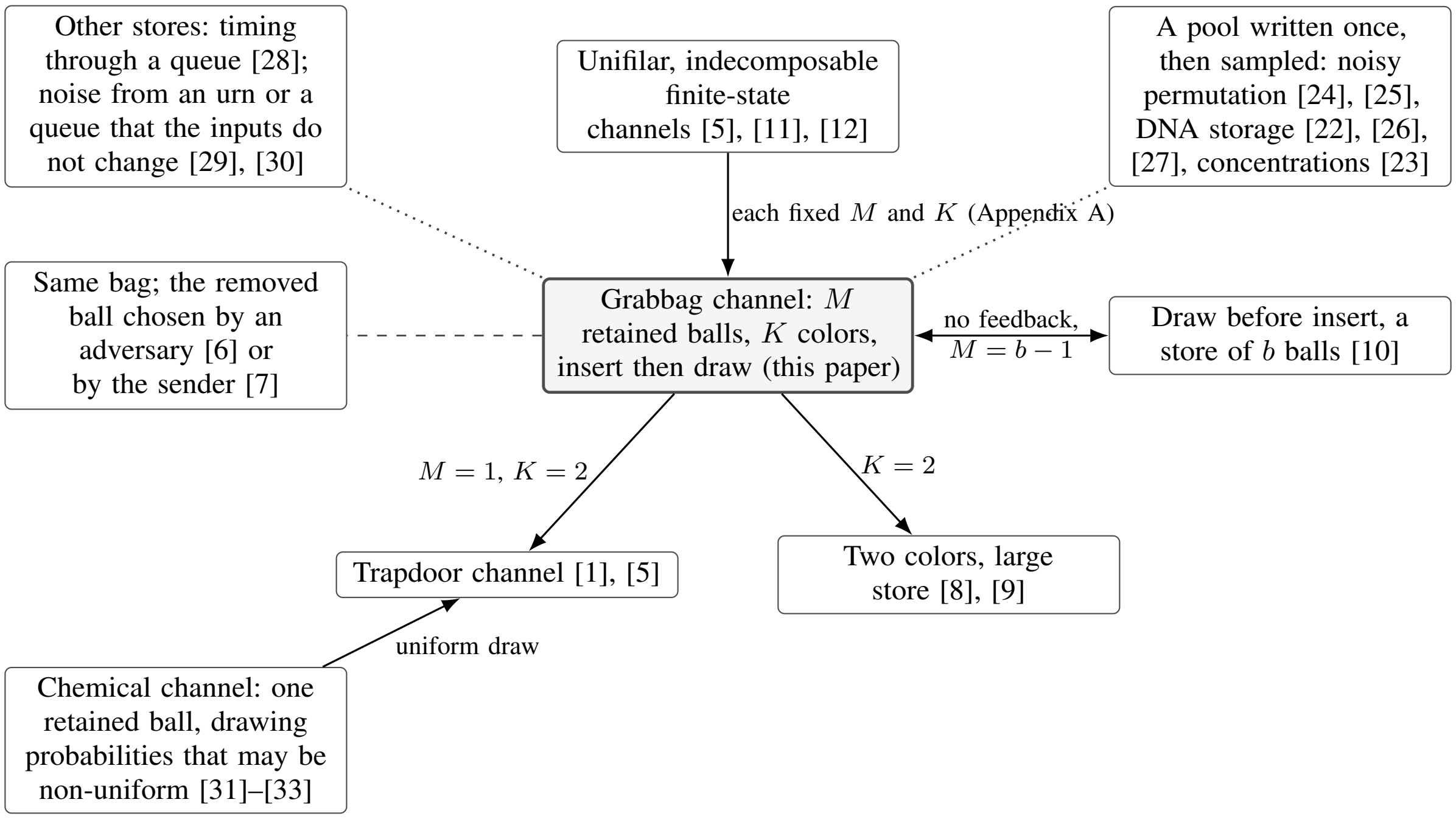
\begin{figure}[t]
\centering
\begin{tikzpicture}[x=1cm,y=1cm,>=Latex,
 model/.style={draw=black!70, rounded corners=2pt, align=center, font=\footnotesize, inner sep=3pt, text width=35mm},
 main/.style={model, text width=38mm, line width=0.9pt, fill=black!4},
 rel/.style={->, line width=0.6pt},
 mech/.style={dashed, line width=0.5pt, black!75},
 anal/.style={dotted, line width=0.7pt, black!75},
 lab/.style={font=\scriptsize, align=center, inner sep=1pt}]
\node[main] (G) at (0,0) {Grabbag channel: $M$ retained balls, $K$ colors, insert then draw (this paper)};
\node[model] (FSC) at (0,2.6) {Unifilar, indecomposable finite-state channels~\cite{bbt,gallager,trapdoorfb}};
\node[model] (P) at (-6.0,0) {Same bag; the removed ball chosen by an adversary~\cite{ahlswede} or by the sender~\cite{kobayashi}};
\node[model] (DI) at (6.0,0) {Draw before insert, a store of $b$ balls~\cite{gastpar}};
\node[model] (Q) at (-6.0,2.6) {Other stores: timing through a queue~\cite{bitsqueues}; noise from an urn or a queue that the inputs do not change~\cite{contagion,qbc}};
\node[model] (NP) at (6.0,2.6) {A pool written once, then sampled: noisy permutation~\cite{makur,tangpolyanskiy}, DNA storage~\cite{shomoronyheckel,weinbergermerhav,tamirshort}, concentrations~\cite{gerzon}};
\node[model] (T) at (-2.4,-2.6) {Trapdoor channel~\cite{blackwell1961,trapdoorfb}};
\node[model] (Ch) at (-6.0,-4.4) {Chemical channel: one retained ball, drawing probabilities that may be non-uniform~\cite{chemical2007,wuanastasopoulos,sabaggraph}};
\node[model] (GH) at (2.4,-2.6) {Two colors, large store~\cite{bergerseminar,gileshajek}};
\draw[rel] (FSC) -- node[lab, right]{each fixed $M$ and $K$ (\cref{app:fsc})} (G);
\draw[rel] (G) -- node[lab, left, xshift=-1mm]{$M=1$, $K=2$} (T);
\draw[rel] (Ch) -- node[lab, below right]{uniform draw} (T);
\draw[rel] (G) -- node[lab, right, xshift=1mm]{$K=2$} (GH);
\draw[rel, <->] (G) -- node[lab, above]{no feedback,} node[lab, below]{$M=b-1$} (DI);
\draw[mech] (G) -- (P);
\draw[anal] (G.north west) -- (Q.south east);
\draw[anal] (G.north east) -- (NP.south west);
\end{tikzpicture}
\caption{Relations between the grabbag channel and other models. A solid arrow points from a class or a model to a special case of it. The two-headed arrow is the correspondence of \cref{subsec:model-variants}, under which the input is shifted by one use. A dashed line joins models with the same kind of store but a changed mechanism, and a dotted line marks an analogy only (\cref{sec:intro,sec:discussion}). The feedback results of~\cite{wuanastasopoulos,sabaggraph} are for an initial state known to both terminals.}
\label{fig:relations}
\end{figure}

The relation of the grabbag channel to other models is drawn in \cref{fig:relations}. Each node cites its sources, and the style of each line gives the kind of relation. The remarks of \cref{sec:discussion} explain them.

\subsection{Contributions}\label{subsec:contributions}
Our contributions are as follows. We prove an upper bound on the information that a feedback code conveys in $n$ uses, linear in $n$ with an additive term that does not depend on $n$, uniformly over the initial state and the code (\cref{thm:converse}). It gives $\CC\leq\CF<9.77(K/(M+1))^{2/3}$ whenever $K\leq M+1$. A nonfeedback code is a feedback code, so the same bound is the converse without feedback. For two colors, a bound with this $n(M+1)^{-2/3}$ dependence that holds for every length requires an additive term of order at least $M^{1/3}$ (\cref{prop:accumulation}).

We also prove a lower bound on the capacity without feedback that holds uniformly over the unknown initial composition (\cref{thm:lower}), and a lower bound that grows like $K^{2/3}$ (\cref{prop:kdependence}). Together, these results give the order $(K/M)^{2/3}$ of both capacities, with absolute constants, when $K\leq M+1$ and $M$ is large (\cref{cor:order}). \Cref{app:fsc} shows that the channel is indecomposable and relates the capacities to the finite-state-channel information formulas. The main results do not use these formulas.

\subsection{Proof Techniques}\label{subsec:techniques}
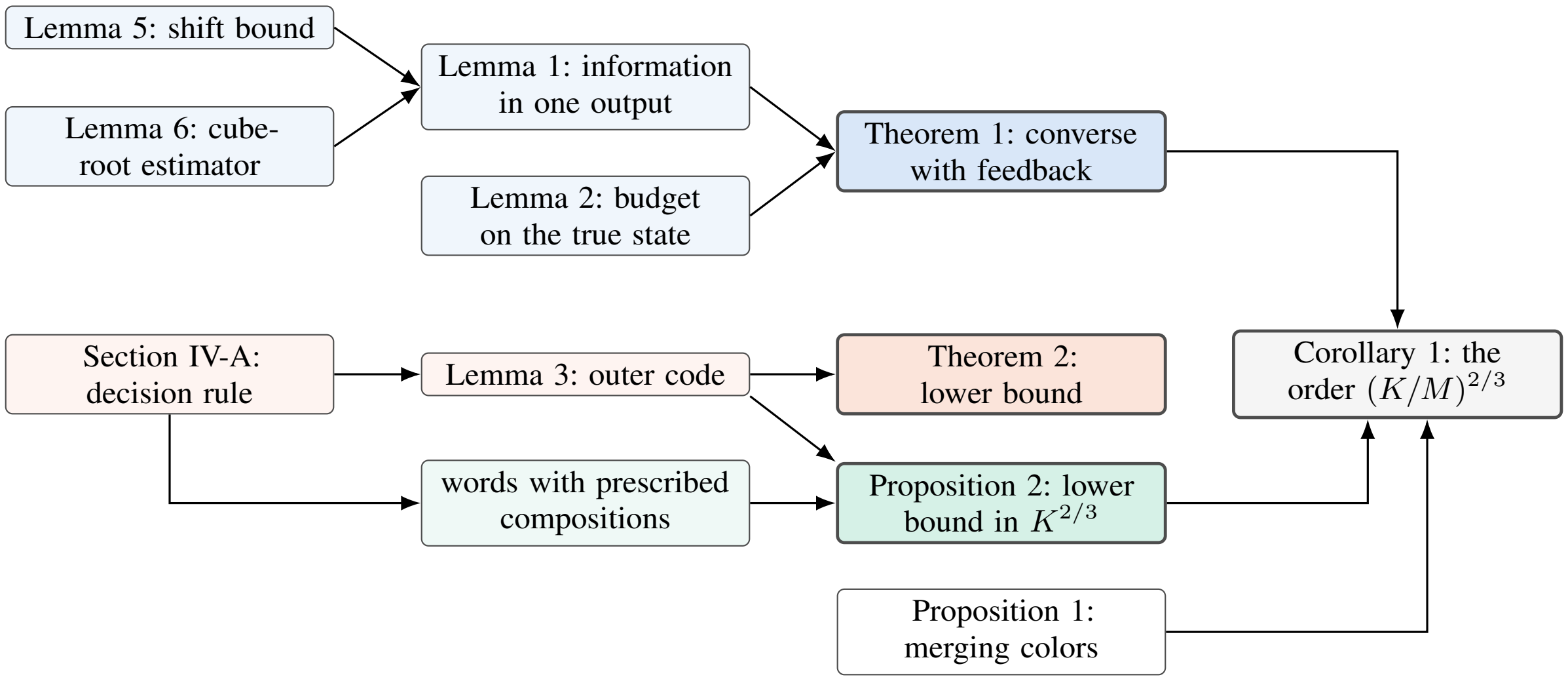
\begin{figure}[t]
\centering
\begin{tikzpicture}[x=1cm,y=1cm,>=Latex,
 box/.style={draw=black!70, rounded corners=2pt, align=center, font=\footnotesize, inner sep=3pt, text width=31mm},
 res/.style={box, line width=0.9pt, fill=black!4},
 arr/.style={->, line width=0.6pt}]
\node[box, fill=converse!7] (L1) at (0,0) {\cref{lem:shift}: shift bound};
\node[box, fill=converse!7] (L2) at (0,-1.2) {\cref{lem:cuberoot}: cube-root estimator};
\node[box, fill=converse!7] (L3) at (4.2,-0.6) {\cref{lem:oneuse}: information in one output};
\node[box, fill=converse!7] (L4) at (4.2,-1.9) {\cref{lem:budget}: budget on the true state};
\node[res, fill=converse!18] (T1) at (8.4,-1.25) {\cref{thm:converse}: converse with feedback};
\draw[arr] (L1.east) -- (L3.west); \draw[arr] (L2.east) -- (L3.west);
\draw[arr] (L3.east) -- (T1.west); \draw[arr] (L4.east) -- (T1.west);
\node[box, fill=lowerbound!7] (A1) at (0,-3.5) {\cref{subsec:inner}: decision rule};
\node[box, fill=lowerbound!7] (A2) at (4.2,-3.5) {\cref{lem:directcoding}: outer code};
\node[res, fill=lowerbound!18] (T2) at (8.4,-3.5) {\cref{thm:lower}: lower bound};
\node[box, fill=kbound!7] (A3) at (4.2,-4.8) {words with prescribed compositions};
\node[res, fill=kbound!18] (P2) at (8.4,-4.8) {\cref{prop:kdependence}: lower bound in $K^{2/3}$};
\node[box] (P1) at (8.4,-6.1) {\cref{prop:merging}: merging colors};
\node[res] (C1) at (12.4,-3.5) {\cref{cor:order}: the order $(K/M)^{2/3}$};
\draw[arr] (A1.east) -- (A2.west); \draw[arr] (A2.east) -- (T2.west);
\draw[arr] (A1.south) |- (A3.west); \draw[arr] (A3.east) -- (P2.west); \draw[arr] (A2.south east) -- (P2.north west);
\draw[arr] (T1.east) -| (C1.north); \draw[arr] (P2.east) -| ([xshift=-3mm]C1.south); \draw[arr] (P1.east) -| ([xshift=3mm]C1.south);
\end{tikzpicture}
\caption{The structure of the proofs. The converse (\cref{sec:converse}) and the lower bounds (\cref{sec:achievability}) are independent, and neither uses the finite-state-channel formulas of \cref{app:fsc}. The shift bound and cube-root estimator, together with
the proofs of the two converse lemmas, appear in
\cref{app:converse-lemmas}.}
\label{fig:roadmap}
\end{figure}

The converse and the lower bounds rest on different ideas. \Cref{fig:roadmap} shows how the lemmas combine. The converse bounds the information that each output carries about the message, given the earlier outputs. The chain rule for mutual information~\cite[Theorem~2.5.2]{covertho} then adds these bounds. Consider a connected unifilar finite-state channel whose initial state is known to both terminals. Its feedback capacity is the optimal average reward of a dynamic program on the receiver's posterior law of the state~\cite[Sec.~V]{trapdoorfb}, and a bounded solution of the Bellman equation of this program determines that reward~\cite[Theorem~5]{trapdoorfb}. A dynamic-programming framework for more general channels with feedback is developed in~\cite{tatikonda}. For the grabbag channel, the state space grows with $M$, and we do not solve such an equation.
Instead, we fix the initial state.
\Cref{lem:oneuse} bounds the information in each output
by the decrease of an explicit potential of the conditional
law of the state given the preceding outputs and the fixed
initial composition, plus three terms.
 These terms are the costs of the insertion and the removal, and the loss of an estimator of the output law. Here, the initial composition is fixed, and the conditional law is taken at an output history of positive probability. The potentials then telescope over the uses.

The potential is the mutual information between the composition and a noisy observation of it. The observation adds two to each color count, takes cube roots, and adds independent Laplace noise (\cref{subsec:converse-ingredients}). Adding independent continuous noise to a discrete random variable is a device used, for example, in~\cite[Problem~8.7]{covertho}. Here, it makes the potential a difference of two differential entropies. With the Laplace density, moving the cube roots changes the differential entropy by at most six times the expected $\ell_1$ distance moved (\cref{lem:shift}). A color is drawn with probability proportional to its count, and the removal of a ball of a common color moves its cube root little. Moreover, the cube roots of the counts after the insertion sum to at most $\min\{K,M+1\}^{2/3}(M+1)^{1/3}$ by Jensen's inequality~\cite[Theorem~2.6.2]{covertho}. This is the origin of the factor $(K/(M+1))^{2/3}$. The remaining information is bounded by comparing the conditional law of the output given the composition with an estimate computed from the noisy observation. The comparison uses the identity of~\cite[Eq.~(10)]{lapidothmoser}. The resulting relative entropy is bounded by a scaled sum of relative entropies between Poisson laws, and \cref{lem:cuberoot} bounds the expectation of each of them.

The insertion cost can be large when the inserted color is rare
in the bag. To control the accumulated cost, we introduce a
second potential: the sum of the cube roots of the retained
color counts, with one added to each count. Inserting a ball
increases this potential by an amount that controls the
insertion cost, while a random draw decreases it by only a
bounded amount in expectation. Since the potential itself is
bounded, summing its changes over time bounds the total
insertion cost by a term proportional to the number of channel
uses, plus a remainder independent of that number
(\cref{lem:budget}).
Together with the bound on the information in each output,
this argument gives an inequality of Bellman type: the
information gained at each use is controlled by changes in
bounded potentials and a remaining cost. This yields a
converse without requiring an explicit solution of the
Bellman equation. Fano's
inequality~\cite[Theorem~2.10.1]{covertho} then converts the
information bound into a capacity upper bound. The bound
holds uniformly over the initial composition, both under
the worst-initial-state criterion and when the initial
state is known to both terminals.

For the first lower bound, a single color is transmitted
for two consecutive intervals of $\tau$ uses, and the
decoder forms a weighted difference of their output
histograms (\cref{subsec:inner}).
 The unknown initial composition contributes to the expectation of each output a term that decays geometrically along the block. The weights cancel this term exactly. Thus, the expectation of the statistic does not depend on the initial composition, although its law does. A martingale representation bounds the second moment of the statistic uniformly in the number of colors. A second-moment bound~\cite[Ch.~IX, Sec.~6]{feller}
then shows that a block of length proportional to $M^{2/3}$
permits a color decision with uniformly small error probability,
regardless of the initial composition. Choosing the
proportionality constant sufficiently large makes this error
probability small enough for the outer code to correct the
remaining errors. The signal-to-noise calculation
in~\cref{eq:snr} explains the $M^{2/3}$ blocklength scale.

An outer code makes the error probability vanish (\cref{lem:directcoding}). The errors of different blocks are not independent, because the composition at the start of a block depends on the earlier outputs. The lemma needs only an error bound for each block that holds given the past. The uniformity of the decision rule over the initial composition provides this bound. The outer code is a random code~\cite[Sec.~5.5]{gallager} decoded within a Hamming radius. Conditional exponential moments bound the number of wrong block decisions, and a bound on the size of a Hamming ball controls the competing codewords. The proof of~\cite[Theorem~15.1.1]{alonspencer} uses the same code and decoder for independent binary errors. Without independence, the exponential moment is bounded by conditioning on the past one block at a time, as in the proof of Azuma's inequality~\cite[Theorem~7.2.1]{alonspencer}.

To use many colors at once, the same word is sent twice in the same order, and the message is carried by the composition of the word. Gilbert's greedy argument~\cite{gilbert1952} gives a number of compositions exponential in $\lfloor K/2\rfloor$ that are pairwise far apart. With the same statistic and outer code, they give the lower bound that grows like $K^{2/3}$ (\cref{prop:kdependence}). Merging colors shows that the capacities do not decrease when colors are added (\cref{prop:merging}). The upper and lower bounds then combine into \cref{cor:order}.

Of the tools above, the chain rule, Fano's inequality, the second-moment bound, random coding and Gilbert's argument are classical. The potential defined through noisy cube roots, the budget on the true state and the statistic whose expectation does not depend on the initial composition are constructed for the grabbag channel. The proofs rest on them.

\subsection{Organization}\label{subsec:organization}
\Cref{sec:main} defines the model and states the results. \Cref{sec:converse} proves the converse, and \cref{sec:achievability} the lower bounds. \Cref{sec:discussion} compares the model with its relatives and explains the exponent heuristically. \Cref{sec:conclusion} lists open problems. \Cref{app:fsc} establishes the finite-state-channel
properties, and \cref{app:converse-lemmas} contains the
technical estimates and proofs supporting the converse.

\section{Model, Definitions and Main Results}\label{sec:main}
\subsection{Notation}
All logarithms are natural, that is, to the base $\mathrm{e}$, and rates are in nats per channel use.  For an integer $K\geq1$, put
\begin{equation}
 [K]\coloneqq\{1,\ldots,K\}.
\end{equation}
For a set $B$ and a point $x$,
\begin{equation}
 \ind_B(x)\coloneqq
 \begin{cases}
  1&\text{if }x\in B,\\
  0&\text{otherwise},
 \end{cases}
\end{equation}
and, for an event $\mathcal E$ with complement $\mathcal E^\complement$, its indicator random variable is
\begin{equation}
 \ind_{\mathcal E}\coloneqq
 \begin{cases}
  1&\text{on }\mathcal E,\\
  0&\text{on }\mathcal E^\complement.
 \end{cases}
\end{equation} For $i\in[K]$, the $i$th standard basis vector $\ee_i$ of $\mathbb R^K$ has the coordinates
\begin{equation}
 (\ee_i)_j\coloneqq\ind_{\{i\}}(j),\qquad j\in[K].
\end{equation}
For $v,w\in\mathbb R^K$,
\begin{equation}
 \|v\|_1\coloneqq\sum_{i=1}^K|v_i|,\qquad
 \|v\|_2\coloneqq\Biggl(\sum_{i=1}^Kv_i^2\Biggr)^{1/2},\qquad
 \langle v,w\rangle\coloneqq\sum_{i=1}^Kv_iw_i.
\end{equation}
For a finite string $y^n=(y_1,\ldots,y_n)$ and $\ell\in\{0,1,\ldots,n\}$,
\begin{equation}
 y^\ell\coloneqq(y_1,\ldots,y_\ell),
\end{equation}
where $y^0$ is the empty string.

Random variables are written in upper case and their realizations
in lower case. We write $H(A)$ and $I(A\wedge B)$ for entropy
and mutual information, respectively, and $H(A\mid C)$ and
$I(A\wedge B\mid C)$ for their conditional counterparts.
Conditioning on $C=c$ refers to the corresponding quantities
under the conditional law given that realization, whenever
$\PP\{C=c\}>0$.
For probability laws $P$ and $Q$ on a countable set $\mathcal X$, and for probability densities $p_1$ and $p_2$ on $\mathbb R^K$, the relative entropies are
\begin{equation}
 D(P\,\Vert\,Q)\coloneqq\sum_{x\in\mathcal X}P(x)\log\frac{P(x)}{Q(x)},\qquad
 D(p_1\,\Vert\,p_2)\coloneqq\int_{\mathbb R^K}p_1(v)\log\frac{p_1(v)}{p_2(v)}\,\mathrm dv,
\end{equation}
the latter whenever the integral exists. In these sums and integrals, the conventions $0\log(0/a)=0$ for $a\geq0$ and $a\log(a/0)=\infty$ for $a>0$ are used~\cite[Ch.~2 and Sec.~8.5]{covertho}. For a random vector $A$ in $\mathbb R^K$ with a density $p$, its differential entropy is
\begin{equation}
 h(A)\coloneqq-\int_{\mathbb R^K}p(v)\log p(v)\,\mathrm dv
\end{equation}
whenever the integral exists. Integrals over $\mathbb R^K$ are with respect to Lebesgue measure. The binary entropy function is
\begin{equation}
 H_{\mathrm b}(u)\coloneqq-u\log u-(1-u)\log(1-u),\qquad u\in[0,1],
\end{equation}
and, for $p\in[0,1]$, the Bernoulli law $\Ber(p)$ is the probability law on $\{0,1\}$ given by
\begin{equation}
 \Ber(p)(1)\coloneqq p,\qquad\Ber(p)(0)\coloneqq1-p.
\end{equation}

\subsection{The channel}
Fix integers $K\geq2$ and $M\geq1$. A \emph{composition} of an integer $N\geq0$ into $K$ parts is a vector $z\in\mathbb Z_{\geq0}^K$ with $\sum_{i=1}^Kz_i=N$; the set of such vectors is
\begin{equation}\label{eq:states}
 \mathcal S_{N,K}\coloneqq\left\{z\in\mathbb Z_{\geq0}^K\;\middle|\;\sum_{i=1}^Kz_i=N\right\},\qquad|\mathcal S_{N,K}|=\binom{N+K-1}{K-1}.
\end{equation}
The \emph{state} of the grabbag channel is the composition of the $M$ retained balls, an element of $\mathcal S_{M,K}$. Given the current state $z\in\mathcal S_{M,K}$ and input color
$x\in[K]$, the output equals $i\in[K]$ with probability
\begin{equation}\label{eq:kernel}
 W(i\mid z,x)\coloneqq\frac{z_i+\ind_{\{i\}}(x)}{M+1}.
\end{equation}
If the realized output is $i$, the next state is
$z+\ee_x-\ee_i$, which belongs to $\mathcal S_{M,K}$ for
every output of positive probability. In words, the input ball is inserted into the bag, which then holds the composition $z+\ee_x\in\mathcal S_{M+1,K}$. One of the $M+1$ balls is drawn uniformly, its color is the output, and it is removed.

\begin{definition}[Codes]\label{def:codes}
Let $n\geq1$ and $J\geq1$ be integers. A \emph{feedback code} of length $n$ with $J$ messages consists of encoding functions $f_t:[J]\times[K]^{t-1}\to[K]$, $t=1,\ldots,n$, and a decoding function $\varphi:[K]^n\to[J]$. A \emph{nonfeedback code} is a feedback code whose encoding functions do not depend on their second argument. The message is a random variable $Q$ taking values in $[J]$. For a feedback code, an initial state $s\in\mathcal S_{M,K}$ and a realization $q\in[J]$ of $Q$, the output string has the law
\begin{equation}\label{eq:law}
 \PP\bigl\{Y^n=y^n\bigm|Q=q,S_0=s\bigr\}=\prod_{t=1}^nW\bigl(y_t\bigm|s_{t-1},x_t\bigr),
\end{equation}
where $x_t\coloneqq f_t(q,y^{t-1})$ and $s_t\coloneqq s_{t-1}+\ee_{x_t}-\ee_{y_t}$ for $t=1,\ldots,n$, with $s_0\coloneqq s$. The product is set to zero from the first factor that vanishes, so that $s_t$ is needed only along strings of positive probability, where it lies in $\mathcal S_{M,K}$. We write $X_t\coloneqq f_t(Q,Y^{t-1})$ and $S_t\coloneqq S_{t-1}+\ee_{X_t}-\ee_{Y_t}$ for the corresponding random inputs and states. A probability, an expectation or an information quantity conditioned on $S_0=s$ is computed under the joint law of $Q$ and $Y^n$ that~\cref{eq:law} defines with the initial state $s$. The \emph{error probability from $s$} and the \emph{error probability} of the code are
\begin{equation}\label{eq:error}
 P_e(s)\coloneqq\frac1J\sum_{q=1}^J\PP\bigl\{\varphi(Y^n)\neq q\bigm|Q=q,S_0=s\bigr\},
 \qquad
 P_e\coloneqq\max_{s\in\mathcal S_{M,K}}P_e(s).
\end{equation}
\end{definition}

In~\cref{eq:law}, the input at use $t$ depends on the message and on the outputs before use $t$ only. Thus, feedback, when present, is noiseless and strictly causal. The initial state is fixed but unknown to the encoder and the decoder: the criterion~\cref{eq:error} takes the worst initial state for the error averaged over equiprobable messages.

\begin{definition}[Capacities]\label{def:capacity}
A rate $R\geq0$ is \emph{achievable without feedback} if there is a sequence of nonfeedback codes, the $n$th of length $n$ with $J_n$ messages, such that $P_e\to0$ and $\liminf_{n\to\infty}n^{-1}\log J_n\geq R$. The \emph{nonfeedback capacity} $\CC$ is the supremum of the rates achievable without feedback. Replacing nonfeedback codes by feedback codes defines rates \emph{achievable with feedback} and the \emph{feedback capacity} $\CF$. For $s\in\mathcal S_{M,K}$, replacing $P_e$ by $P_e(s)$, and allowing the codes to depend on $s$, defines the \emph{feedback capacity with the initial state $s$ known to both terminals}, $C^{\mathrm{fb,known}}_{M,K}(s)$.
\end{definition}

Since a nonfeedback code is a feedback code, $\CC\leq\CF$. Moreover, $\CF\leq\log K$ (\cref{rem:converse-largeK}). In \cref{def:capacity}, the blocklength tends to infinity with $M$ and $K$ fixed. The results below describe how the capacities depend on $M$ and $K$.

\begin{example}[One retained ball]\label{ex:trapdoor}
Let $M=1$ and $K=2$, so that the state is the color of the one retained ball. By~\cref{eq:kernel}, the transitions are
\begin{center}
\renewcommand{\arraystretch}{1.15}
\begin{tabular}{@{}cccc@{}}
\toprule
Retained color & Input $x$ & Output $y$ (probability) & Next retained color\\
\midrule
color $i$ & color $i$ & color $i$ (1) & color $i$\\
color $i$ & color $j\neq i$ & color $i$ ($1/2$) & color $j$\\
color $i$ & color $j\neq i$ & color $j$ ($1/2$) & color $i$\\
\bottomrule
\end{tabular}
\end{center}
When the input matches the retained ball, the output is that color with certainty. Otherwise, each of the two balls is drawn with probability $1/2$, and the other one stays. This is Blackwell's trapdoor channel~\cite{blackwell1961}.
\end{example}

\begin{example}[Two colors]\label{ex:twocolors}
For $K=2$, the state is determined by the number $S_{t,1}\in\{0,\ldots,M\}$ of retained balls of color~1. By~\cref{eq:kernel}, the output is color~1 with probability $\bigl(S_{t-1,1}+\ind_{\{X_t=1\}}\bigr)/(M+1)$, and $S_{t,1}=S_{t-1,1}+\ind_{\{X_t=1\}}-\ind_{\{Y_t=1\}}$. If the bag holds only balls of color~2 and the input is always color~2, the output is always color~2. If the input then switches to color~1, an output of color~1 has probability $(S_{t-1,1}+1)/(M+1)$, which grows only slowly as balls of color~1 accumulate. A receiver that does not know $S_{0,1}$ cannot tell whether a run of outputs of one color comes from the inputs or from the initial composition. The decision rule of \cref{subsec:inner} overcomes this with a number of uses per color of order $(M+1)^{2/3}$.
\end{example}

\subsection{Main results}
Put
\begin{equation}\label{eq:gamma}
 \gamma\coloneqq\frac{11}{12}-\frac{H_{\mathrm b}(1/12)}{\log2},\qquad
 \eta\coloneqq\log2-H_{\mathrm b}\left(\frac14\right),\qquad
 c_0\coloneqq2^{-46/3}\,\frac{\gamma\eta}{72}.
\end{equation}
Numerically, $\gamma\eta/72>9\times10^{-4}$ and $c_0>2.2\times10^{-8}$. These constants come from the lower bounds. For the outer code of \cref{lem:directcoding} with $K$ letters and $p_0=1/12$, the numerator of the rate in~\cref{eq:directcoding-rate} is at least $\gamma\log K$, by~\cref{eq:positiveconstant}. The exponent $\eta$ is that of Gilbert's bound for binary words at relative distance $1/4$, which counts the compositions used in \cref{prop:kdependence} (see~\cref{eq:kdep-gv}). The constant $c_0$ adds the loss of passing to the uniform bound of \cref{cor:order}, explained after it.

The first result is a converse for feedback codes. It holds for every number of colors and every size of the bag, and it applies to nonfeedback codes as well.

\begin{theorem}[Converse]\label{thm:converse}
For every $K\geq2$, every $M\geq1$, every feedback code of length $n$ with $J$ messages, every probability law of the message $Q$ on $[J]$ and every initial state $s\in\mathcal S_{M,K}$,
\begin{equation}\label{eq:converse-finite-thm}
 I\bigl(Q\wedge Y^n\bigm|S_0=s\bigr)\leq n\bigl(4\lambda_{K,M}+B_{K,M}\bigr)+6\varrho_{K,M},
\end{equation}
where
\begin{align}
 \lambda_{K,M}&\coloneqq\left(\frac{\min\{K,M+1\}}{M+1}\right)^{2/3},\nonumber\\
 B_{K,M}&\coloneqq\frac{95}{24}\,\frac{K^{2/3}(M+1+2K)^{1/3}}{M+1},\nonumber\\
 \varrho_{K,M}&\coloneqq K^{2/3}(M+K)^{1/3}-(M+1)^{1/3}-(K-1).
\end{align}
Consequently, $\CC\leq\CF\leq4\lambda_{K,M}+B_{K,M}$, and, if $K\leq M+1$,
\begin{equation}\label{eq:converse-capacity}
 \CC\leq\CF\leq4\bigl(1+3^{1/3}\bigr)\left(\frac K{M+1}\right)^{2/3}<9.77\left(\frac K{M+1}\right)^{2/3}.
\end{equation}
The same bounds hold for the capacity with the initial state known to both terminals.
\end{theorem}

In~\cref{eq:converse-finite-thm}, $2\lambda_{K,M}$ bounds the cost of the draw in each use. By \cref{lem:budget}, the insertions cost at most another $2\lambda_{K,M}$ per use on average, up to the remainder $6\varrho_{K,M}$, where $\varrho_{K,M}$ bounds the range of a potential that pays for them. The term $B_{K,M}$ bounds the expected relative-entropy loss of an estimator of the output law. Its factor $95/24$ comes from \cref{lem:cuberoot} (\cref{sec:converse}).

For $K=2$, \cref{eq:converse-capacity} reads
\begin{equation}
 C^{\mathrm{fb}}_{M,2}\leq\bigl(4\cdot12^{1/3}+2^{8/3}\bigr)(M+1)^{-2/3}<15.51\,(M+1)^{-2/3}
\end{equation}
for every $M\geq1$.

The second result is a lower bound without feedback, uniform over the unknown initial state.

\begin{theorem}[Lower bound]\label{thm:lower}
Let $K\geq2$, and let $\tau\coloneqq\bigl\lceil8(M+1)^{2/3}\bigr\rceil$ satisfy $\tau\leq(M+1)/4$. Then
\begin{equation}\label{eq:lower-exact}
 \CF\geq\CC\geq\frac{\log K-H_{\mathrm b}(1/12)-\frac1{12}\log(K-1)}{2\tau}.
\end{equation}
In particular,
\begin{equation}\label{eq:lower-asym}
 \CF\geq\CC\geq\frac\gamma{18}\,(\log K)\,(M+1)^{-2/3}.
\end{equation}
\end{theorem}

\begin{remark}\label{rem:threshold}
The condition $\tau\leq(M+1)/4$ holds for every
$M\geq2^{16}$. Indeed,
\begin{equation}
 \tau=\bigl\lceil8(M+1)^{2/3}\bigr\rceil
 \leq9(M+1)^{2/3}
 \leq\frac{M+1}{4},
\end{equation}
where the last inequality follows from
$(M+1)^{1/3}\geq36$.
\end{remark}


The third result shows that the capacities grow like a power of $K$ when $K$ is small compared with $M$. It uses the monotonicity in the number of colors.

\begin{proposition}[Merging colors]\label{prop:merging}
For $2\leq k\leq K$ and every $M\geq1$, $C_{M,k}\leq C_{M,K}$ and $C^{\mathrm{fb}}_{M,k}\leq C^{\mathrm{fb}}_{M,K}$.
\end{proposition}

\begin{proposition}[A lower bound that grows like $K^{2/3}$]\label{prop:kdependence}
Let $K\geq2$, $r\coloneqq\lfloor K/2\rfloor$ and $M+1\geq2^{22}r$. Then
\begin{equation}\label{eq:kdep}
 \CF\geq\CC\geq\frac{\gamma\eta}{72}\left(\frac r{M+1}\right)^{2/3}.
\end{equation}
\end{proposition}

The condition $M+1\geq2^{22}r$ plays the role that $M\geq2^{16}$ plays for \cref{thm:lower} (\cref{rem:threshold}). It makes the blocks of the proof satisfy $\tau\leq(M+1)/4$.

\begin{corollary}[The order]\label{cor:order}
For $M+1\geq2^{22}$ and $2\leq K\leq M+1$,
\begin{equation}\label{eq:orderK}
 c_0\left(\frac K{M+1}\right)^{2/3}\leq\CC\leq\CF\leq9.77\left(\frac K{M+1}\right)^{2/3}.
\end{equation}
In particular, for every fixed $K\geq2$, both capacities are $\Theta(M^{-2/3})$ as $M\to\infty$.
\end{corollary}

The constants in~\cref{eq:orderK} differ by a factor of about $4.4\times10^8$. For two colors, \cref{eq:lower-exact}, \cref{rem:threshold} and~\cref{eq:converse-capacity} give, for every $M\geq2^{16}$,
\begin{equation}\label{eq:twocolor-gap}
 0.0253\,(M+1)^{-2/3}\leq C_{M,2}\leq C^{\mathrm{fb}}_{M,2}\leq15.51\,(M+1)^{-2/3},
\end{equation}
where the lower constant uses $\tau\leq8(M+1)^{2/3}+1\leq8.001(M+1)^{2/3}$. The ratio of the constants is near $613$ (\cref{fig:bounds}). The lower-bound constants reflect the block length $2\tau$, the second-moment error estimate and the outer-code threshold $p_0=1/12$. In passing from \cref{prop:kdependence} to the uniform bound of \cref{cor:order}, the reduction to fewer colors by merging (\cref{prop:merging}) introduces the factor $2^{-46/3}$.

\begin{remark}\label{rem:scope}
The results determine the order of both capacities, with absolute constants, when $K\leq M+1$ and $M$ is large. They do not determine the leading constants, and they do not determine whether $\CF/\CC\to1$. By~\cref{eq:orderK}, feedback can increase the capacity by at most a constant factor in this range. For $K>M+1$, \cref{thm:converse} gives no bound of order $(K/M)^{2/3}$ (\cref{rem:converse-largeK}). For $K=1$, both capacities are zero.
\end{remark}

The exponent $-2/3$ has a signal-to-noise interpretation. Repeating a color for $\tau$ uses changes its output probability by at most of order $\tau/M$. For a difference of two suitably weighted output histograms, the contribution of the unknown initial composition to the expectation cancels. Its expectations for two different colors are separated by order $\tau^2/M$, whereas its standard deviation is at most of order $\sqrt\tau$. The squared separation divided by this bound on the variance is of order
\begin{equation}\label{eq:snr}
 \frac{(\tau^2/M)^2}\tau=\frac{\tau^3}{M^2}.
\end{equation}
A constant-quality color decision is therefore possible when $\tau$ is of order $M^{2/3}$, and the resulting rate is of order $(\log K)/\tau$. This explains the lower bound. The converse is a different argument, outlined at the start of \cref{sec:converse}.
\begin{figure}[t]
\centering
\begin{tikzpicture}
\begin{loglogaxis}[name=main,
 width=0.82\textwidth, height=0.46\textwidth,
 xmin=1.5, xmax=1e13, ymin=1e-11, ymax=3,
 xlabel={$M+1$ (number of balls after insertion)},
 ylabel={rate (nats per use)},
 xtick={1e1,1e3,1e5,1e7,1e9,1e11,1e13},
 ytick={1e-10,1e-8,1e-6,1e-4,1e-2,1},
 legend pos=south west, legend cell align=left,
 legend style={font=\footnotesize,draw=black!40},
 tick label style={font=\footnotesize}, label style={font=\small},
 grid=major, grid style={black!10}]
\addplot[name path=bandup, domain=65537:1e13, samples=60, draw=none, forget plot] {15.51*x^(-2/3)};
\addplot[name path=bandlo, domain=65537:1e13, samples=60, draw=none, forget plot] {0.0253*x^(-2/3)};
\addplot[black!12, forget plot] fill between[of=bandup and bandlo];
\addplot[domain=2:1e13, samples=60, conv] {15.51*x^(-2/3)};
\addlegendentry{upper bound (\cref{thm:converse})}
\addplot[domain=65537:1e13, samples=60, lowb] {0.0253*x^(-2/3)};
\addlegendentry{lower bound (\cref{thm:lower}, $M\geq2^{16}$)}
\addlegendimage{area legend, draw=none, fill=black!12}
\addlegendentry{band holding $C_{M,2}$ and $C^{\mathrm{fb}}_{M,2}$}
\addplot[domain=1.5:1e13, samples=2, thin, densely dotted, black] {ln(2)};
\addlegendentry{trivial bound $\log2$}
\addplot[only marks, mark=*, mark size=1.4pt, black] coordinates {(2,0.481212)};
\addlegendentry{trapdoor, feedback, initial state known~\cite{trapdoorfb}}
\addplot[only marks, mark=-, mark size=2.5pt, black, error bars/.cd, y dir=both, y explicit] coordinates {(2,0.400973) +- (0,0.004493)};
\addlegendentry{trapdoor, no feedback: analytical bounds~\cite{computable}}
\draw[densely dotted, thick, lowerbound] (axis cs:65537,1e-6) -- (axis cs:65537,3);
\node[font=\footnotesize, align=left, anchor=north west] at (axis cs:9e4,2e-1) {\cref{thm:lower} needs $\tau\leq(M+1)/4$.\\\Cref{rem:threshold} guarantees it for $M\geq2^{16}$.};
\end{loglogaxis}
\node[font=\small, anchor=north] (capa) at (main.outer south) {(a) two colors};
\begin{axis}[name=trap, at={($(capa.south)+(0,-0.5cm)$)}, anchor=north,
 width=0.82\textwidth, height=0.22\textwidth,
 xmin=0.39, xmax=0.49, ymin=0.4, ymax=3.6, ytick=\empty,
 xtick={0.39,0.41,0.43,0.45,0.47,0.49},
 xticklabel style={/pgf/number format/fixed, /pgf/number format/precision=2},
 xlabel={rate for $M=1$, $K=2$ (nats per use)},
 tick label style={font=\footnotesize}, label style={font=\small},
 xmajorgrids, grid style={black!10}]
\addplot[only marks, mark=*, mark size=1.6pt, black] coordinates {(0.481212,3)};
\node[font=\footnotesize, anchor=east] at (axis cs:0.4785,3) {feedback, initial state known~\cite{trapdoorfb}};
\addplot[only marks, mark=*, mark size=1.6pt, black] coordinates {(0.405465,2)};
\node[font=\footnotesize, anchor=west] at (axis cs:0.4085,2) {feedback delayed by two uses, initial state known~\cite{delayed}};
\addplot[black, line width=0.9pt, mark=|, mark size=3pt] coordinates {(0.396480,1) (0.405465,1)};
\addplot[only marks, mark=|, mark size=5pt, line width=0.9pt, black] coordinates {(0.399599,1)};
\node[font=\footnotesize, anchor=west] at (axis cs:0.4085,1) {no feedback: analytical bounds~\cite{computable}, numerical upper bound~\cite{delayed} (long tick)};
\end{axis}
\node[font=\small, anchor=north] at (trap.outer south) {(b) the trapdoor channel};
\end{tikzpicture}
\caption{(a) The bounds of~\cref{eq:twocolor-gap} for two colors against $M+1$ on logarithmic axes, with the trivial bound $\log2$ and the values for the trapdoor channel at $M+1=2$. Both bounds decay with slope $-2/3$. For $M\geq2^{16}$, both capacities lie in the shaded band, whose vertical extent is the factor of about $613$ between the constants. The upper bound is below $\log2$ for $M+1\geq106$. The lower bound is shown from $M=2^{16}$ (the dotted line), the sufficient range of \cref{rem:threshold} in which~\cref{eq:twocolor-gap} gives its coefficient $0.0253$. This is not the smallest range of \cref{thm:lower}. Its hypothesis $\tau\leq(M+1)/4$ fails for $M+1<2^{15}$, where $\tau\geq8(M+1)^{2/3}>(M+1)/4$. (b) The values of \cref{tab:prior} for the trapdoor channel ($M=1$, $K=2$) on a linear scale: the feedback capacity and the capacity with feedback delayed by two uses, both with the initial state known to both terminals, and, without feedback, the interval between the analytical bounds with the numerical upper bound inside it. The curves and marks are proved bounds or known values, not estimates of the capacities.}
\label{fig:bounds}
\end{figure}

\section{Proof of the Converse}\label{sec:converse}
This section proves \cref{thm:converse} by combining a bound
on the information in each output with a bound on the
accumulated insertion cost. Throughout this section,
$K\geq2$ and $M\geq1$ are fixed integers.


\subsection{Two ingredients for the converse}
\label{subsec:converse-ingredients}

The converse rests on two estimates. The first bounds the
information in one output by the decrease of a potential of
the state distribution, an insertion cost, and a term that
depends only on $M$ and $K$. The second controls the accumulated
insertion cost along any feedback code. We state these
estimates here and combine them in
\cref{subsec:converse-proof}. Their proofs are given in
\cref{app:converse-lemmas}.

Let $Z=(Z_1,\ldots,Z_K)$ have independent coordinates with
density $3\mathrm e^{-6|v|}$, $v\in\mathbb R$. Its joint density is
\begin{equation}\label{eq:laplace}
 p_Z(v)=\prod_{i=1}^K3\mathrm e^{-6|v_i|},
 \qquad v\in\mathbb R^K.
\end{equation}
Define the cube-root map
\begin{equation}\label{eq:cuberootmap}
 g(z)\coloneqq
 \bigl((z_1+2)^{1/3},\ldots,(z_K+2)^{1/3}\bigr),
 \qquad z\in\mathbb Z_{\geq0}^K.
\end{equation}
For a finitely supported probability law $\nu$ on
$\mathbb Z_{\geq0}^K$, let $R\sim\nu$ be independent of $Z$,
put $V_\nu\coloneqq g(R)+Z$, and define
\begin{equation}\label{eq:potential}
 \Phi(\nu)\coloneqq I(R\wedge V_\nu)
 =h(V_\nu)-h(Z).
\end{equation}
The equality follows because, conditional on $R$, the random
vector $V_\nu$ is a translate of $Z$. Here, $I(R\wedge V_\nu)$ is the mutual information of the discrete $R$ and the continuous $V_\nu$, the average $\sum_{z\in\mathbb Z_{\geq0}^K:\,\nu(z)>0}\nu(z)D(p_z\,\Vert\,p_\nu)$ of relative entropies between densities written out in~\cref{eq:potential-divergence} of \cref{app:converse-lemmas}. The differential
entropies in \cref{eq:potential} are finite, as verified in
\cref{app:converse-lemmas}. Thus, $\Phi(\nu)$ measures how much
information about the count vector remains in its noisy
cube-root observation. The elementary bounds on mutual
information give
\begin{equation}\label{eq:potential-range}
 0\leq\Phi(\nu)\leq H(R).
\end{equation}
In particular, $\Phi(\nu)=0$ whenever $\nu$ is a point mass.

The shift by two in $g$ and the scale of $Z$ have separate roles. Adding two to each count keeps every coordinate of $g$ at least $2^{1/3}$, so that the estimator of \cref{lem:cuberoot}, which needs $\theta\geq1$, applies to every color, including the colors absent from the bag. The scale $1/6$ of the Laplace noise gives the factor $6$ in the shift bound of \cref{lem:shift}. This factor multiplies the costs of the insertion and the draw. The moments of this noise give the factor $95/24$ in $B_{K,M}$ (\cref{lem:cuberoot}).

To state the first estimate, consider one channel use.
Let $\pi$ be a probability law of a pair $(S,X)$ on
$\mathcal S_{M,K}\times[K]$, and let
$U\coloneqq S+\ee_X\in\mathcal S_{M+1,K}$ be the composition
after insertion and before removal. Given $(S,X)$, generate
the output $Y$ according to
\begin{equation}\label{eq:oneuse-law}
 \PP\{Y=i\mid S=s,X=x\}
 =\frac{s_i+\ind_{\{i\}}(x)}{M+1},
 \qquad i\in[K].
\end{equation}
This conditional law depends on $(S,X)$ only through $U$.
Consequently, $Y$ and $(S,X)$ are conditionally independent
given $U$, and
$\PP\{Y=i\mid U\}=U_i/(M+1)$.

Write $\alpha$ for the law of $S$ and
$q_i\coloneqq\PP\{Y=i\}$ for the output probabilities.
For each $i$ with $q_i>0$, let $\alpha^i$ be the conditional
law of the next retained composition $U-\ee_i$ given $Y=i$.
Thus, $\Phi(\alpha)$ is the potential before the channel use,
whereas $\sum_{i\in[K]:\,q_i>0}q_i\Phi(\alpha^i)$ is its expectation
after observing the output and removing the drawn ball.
Put
 \begin{align}
 a(\pi)
 &\coloneqq\E\bigl[(S_X+2)^{-2/3}\bigr],\nonumber\\
 \lambda_{K,M}
 &=
 \left(\frac{\min\{K,M+1\}}{M+1}\right)^{2/3},\nonumber\\
 B_{K,M}
 &=
 \frac{95}{24}\,
 \frac{K^{2/3}(M+1+2K)^{1/3}}{M+1}. \label{eq:oneuse-terms}
 \end{align}
Here, $S_X$ is the retained count of the inserted color
before insertion, and $\lambda_{K,M}$ and $B_{K,M}$ are the constants of \cref{thm:converse}.

\begin{lemma}[Information in one output]\label{lem:oneuse}
For every probability law $\pi$ of $(S,X)$ on
$\mathcal S_{M,K}\times[K]$,
\begin{equation}\label{eq:oneuse}
 I(U\wedge Y)
 \leq
 \Phi(\alpha)
 -\sum_{i\in[K]:\,q_i>0}q_i\Phi(\alpha^i)
 +2a(\pi)+2\lambda_{K,M}+B_{K,M}.
\end{equation}
\end{lemma}

When \cref{lem:oneuse} is applied conditionally on the
preceding outputs, its potential terms telescope after
averaging and summing over channel uses. The remaining
difficulty is the insertion cost $2a(\pi)$, which can be
large when the inserted color is rare in the bag.
The next lemma controls its accumulated contribution by
tracking a second potential, this time evaluated at the
actual retained composition.

\begin{lemma}[A budget on the true state]\label{lem:budget}
Fix a feedback code of length $n$, a probability law of the
message $Q$, and an initial composition
$s\in\mathcal S_{M,K}$. Define
\begin{equation}\label{eq:state-budget-potential}
 F(z)\coloneqq\sum_{i=1}^K(z_i+1)^{1/3},
 \qquad z\in\mathcal S_{M,K}.
\end{equation}
Then
 \begin{align}
 \sum_{t=1}^n
 \E\Bigl[\bigl(S_{t-1,X_t}+2\bigr)^{-2/3}\Bigr]
 &\leq
 n\lambda_{K,M}
 +3\bigl(\E[F(S_n)]-F(s)\bigr)\nonumber\\
 &\leq
 n\lambda_{K,M}+3\varrho_{K,M},\label{eq:budget}
 \end{align}
where $S_{t-1,X_t}$ denotes the coordinate $X_t$ of
$S_{t-1}$, all expectations are taken with $S_0=s$, and, as in \cref{thm:converse},
\begin{equation}\label{eq:varrho}
 \varrho_{K,M}
 =
 K^{2/3}(M+K)^{1/3}-(M+1)^{1/3}-(K-1).
\end{equation}
\end{lemma}

The bound in \cref{eq:budget} consists of a term proportional
to the number of uses and a remainder independent of that
number. The cube roots in $F$ add one, not two, to each count. Then inserting a ball of color $x$ into a composition $z$ raises $F$ by at least $\frac13(z_x+2)^{-2/3}$, one third of the quantity averaged in $a(\pi)$ (\cref{app:budget-proof}). Together, \cref{lem:oneuse,lem:budget} provide the
two estimates needed for the converse.

\subsection{Proof of \texorpdfstring{\cref{thm:converse}}{Theorem \getrefnumber{thm:converse}}}\label{subsec:converse-proof}
\begin{IEEEproof}
Fix a feedback code of length $n$ with $J$ messages and an initial composition $s\in\mathcal S_{M,K}$, and let the message $Q$ have any probability law on $[J]$. All probabilities in this proof are taken with $S_0=s$. Since the encoder is deterministic and $s$ is fixed, the composition $U_t\coloneqq S_{t-1}+\ee_{X_t}$ is a function of $(Q,Y^{t-1})$, and, given $(Q,Y^{t-1})$, the output $Y_t$ equals $i$ with probability $U_{t,i}/(M+1)$. Hence, for each $y^{t-1}$ of positive probability, $Q\to U_t\to Y_t$ is a Markov chain given $Y^{t-1}=y^{t-1}$, with $U_t$ a function of $Q$, and
\begin{equation}\label{eq:converse-peruse}
 I(Q\wedge Y_t\mid Y^{t-1}=y^{t-1})=I(U_t\wedge Y_t\mid Y^{t-1}=y^{t-1})
\end{equation}
by the data-processing inequality applied in both directions~\cite[Theorem~2.8.1]{covertho}.

Let $\pi_t(y^{t-1})$ be the conditional law of $(S_{t-1},X_t)$ given $Y^{t-1}=y^{t-1}$, and let $\alpha_t(y^{t-1})$ be the conditional law of $S_{t-1}$ given $Y^{t-1}=y^{t-1}$. Apply \cref{lem:oneuse} to $\pi=\pi_t(y^{t-1})$. Its law $\alpha$ is $\alpha_t(y^{t-1})$. Its law $\alpha^i$ is the conditional law of $S_t=U_t-\ee_{Y_t}$ given $(Y^{t-1},Y_t)=(y^{t-1},i)$, which is $\alpha_{t+1}(y^{t-1},i)$. Its $q_i$ is $\PP\{Y_t=i\mid Y^{t-1}=y^{t-1}\}$. Moreover, $\E\bigl[a\bigl(\pi_t(Y^{t-1})\bigr)\bigr]=\E\Bigl[\bigl(S_{t-1,X_t}+2\bigr)^{-2/3}\Bigr]$. Averaging~\cref{eq:oneuse} over $Y^{t-1}$ and using~\cref{eq:converse-peruse}, we obtain
\begin{equation}\label{eq:converse-peruse-bound}
 I(Q\wedge Y_t\mid Y^{t-1})\leq\E\bigl[\Phi(\alpha_t(Y^{t-1}))\bigr]-\E\bigl[\Phi(\alpha_{t+1}(Y^{t-1},Y_t))\bigr]+2\,\E\Bigl[\bigl(S_{t-1,X_t}+2\bigr)^{-2/3}\Bigr]+2\lambda_{K,M}+B_{K,M}.
\end{equation}
Therefore,
\begin{align}
 I(Q\wedge Y^n)
 &\overset{\text{(a)}}{=}\sum_{t=1}^nI(Q\wedge Y_t\mid Y^{t-1})\nonumber\\
 &\overset{\text{(b)}}{\leq}\Phi(\alpha_1)-\E\bigl[\Phi(\alpha_{n+1}(Y^n))\bigr]+2\sum_{t=1}^n\E\Bigl[\bigl(S_{t-1,X_t}+2\bigr)^{-2/3}\Bigr]+n\bigl(2\lambda_{K,M}+B_{K,M}\bigr)\nonumber\\
 &\overset{\text{(c)}}{\leq}2\sum_{t=1}^n\E\Bigl[\bigl(S_{t-1,X_t}+2\bigr)^{-2/3}\Bigr]+n\bigl(2\lambda_{K,M}+B_{K,M}\bigr)\nonumber\\
 &\overset{\text{(d)}}{\leq}n\bigl(4\lambda_{K,M}+B_{K,M}\bigr)+6\varrho_{K,M},\label{eq:converse-finite}
\end{align}
where (a) follows from the chain rule for mutual information~\cite[Theorem~2.5.2]{covertho}; (b) follows from~\cref{eq:converse-peruse-bound}, whose potential terms telescope; (c) follows from $\Phi(\alpha_1)=0$, since $\alpha_1$ is the point mass at $s$, and from $\Phi\geq0$ by~\cref{eq:potential-range}; and (d) follows from \cref{lem:budget}. This proves~\cref{eq:converse-finite-thm}.

Now let $Q$ be uniform on $[J]$, and let $P_e$ be the error probability of the code in \cref{def:codes}, which bounds its error probability from every initial composition. Fano's inequality~\cite[Theorem~2.10.1]{covertho} and $H(Q)=\log J$ give
\begin{equation}\label{eq:converse-fano}
 (1-P_e)\log J\leq I(Q\wedge Y^n)+\log2\leq n\bigl(4\lambda_{K,M}+B_{K,M}\bigr)+6\varrho_{K,M}+\log2.
\end{equation}
For a sequence of codes with $P_e\to0$, dividing~\cref{eq:converse-fano} by $n$ and letting $n\to\infty$ with $M$ and $K$ fixed shows that every achievable rate is at most $4\lambda_{K,M}+B_{K,M}$. Since~\cref{eq:converse-finite} holds for every code and every initial composition, the same argument applies to codes designed for an initial composition known to both terminals. A nonfeedback code is a feedback code whose encoding functions do not depend on the past outputs, so $C_{M,K}\leq C^{\mathrm{fb}}_{M,K}$.

If $K\leq M+1$, then $\lambda_{K,M}=(K/(M+1))^{2/3}$ and $M+1+2K\leq3(M+1)$, so that
\begin{equation}\label{eq:converse-constant}
 4\lambda_{K,M}+B_{K,M}\overset{\text{(a)}}{\leq}4\left(\frac{K}{M+1}\right)^{2/3}+4\cdot\frac{K^{2/3}(3(M+1))^{1/3}}{M+1}=4\bigl(1+3^{1/3}\bigr)\left(\frac{K}{M+1}\right)^{2/3},
\end{equation}
where (a) uses $95/24<4$. This proves~\cref{eq:converse-capacity} and completes the proof of \cref{thm:converse}.
\end{IEEEproof}

\begin{remark}\label{rem:converse-largeK}
When $K>M+1$, the term $B_{K,M}$ grows like $K/(M+1)$, and \cref{thm:converse} gives no bound of order $(K/(M+1))^{2/3}$. The trivial bound $\CF\leq\log K$ follows from $I(Q\wedge Y^n)\leq H(Y^n)\leq n\log K$ and Fano's inequality. For each fixed $M$, it is smaller than $4\lambda_{K,M}+B_{K,M}$ when $K$ is large, whereas when $M$ and $K$ grow together either bound may be the smaller.
\end{remark}

\subsection{The finite-length remainder}\label{subsec:remainder}
The additive term $6\varrho_{K,M}$ in~\cref{eq:converse-finite-thm} is at most $6\cdot2^{1/3}K^{2/3}(M+1)^{1/3}$ when $K\leq M+1$. For two colors it is of order $M^{1/3}$, and the following proposition shows that a term of this order is necessary in a bound that holds for every length.

\begin{proposition}[Necessity of the finite-length remainder]\label{prop:accumulation}
Let $K=2$, and let the initial bag hold $M$ balls of color~$2$. For $0<\xi<1$ and $n=\lfloor\xi (M+1)\rfloor$, there are nonfeedback codes of length $n$ such that
\begin{equation}\label{eq:accumulation}
 I(Q\wedge Y^n)\geq\left(\frac7{256}\xi^{2/3}-o(1)\right)(M+1)^{1/3}\qquad\text{as }M\to\infty.
\end{equation}
Consequently, for every $c_1>0$, a bound $I(Q\wedge Y^n)\leq c_1n(M+1)^{-2/3}+R(M)$ that holds for every $n$ and every code with a known initial composition requires $R(M)\geq cM^{1/3}$ for some $c>0$ and all large $M$.
\end{proposition}

\begin{IEEEproof}
Put $\tau\coloneqq\Bigl\lceil8\bigl((M+1)n\bigr)^{1/3}\Bigr\rceil$ and $L\coloneqq\lfloor n/(2\tau)\rfloor$, and let the message be $Q=(Q_1,\ldots,Q_L)$, uniform on $\{0,1\}^L$. Block $k$ consists of the uses $2(k-1)\tau+1,\ldots,2k\tau$. If $Q_k=1$, the block inserts color~1 for $\tau$ uses and then color~2 for $\tau$ uses. If $Q_k=0$, the order is reversed. The remaining uses insert color~2. Each block inserts $\tau$ balls of color~1, so the number $Z^{(k)}$ of balls of color~1 at the start of block $k$ is a function of the earlier outputs, and $Z^{(k)}+\tau\leq L\tau\leq n/2$. For large $M$, also $\tau\leq (M+1)/2$.

Fix $k$, condition on $Z^{(k)}=z$ and $Q_k=a$, and let $T$ be the number of outputs of color~1 in the first half of block $k$. Put $\rho\coloneqq1-1/(M+1)$. At the $r$th use of this half, let $V_r$ be the number of balls of color~1 after the insertion, and let $\Delta_r\coloneqq\ind_{\{Y=1\}}-V_r/(M+1)$ for the output $Y$ of that use. The $\Delta_r$ are martingale differences with conditional variances at most $V_r/(M+1)\leq(z+\tau)/(M+1)$, and the number of balls of color~1 after the $r$th draw is $\rho V_r-\Delta_r$. Here, the conditional expectations and variances are taken given $Z^{(k)}=z$, $Q_k=a$ and the outputs preceding the current draw in this half-block. Taking expectations and summing,
\begin{equation}\label{eq:accum-mean}
 \E[T]=(1-\rho^\tau)z+a\,d_\tau,\qquad d_\tau\coloneqq\sum_{j=1}^\tau(1-\rho^j)\overset{\text{(a)}}{\geq}\sum_{j=1}^\tau\frac j{2(M+1)}\geq\frac{\tau^2}{4(M+1)},
\end{equation}
where (a) follows from $1-\rho^j\geq j\rho^{(M+1)/2}/(M+1)\geq j/(2(M+1))$ for $j\leq (M+1)/2$. Unrolling the recursion gives $T-\E[T]=\sum_{r=1}^\tau\rho^{\tau-r}\Delta_r$, so that
\begin{equation}\label{eq:accum-var}
 \E\Bigl[\bigl(T-\E[T]\bigr)^2\Bigr]\overset{\text{(a)}}{=}\sum_{r=1}^\tau\rho^{2(\tau-r)}\E\bigl[\Delta_r^2\bigr]\overset{\text{(b)}}{\leq}\frac{\tau(z+\tau)}{M+1}\overset{\text{(c)}}{\leq}\frac{\tau n}{2(M+1)},
\end{equation}
where (a) follows from the fact that the $\Delta_r$ are martingale differences, so that $\E[\Delta_r\Delta_{r'}]=0$ for $r\neq r'$; (b) follows from $\rho\leq1$ and $\E[\Delta_r^2]\leq(z+\tau)/(M+1)$; and (c) follows from $z+\tau\leq n/2$.
Thus, $d_\tau$ is the difference between the conditional expectations of $T$ for $a=1$ and $a=0$. The decoder decides that $Q_k=1$ if and only if $T>(1-\rho^\tau)Z^{(k)}+d_\tau/2$. By Chebyshev's inequality, \cref{eq:accum-mean} and~\cref{eq:accum-var}, its error probability is at most
\begin{equation}\label{eq:accum-error}
 \frac{4\,\E\bigl[(T-\E[T])^2\bigr]}{d_\tau^2}\leq\frac{4\cdot\tau n/(2(M+1))}{\tau^4/(16(M+1)^2)}=\frac{32\,(M+1)n}{\tau^3}\leq\frac1{16}
\end{equation}
for every past and both possible realizations of $Q_k$, since $\tau^3\geq8^3(M+1)n$. The bit $Q_k$ is independent of $Q_1,\ldots,Q_{k-1}$ and of the outputs before block~$k$. Hence
\begin{equation}\label{eq:accum-info}
 I(Q\wedge Y^n)\overset{\text{(a)}}{\geq}\sum_{k=1}^LI\bigl(Q_k\wedge Y^n\bigm|Q^{k-1}\bigr)\overset{\text{(b)}}{\geq}L\Bigl(\log2-H_{\mathrm b}\bigl(\tfrac1{16}\bigr)\Bigr)\geq\frac{7L}{16},
\end{equation}
where (a) follows from the chain rule~\cite[Theorem~2.5.2]{covertho} and $H(Q)=\sum_{k=1}^LH(Q_k\mid Q^{k-1})$, and (b) follows from Fano's inequality in the form of the corollary of~\cite[Theorem~2.10.1]{covertho}, Eq.~(2.140), which for a binary $Q_k$ and an estimate with values in $\{0,1\}$ gives $H(Q_k\mid Y^n,Q^{k-1})\leq H_{\mathrm b}(1/16)$, the decision erring with probability at most $1/16$. Finally, $L\geq n/(2\tau)-1=(\xi^{2/3}/16-o(1))(M+1)^{1/3}$, which gives~\cref{eq:accumulation}. For the second claim, note that $c_1n(M+1)^{-2/3}\leq c_1\xi (M+1)^{1/3}$ and that $\frac7{256}\xi^{2/3}>c_1\xi$ when $\xi<(7/(256c_1))^3$. For such a $\xi$, \cref{eq:accumulation} and the assumed bound give $R(M)\geq\bigl(\frac7{256}\xi^{2/3}-c_1\xi-o(1)\bigr)(M+1)^{1/3}$, which completes the proof of \cref{prop:accumulation}.
\end{IEEEproof}

The code builds up the color that is absent at the start. While that color is rare, its output indicator has small
conditional variance. Moreover, the receiver can determine
its retained count at the start of each block: every block inserts $\tau$ balls of each color, the initial composition is known, and the earlier outputs are observed. For a fixed $M$, an additive term of order $M^{1/3}$ does not grow with $n$, so it does not affect the capacity.
\section{Proof of the Lower Bounds}\label{sec:achievability}
This section proves \cref{thm:lower}, \cref{prop:merging}, \cref{prop:kdependence} and \cref{cor:order}. Except in \cref{subsec:merging}, the codes of this section are nonfeedback codes. Since a nonfeedback code is a feedback code (\cref{def:codes}), every rate achievable without feedback is achievable with feedback, so the lower bounds also hold for $\CF$.

\subsection{A color decision rule that does not use the initial composition}\label{subsec:inner}
\begin{figure}[t]
\centering
\begin{tikzpicture}[x=1cm,y=1cm,>=Latex,
 blk/.style={draw=black!70, minimum height=6mm, font=\footnotesize, inner sep=2pt},
 lab/.style={font=\footnotesize, align=center}]
\node[blk, minimum width=50mm, fill=lowerbound!10] (h1) at (0,0) {inputs $c,\ldots,c$ ($\tau$ uses)};
\node[blk, minimum width=50mm, fill=lowerbound!10, right=0mm of h1] (h2) {inputs $c,\ldots,c$ ($\tau$ more uses)};
\node[lab, above=1mm of h1] {uses $1,\ldots,\tau$: weight $w_j=1$};
\node[lab, above=1mm of h2] {uses $\tau+1,\ldots,2\tau$: weight $w_j=-\rho^{-\tau}$};
\node[lab, below=7mm of h1] (o1) {$\displaystyle\sum_{j=1}^{\tau}\ee_{Y_j}$};
\node[lab, below=7mm of h2] (o2) {$\displaystyle\sum_{j=\tau+1}^{2\tau}\ee_{Y_j}$};
\draw[->] (h1) -- (o1);
\draw[->] (h2) -- (o2);
\node[lab, below=10mm of $(o1)!0.5!(o2)$] (T) {$T=\displaystyle\sum_{j=1}^{\tau}\ee_{Y_j}-\rho^{-\tau}\sum_{j=\tau+1}^{2\tau}\ee_{Y_j}$};
\draw[->] (o1) -- (T);
\draw[->] (o2) -- (T);
\node[lab, below=1mm of T] {$\E[T]=-d\,\ee_c$ for every initial composition\\the decision $\widehat C$ is a color $i$ with the smallest $T_i$};
\end{tikzpicture}
\caption{The statistic of \cref{subsec:inner}. A block inserts the color $c$ for $2\tau$ uses, and the decoder subtracts the output histogram of the second half, weighted by $\rho^{-\tau}$ with $\rho=M/(M+1)$, from that of the first half. The contribution of the unknown initial composition to the expectation of each output decays like $\rho^{j-1}$, and the weights cancel it exactly: $\E[T]=-d\,\ee_c$ for every initial composition, with $d=\tau\bigl(\rho^{-\tau}-1\bigr)$. The law of $T$ still depends on the initial composition. \Cref{subsec:kdependence} sends a word of several colors twice in the same order, which also cancels the effect of the order of its letters on the expectation of $T$.}
\label{fig:statistic}
\end{figure}
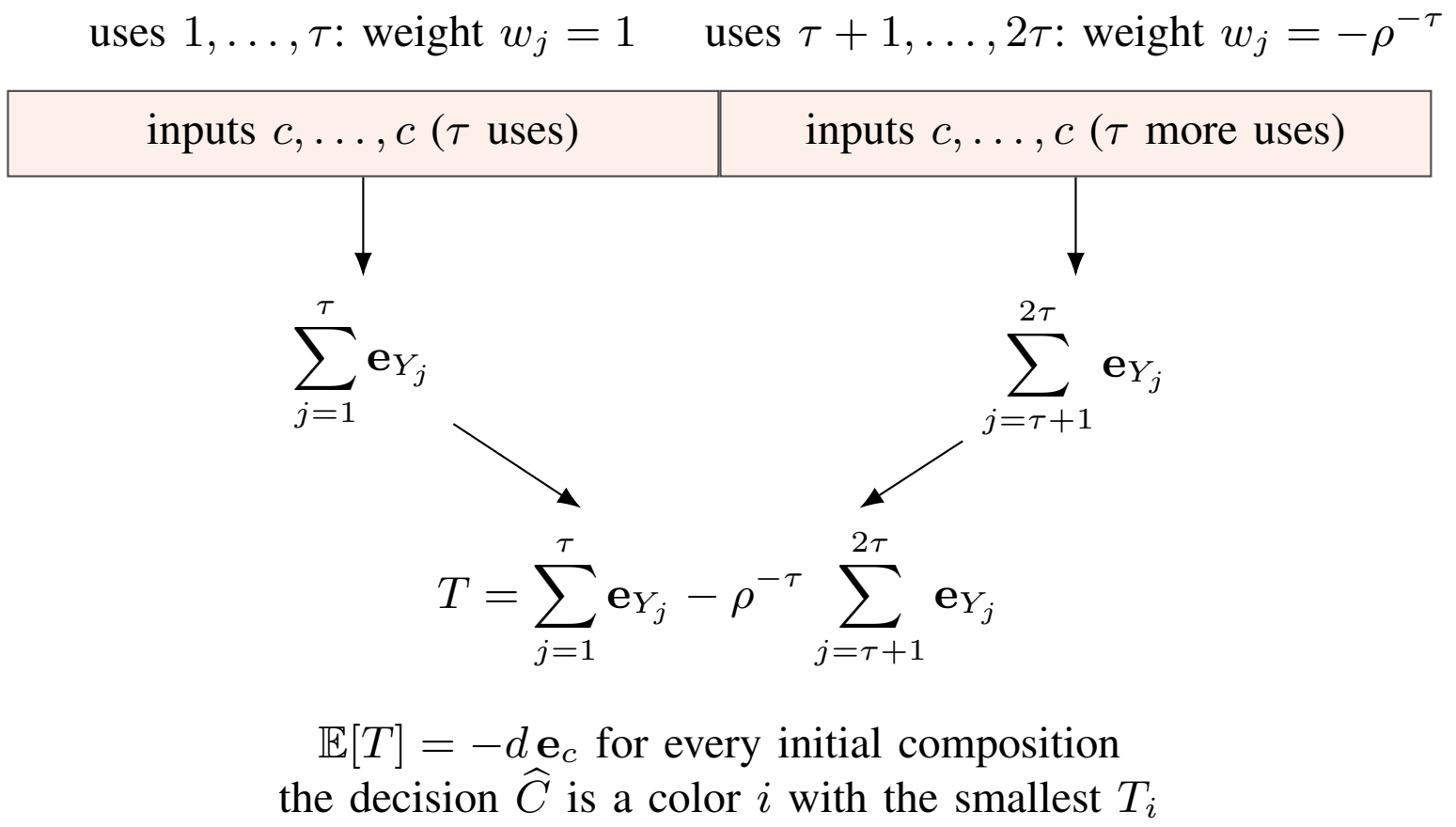

We first construct a block transmission scheme that permits a single color to be decoded with uniformly small error, regardless of the initial composition of the bag. The initial composition affects the output distribution and is not known to the decoder. We combine two output histograms so that their initial-composition contributions cancel exactly in expectation. We then control the fluctuations of this statistic using martingale differences, without assuming that successive outputs are independent. \Cref{fig:statistic} illustrates the statistic.

Fix integers $M\geq1$ and $\tau\geq1$, a color $c\in[K]$ and an initial composition $S_0=s\in\mathcal S_{M,K}$. Transmit
\begin{equation}\label{eq:inner-constant-input}
 X_j=c,\qquad j=1,\ldots,2\tau.
\end{equation}
Throughout the analysis of this block, $s$ and $c$ are fixed, and all expectations and probabilities refer to the law of the outputs given $S_0=s$ under the inputs~\cref{eq:inner-constant-input}. Define
\begin{equation}\label{eq:inner-rho}
 \rho\coloneqq\frac M{M+1}=1-\frac1{M+1},
\end{equation}
so that $0<\rho<1$. It is the probability that a given one of the $M+1$ balls in the bag is not drawn in one use. Let $\mathcal F_j$ be the $\sigma$-field generated by $Y_1,\ldots,Y_j$, with $\mathcal F_0$ trivial. Since $(Y_1,\ldots,Y_j)$ takes finitely many values, the conditional expectation $\E[\cdot\mid\mathcal F_j]$ given $\mathcal F_j$ is the conditional expectation given $(Y_1,\ldots,Y_j)$, and $\E[\cdot\mid\mathcal F_0]$ is the expectation. By the state update in \cref{def:codes}, $S_{j-1}$ is determined by the output history:
\begin{equation}\label{eq:inner-state-history}
 S_{j-1}=s+(j-1)\ee_c-\sum_{t=1}^{j-1}\ee_{Y_t}.
\end{equation}
Therefore, by~\cref{eq:kernel},
\begin{equation}\label{eq:inner-output-distribution}
 \E\bigl[\ee_{Y_j}\bigm|\mathcal F_{j-1}\bigr]=\frac{S_{j-1}+\ee_c}{M+1}.
\end{equation}
Define the random vectors
\begin{equation}\label{eq:noise}
 N_j\coloneqq\ee_{Y_j}-\frac{S_{j-1}+\ee_c}{M+1},\qquad j=1,\ldots,2\tau.
\end{equation}
Each $N_j$ is $\mathcal F_j$-measurable and satisfies
\begin{equation}\label{eq:inner-martingale-mean}
 \E\bigl[N_j\bigm|\mathcal F_{j-1}\bigr]=0
\end{equation}
by~\cref{eq:inner-output-distribution}. Thus, $(N_j)_{j=1}^{2\tau}$ is a sequence of vector martingale differences. Its conditional second moment has a bound independent of $K$. Writing $P_j\coloneqq(S_{j-1}+\ee_c)/(M+1)$ for the conditional output law, a random probability vector with coordinates $P_{j,i}$, we have
\begin{align}
 \E\bigl[\|N_j\|_2^2\bigm|\mathcal F_{j-1}\bigr]
 &=\sum_{i=1}^KP_{j,i}\,\|\ee_i-P_j\|_2^2\nonumber\\
 &\overset{\text{(a)}}{=}\sum_{i=1}^KP_{j,i}\bigl(1-2P_{j,i}+\|P_j\|_2^2\bigr)\nonumber\\
 &\overset{\text{(b)}}{=}1-\sum_{i=1}^KP_{j,i}^2\leq1,\label{eq:noisevariance}
\end{align}
where (a) follows from $\|\ee_i\|_2^2=1$ and $\langle\ee_i,P_j\rangle=P_{j,i}$, and (b) follows from $\sum_{i=1}^KP_{j,i}=1$. Bounding the total Euclidean second moment in this way avoids a factor of $K$.

We next determine the expectation of the response to the repeated input. Substituting $\ee_{Y_j}=(S_{j-1}+\ee_c)/(M+1)+N_j$ into the state update $S_j=S_{j-1}+\ee_c-\ee_{Y_j}$ gives
\begin{equation}\label{eq:linearstate}
 S_j=S_{j-1}+\ee_c-\frac{S_{j-1}+\ee_c}{M+1}-N_j=\rho S_{j-1}+\rho\,\ee_c-N_j.
\end{equation}
Taking expectations and using~\cref{eq:inner-martingale-mean}, we obtain $\E[S_j]=\rho\,\E[S_{j-1}]+\rho\,\ee_c$. The vector $M\ee_c$ is a fixed point of this recursion because $\rho(M+1)=M$, so that $\E[S_j]-M\ee_c=\rho(\E[S_{j-1}]-M\ee_c)$. Iterating from $S_0=s$ and applying~\cref{eq:inner-output-distribution} give
\begin{equation}\label{eq:means}
 \E[S_j]=M\ee_c+\rho^j(s-M\ee_c),
 \qquad
 \E\bigl[\ee_{Y_j}\bigr]=\ee_c+\frac{\rho^{j-1}}{M+1}(s-M\ee_c).
\end{equation}
The expectation of the output consists of a term that depends only on the transmitted color and a geometrically decaying term that contains the unknown initial composition.

To eliminate the latter term, weight the outputs of the second half of the block by $\rho^{-\tau}$ and subtract them from those of the first half:
\begin{equation}\label{eq:weights}
 w_j\coloneqq
 \begin{cases}
 1,&1\leq j\leq\tau,\\
 -\rho^{-\tau},&\tau<j\leq2\tau,
 \end{cases}
 \qquad
 T\coloneqq\sum_{j=1}^{2\tau}w_j\ee_{Y_j}.
\end{equation}
The weights depend only on $M$ and $\tau$, which are known to the decoder. The cancellation is exact:
\begin{equation}\label{eq:cancellation}
 \sum_{j=1}^{2\tau}w_j\rho^{j-1}=\sum_{j=1}^\tau\rho^{j-1}-\rho^{-\tau}\rho^\tau\sum_{j=1}^\tau\rho^{j-1}=0.
\end{equation}
Hence,
\begin{equation}\label{eq:statmean}
 \E[T]\overset{\text{(a)}}{=}\left(\sum_{j=1}^{2\tau}w_j\right)\ee_c+\frac{s-M\ee_c}{M+1}\sum_{j=1}^{2\tau}w_j\rho^{j-1}\overset{\text{(b)}}{=}-d\,\ee_c,
 \qquad d\coloneqq\tau\bigl(\rho^{-\tau}-1\bigr),
\end{equation}
where (a) follows from~\cref{eq:means}, and (b) follows from~\cref{eq:cancellation} and $\sum_{j=1}^{2\tau}w_j=\tau\bigl(1-\rho^{-\tau}\bigr)$. Moreover,
\begin{equation}\label{eq:inner-separation-lower}
 d=\tau\bigl(\rho^{-\tau}-1\bigr)\overset{\text{(a)}}{\geq}\tau\Biggl[\left(1+\frac1{M+1}\right)^\tau-1\Biggr]\overset{\text{(b)}}{\geq}\frac{\tau^2}{M+1},
\end{equation}
where (a) follows from $\rho^{-1}=1+1/M\geq1+1/(M+1)$, and (b) follows from Bernoulli's inequality $(1+x)^\tau\geq1+\tau x$ for $x\geq0$. Thus, the possible expectations of $T$ are the $K$ known vectors $-d\ee_1,\ldots,-d\ee_K$, whatever $s$ is. The law of $T$ can still depend on $s$. We next show that its fluctuations admit a bound uniform over $s$ as well.

Subtracting the recursion for the expectations from~\cref{eq:linearstate} gives $S_j-\E[S_j]=\rho(S_{j-1}-\E[S_{j-1}])-N_j$ with $S_0-\E[S_0]=0$, so that
\begin{equation}\label{eq:statefluct}
 S_{j-1}-\E[S_{j-1}]=-\sum_{t=1}^{j-1}\rho^{j-1-t}N_t.
\end{equation}
By~\cref{eq:noise} and~\cref{eq:statefluct},
\begin{equation}\label{eq:outputfluct}
 \ee_{Y_j}-\E\bigl[\ee_{Y_j}\bigr]=N_j+\frac{S_{j-1}-\E[S_{j-1}]}{M+1}=N_j-\frac1{M+1}\sum_{t=1}^{j-1}\rho^{j-1-t}N_t.
\end{equation}
Multiplying~\cref{eq:outputfluct} by $w_j$, summing over the block and interchanging the order of the finite double sum, we obtain
\begin{equation}\label{eq:martrepr}
 T-\E[T]=\sum_{t=1}^{2\tau}\kappa_tN_t,
 \qquad
 \kappa_t\coloneqq w_t-\frac1{M+1}\sum_{j=t+1}^{2\tau}w_j\rho^{j-1-t}.
\end{equation}
All coefficients $\kappa_t$ are deterministic. This representation accounts explicitly for the dependence among the outputs.

Assume now that $M+1\geq4$ and $\tau\leq(M+1)/4$. Then
\begin{equation}\label{eq:inner-weight-magnitude}
 \rho^{-\tau}=\exp\bigl(-\tau\log\rho\bigr)\overset{\text{(a)}}{\leq}\exp\left(\frac\tau M\right)\overset{\text{(b)}}{\leq}\mathrm{e}^{1/3}<2,
\end{equation}
where (a) follows from $-\log(1-x)\leq x/(1-x)$ for $0<x<1$, applied with $x=1/(M+1)$, which gives $-\log\rho\leq1/M$, and (b) follows from $\tau/M\leq(M+1)/(4M)\leq1/3$, which uses $M\geq3$. Hence, $|w_j|\leq2$ for every $j$, and
\begin{align}
 |\kappa_t|
 &\overset{\text{(a)}}{\leq}|w_t|+\frac1{M+1}\sum_{j=t+1}^{2\tau}|w_j|\rho^{j-1-t}\nonumber\\
 &\overset{\text{(b)}}{\leq}2+\frac{2(2\tau-t)}{M+1}\nonumber\\
 &\overset{\text{(c)}}{\leq}2+\frac{4\tau}{M+1}\nonumber\\
 &\overset{\text{(d)}}{\leq}3,\label{eq:weightbound}
\end{align}
where (a) follows from the definition of $\kappa_t$ in~\cref{eq:martrepr} and the triangle inequality; (b) follows from $|w_j|\leq2$, from $\rho^{j-1-t}\leq1$, which holds because $0<\rho<1$ and $j-1-t\geq0$, and from the fact that the sum has $2\tau-t$ terms; (c) follows from $t\geq1$; and (d) follows from $\tau\leq(M+1)/4$.
For $t<t'$, the vector $N_t$ is $\mathcal F_{t'-1}$-measurable, so~\cref{eq:inner-martingale-mean} gives
\begin{equation}\label{eq:inner-orthogonality}
 \E\bigl[\langle N_t,N_{t'}\rangle\bigr]=\E\Bigl[\bigl\langle N_t,\E[N_{t'}\mid\mathcal F_{t'-1}]\bigr\rangle\Bigr]=0.
\end{equation}
Therefore,
\begin{equation}\label{eq:statvariance}
 \E\bigl[\|T-\E[T]\|_2^2\bigr]\overset{\text{(a)}}{=}\sum_{t=1}^{2\tau}\kappa_t^2\,\E\bigl[\|N_t\|_2^2\bigr]\overset{\text{(b)}}{\leq}9\cdot2\tau=18\tau,
\end{equation}
where (a) follows from~\cref{eq:martrepr} and~\cref{eq:inner-orthogonality}, and (b) follows from~\cref{eq:noisevariance} and~\cref{eq:weightbound}. This bound is uniform over $s$, $c$ and $K$.

Given the observed statistic $T$, the decoder chooses the nearest of its possible expectations,
\begin{equation}\label{eq:inner-decoder}
 \widehat C\in\left\{c\in[K]\;\middle|\;\|T+d\ee_c\|_2=\min_{i\in[K]}\|T+d\ee_i\|_2\right\},
\end{equation}
with ties resolved by a fixed deterministic rule. Since $\|T+d\ee_i\|_2^2=\|T\|_2^2+2dT_i+d^2$ and $d>0$, this decoder selects a coordinate of $T$ with the smallest value, and it needs no knowledge of $s$. If the decoder selects a color $i\neq c$, then $\|T+d\ee_i\|_2\leq\|T+d\ee_c\|_2$, and
\begin{equation}\label{eq:inner-error-implies-deviation}
 \sqrt2\,d=\|{-d\ee_i}+d\ee_c\|_2\overset{\text{(a)}}{\leq}\|T+d\ee_i\|_2+\|T+d\ee_c\|_2\leq2\|T+d\ee_c\|_2,
\end{equation}
where (a) follows from the triangle inequality. Thus, an error implies $\|T+d\ee_c\|_2^2\geq d^2/2$, and
\begin{equation}\label{eq:innererror}
 \PP\bigl\{\widehat C\neq c\bigr\}\leq\PP\left\{\|T+d\ee_c\|_2^2\geq\frac{d^2}2\right\}\overset{\text{(a)}}{\leq}\frac{2\,\E\bigl[\|T+d\ee_c\|_2^2\bigr]}{d^2}\overset{\text{(b)}}{\leq}\frac{36\tau}{d^2}\overset{\text{(c)}}{\leq}\frac{36(M+1)^2}{\tau^3},
\end{equation}
where (a) follows from Markov's inequality, (b) follows from~\cref{eq:statmean} and~\cref{eq:statvariance}, and (c) follows from~\cref{eq:inner-separation-lower}. No union bound over the competing colors is needed, because every incorrect decision implies the same Euclidean deviation event.

Finally, choose $\tau\coloneqq\bigl\lceil8(M+1)^{2/3}\bigr\rceil$. Then, $\tau^3\geq8^3(M+1)^2$, and, whenever $M+1\geq4$ and $\tau\leq(M+1)/4$,
\begin{equation}\label{eq:inner-uniform-error}
 \max_{s\in\mathcal S_{M,K}}\max_{c\in[K]}\PP\bigl\{\widehat C\neq c\bigm|S_0=s\bigr\}\leq\frac{36}{8^3}=\frac9{128}<\frac1{12}.
\end{equation}
The condition $\tau\leq(M+1)/4$ holds for every $M\geq2^{16}$ (\cref{rem:threshold}).

\subsection{A direct coding argument}\label{subsec:directcoding}
The decision rule of \cref{subsec:inner} has a uniformly small error probability that does not tend to zero. An outer code makes the error probability vanish. The following lemma uses only the conditional form of~\cref{eq:inner-uniform-error}.

\begin{lemma}[A direct coding argument]\label{lem:directcoding}
Let $\mathcal A$ be a finite set with $|\mathcal A|\geq2$, and let $\ell\geq1$ be an integer. Suppose that each $a\in\mathcal A$ is assigned an input word $x(a)\in[K]^\ell$, and that there are a map $\psi:[K]^\ell\to\mathcal A$ and a constant $p_\ast$ such that, for every $z\in\mathcal S_{M,K}$ and every $a\in\mathcal A$, the outputs $Y^\ell$ of $\ell$ uses with the input word $x(a)$ and the initial composition $S_0=z$ satisfy
\begin{equation}\label{eq:directcoding-hyp}
 \PP\bigl\{\psi(Y^\ell)\neq a\bigm|S_0=z\bigr\}\leq p_\ast.
\end{equation} Let $p_0$ satisfy $p_\ast<p_0\leq1-1/|\mathcal A|$ (replacing $p_\ast$ by a larger constant keeps the hypothesis, so we may and do assume $p_\ast>0$), and put $f_{|\mathcal A|}(u)\coloneqq H_{\mathrm b}(u)+u\log(|\mathcal A|-1)$. Then, every rate
\begin{equation}\label{eq:directcoding-rate}
 R<\frac{\log|\mathcal A|-f_{|\mathcal A|}(p_0)}{\ell}
\end{equation}
is achievable without feedback in the sense of \cref{def:capacity}.
\end{lemma}

\begin{figure}[t]
\centering
\begin{tikzpicture}[x=1cm,y=1cm,>=Latex,
 blk/.style={draw=black!70, minimum height=6.5mm, minimum width=27mm, font=\footnotesize, inner sep=2pt, fill=lowerbound!10},
 dec/.style={draw=black!70, rounded corners=2pt, minimum height=6.5mm, minimum width=27mm, font=\footnotesize, inner sep=2pt},
 lab/.style={font=\footnotesize, align=center},
 arr/.style={->, line width=0.6pt}]
\node[lab] (A) at (5.1,1.55) {message $q$ $\;\longmapsto\;$ outer codeword $\mathbf a(q)=(a_1,\ldots,a_L)\in\mathcal A^L$};
\node[blk] (b1) at (0,0) {block 1: word $x(a_1)$};
\node[blk] (b2) at (3.4,0) {block 2: word $x(a_2)$};
\node[lab] (bd) at (6.1,0) {$\cdots$};
\node[blk] (bL) at (8.8,0) {block $L$: word $x(a_L)$};
\draw[arr] (b1.east) -- node[lab, above, yshift=-0.5mm]{\scriptsize $S_{\ell}$} (b2.west);
\draw[arr] (b2.east) -- node[lab, above, yshift=-0.5mm]{\scriptsize $S_{2\ell}$} (bd.west);
\draw[arr] (bd.east) -- node[lab, above, yshift=-0.5mm]{\scriptsize $S_{(L-1)\ell}$} (bL.west);
\draw[arr] (A.south -| b1) -- (b1.north);
\draw[arr] (A.south -| b2) -- (b2.north);
\draw[arr] (A.south -| bL) -- (bL.north);
\node[dec] (d1) at (0,-1.45) {$\widehat A_1=\psi(\mathbf Y_1)$};
\node[dec] (d2) at (3.4,-1.45) {$\widehat A_2=\psi(\mathbf Y_2)$};
\node[lab] (dd) at (6.1,-1.45) {$\cdots$};
\node[dec] (dL) at (8.8,-1.45) {$\widehat A_L=\psi(\mathbf Y_L)$};
\draw[arr] (b1) -- node[lab, right]{\scriptsize $\mathbf Y_1$} (d1);
\draw[arr] (b2) -- node[lab, right]{\scriptsize $\mathbf Y_2$} (d2);
\draw[arr] (bL) -- node[lab, right]{\scriptsize $\mathbf Y_L$} (dL);
\node[lab, draw=black!70, rounded corners=2pt, inner sep=3pt, text width=112mm, align=center] (H) at (4.4,-3.0) {decoder: the unique message $q'$ whose codeword $\mathbf a(q')$ is within Hamming distance $Lp_0$ of $(\widehat A_1,\ldots,\widehat A_L)$, or $1$ if there is none or more than one};
\draw[arr] (d1.south) -- (d1.south |- H.north);
\draw[arr] (d2.south) -- (d2.south |- H.north);
\draw[arr] (dL.south) -- (dL.south |- H.north);
\end{tikzpicture}
\caption{The code of \cref{lem:directcoding}. A message selects a string of $L$ letters of $\mathcal A$, and block $k$ sends the input word $x(a_k)$ of $\ell$ uses. The bag is not reset between blocks: the composition $S_{(k-1)\ell}$ at the start of block $k$ depends on the earlier outputs. Whatever that composition is, the decision $\widehat A_k$ errs with conditional probability at most $p_\ast$ given the earlier outputs. The decoder outputs the unique message whose codeword is within Hamming distance $Lp_0$ of the decisions. Otherwise, it outputs $1$. The error bound counts every case with no such message or more than one as an error. For \cref{thm:lower}, $\mathcal A=[K]$ and $x(c)$ repeats the color $c$ for $\ell=2\tau$ uses. For \cref{prop:kdependence}, $\mathcal A=\Omega$ and $x(\omega)$ is the word $u(\omega)$ sent twice.}
\label{fig:outercode}
\end{figure}
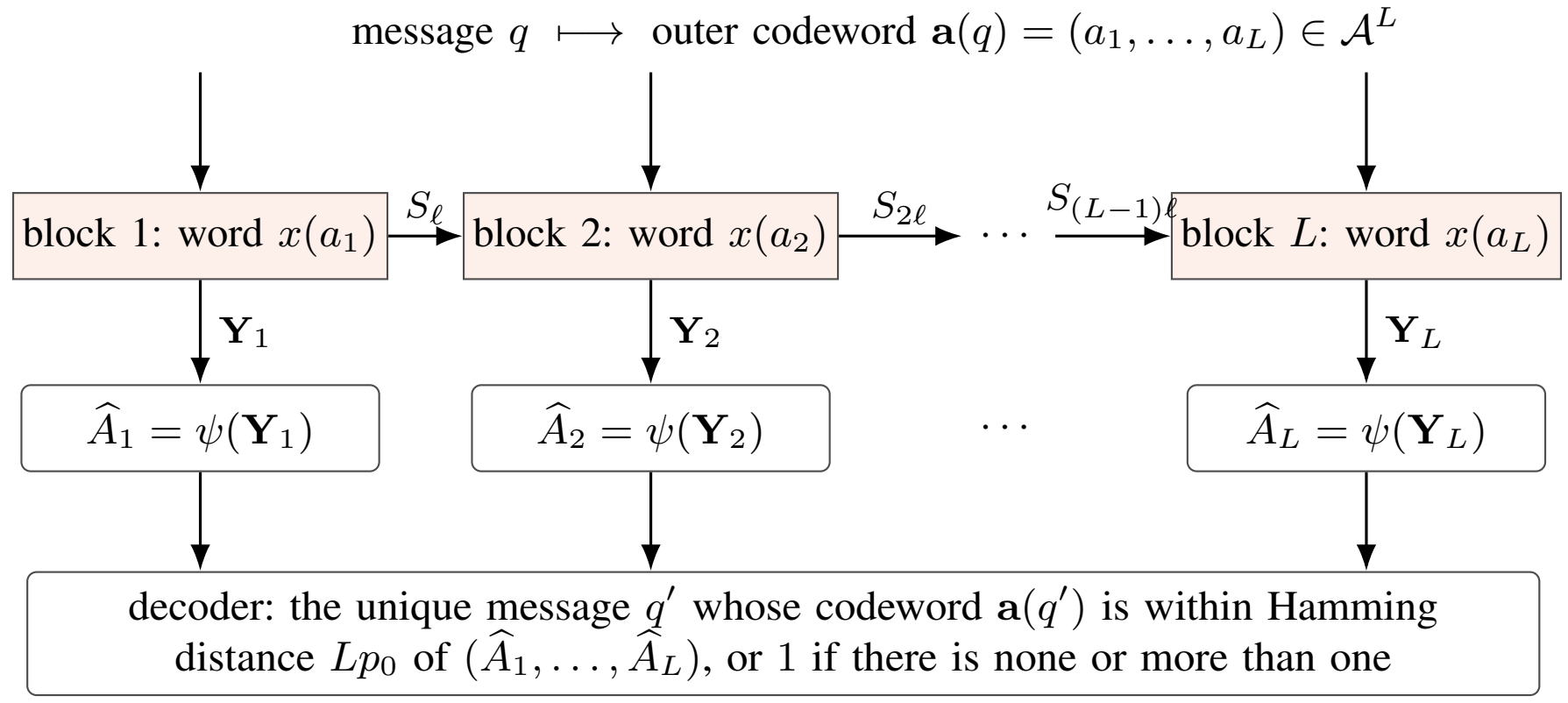

In the lemma, $p_\ast$ bounds the probability that one block is decoded wrongly, whatever the composition at its start. The decoder in the proof outputs the unique message whose codeword disagrees with at most a fraction $p_0>p_\ast$ of the block decisions, and it outputs $1$ when there is no such message or more than one (\cref{fig:outercode}).

\begin{IEEEproof}
Fix $L\geq1$ and a codebook $\mathbf c=(\mathbf a(1),\ldots,\mathbf a(J))$ of strings $\mathbf a(q)=(a_1(q),\ldots,a_L(q))\in\mathcal A^L$, $q\in[J]$. Message $q$ is sent as the concatenation of the words $x(a_1(q)),\ldots,x(a_L(q))$. The receiver applies $\psi$ to each block of $\ell$ outputs and obtains decisions $\widehat A_1,\ldots,\widehat A_L$. For $\mathbf b\in\mathcal A^L$, let
\begin{equation}
 \mathcal B(\mathbf b)\coloneqq\Bigl\{\mathbf a\in\mathcal A^L\Bigm\vert\bigl|\{k\in[L]\mid a_k\neq b_k\}\bigr|\leq Lp_0\Bigr\}
\end{equation}
be the Hamming ball of radius $\lfloor Lp_0\rfloor$ around $\mathbf b$. The receiver outputs the unique $q$ with $\mathbf a(q)\in\mathcal B(\widehat A^L)$. If there is no such $q$ or more than one, it outputs $q=1$, and the bound below counts every such case as an error. This defines a nonfeedback code of length $n\coloneqq\ell L$ with $J$ messages, whose decoding function we denote by $\varphi_{\mathbf c}$, and whose error probability from $s\in\mathcal S_{M,K}$ in~\cref{eq:error} we denote by $P_e(\mathbf c,s)$.

Fix an initial composition $s\in\mathcal S_{M,K}$ and a transmitted string $\mathbf a\in\mathcal A^L$, and let $E_k\coloneqq\ind_{\{\widehat A_k\neq a_k\}}$. In this paragraph, $\PP$ and $\E$ refer to the law of the outputs when the words $x(a_1),\ldots,x(a_L)$ are sent from $S_0=s$. The composition at the start of block $k$ is determined by $s$, $\mathbf a$ and the outputs $\mathbf Y^{k-1}\coloneqq(\mathbf Y_1,\ldots,\mathbf Y_{k-1})$ of the earlier blocks, where $\mathbf Y_j\in[K]^\ell$ is the output word of block $j$. Given these, the draws of block $k$ are fresh, so~\cref{eq:directcoding-hyp} gives
\begin{equation}\label{eq:directcoding-conditional}
 \PP\bigl\{E_k=1\bigm|\mathbf Y^{k-1}\bigr\}\leq p_\ast
\end{equation}
for every history $\mathbf Y^{k-1}$ of positive probability. (A bound on each marginal probability $\PP\{E_k=1\}$ alone would not suffice, since the errors could then coincide in all blocks.) For every $\zeta>0$,
\begin{equation}\label{eq:directcoding-mgf}
 \E\Bigl[\mathrm{e}^{\zeta\sum_{j=1}^kE_j}\Bigr]
 \overset{\text{(a)}}{=}\E\Bigl[\mathrm{e}^{\zeta\sum_{j=1}^{k-1}E_j}\,\E\bigl[\mathrm{e}^{\zeta E_k}\bigm|\mathbf Y^{k-1}\bigr]\Bigr]
 \overset{\text{(b)}}{\leq}\bigl(1-p_\ast+p_\ast \mathrm{e}^\zeta\bigr)\,\E\Bigl[\mathrm{e}^{\zeta\sum_{j=1}^{k-1}E_j}\Bigr],
\end{equation}
where (a) follows from the fact that $E_1,\ldots,E_{k-1}$ are functions of $\mathbf Y^{k-1}$, and (b) follows from~\cref{eq:directcoding-conditional}. By induction on $k$, as in the proof of~\cite[Theorem~7.2.1]{alonspencer},
\begin{equation}
 \E\left[\mathrm{e}^{\zeta\sum_{k=1}^LE_k}\right]\leq\bigl(1-p_\ast+p_\ast\mathrm{e}^\zeta\bigr)^L.
\end{equation}
Markov's inequality with
\begin{equation}
 \zeta=\log\frac{p_0(1-p_\ast)}{p_\ast(1-p_0)}>0,
\end{equation}
the value that minimizes the resulting bound, then gives
\begin{equation}\label{eq:directcoding-tail}
 \PP\left\{\sum_{k=1}^LE_k>Lp_0\right\}\leq \mathrm{e}^{-L\,D(\Ber(p_0)\,\Vert\,\Ber(p_\ast))},
\end{equation}
where $D(\Ber(p_0)\,\Vert\,\Ber(p_\ast))>0$ is the relative entropy between the Bernoulli laws with expectations $p_0$ and $p_\ast$.

Each ball $\mathcal B(\mathbf b)$ has at most $\mathrm{e}^{Lf_{|\mathcal A|}(p_0)}$ points. To see this, consider the product law under which each coordinate $k$ equals $b_k$ with probability $1-p_0$ and each other symbol with probability $p_0/(|\mathcal A|-1)$. Since $p_0/(|\mathcal A|-1)\leq1-p_0$, a string at Hamming distance $j\leq Lp_0$ from $\mathbf b$ has probability at least $(1-p_0)^{L(1-p_0)}(p_0/(|\mathcal A|-1))^{Lp_0}=\mathrm{e}^{-Lf_{|\mathcal A|}(p_0)}$, and the probabilities of the ball's points sum to at most one.

Now let $\mathbf C=(\mathbf A(1),\ldots,\mathbf A(J))$ be a random codebook: its strings are independent, each uniformly distributed on $\mathcal A^L$, and independent of the channel's draws, as in the proof of~\cite[Theorem~15.1.1]{alonspencer}. Given $\mathbf C=\mathbf c$, the outputs have the law of the code with the codebook $\mathbf c$. Fix an initial composition $s\in\mathcal S_{M,K}$, and let the message $Q$ be uniformly distributed on $[J]$, independent of $\mathbf C$ and of the channel's draws. Fix a realization $q\in[J]$ of $Q$. The values $q$ and $s$ are fixed, not random variables. The conditional error probability $\PP\bigl\{\varphi_{\mathbf C}(Y^n)\neq q\bigm|\mathbf C,\,Q=q,\,S_0=s\bigr\}$ is a random variable, a function of $\mathbf C$, and its expectation below is taken with respect to $\mathbf C$. Let $E_1,\ldots,E_L$ be defined as above with $\mathbf a=\mathbf A(q)$. Then
 \begin{align}
  \E\Bigl[\PP\bigl\{\varphi_{\mathbf C}(Y^n)\neq q\bigm|\mathbf C,\,Q=q,\,S_0=s\bigr\}\Bigr]
  &\overset{\text{(a)}}{=}\PP\bigl\{\varphi_{\mathbf C}(Y^n)\neq q\bigm|Q=q,\,S_0=s\bigr\}\nonumber\\
  &\overset{\text{(b)}}{\leq}\PP\Biggl\{\sum_{k=1}^LE_k>Lp_0\Biggm|Q=q,\,S_0=s\Biggr\}\nonumber\\
  &\qquad+\sum_{q'\in[J]\setminus\{q\}}\PP\bigl\{\mathbf A(q')\in\mathcal B(\widehat A^L)\bigm|Q=q,\,S_0=s\bigr\}\nonumber\\
  &\overset{\text{(c)}}{\leq}\mathrm{e}^{-L\,D(\Ber(p_0)\,\Vert\,\Ber(p_\ast))}+J\,\mathrm{e}^{-L(\log|\mathcal A|-f_{|\mathcal A|}(p_0))},\label{eq:directcoding-randomcode}
 \end{align}
where (a) follows from the law of total probability, (b) follows from the fact that the decoder outputs $q$ whenever $\sum_{k=1}^LE_k\leq Lp_0$, which means $\mathbf A(q)\in\mathcal B(\widehat A^L)$, and no other string of $\mathbf C$ lies in $\mathcal B(\widehat A^L)$, together with the union bound, and (c) follows from~\cref{eq:directcoding-tail}, which holds given each realization of $\mathbf A(q)$, and from the fact that each $\mathbf A(q')$ with $q'\neq q$ is independent of $\widehat A^L$ and uniformly distributed on $\mathcal A^L$, because $\widehat A^L$ depends only on $\mathbf A(q)$ and on the channel's draws. Hence, $\mathbf A(q')$ lies in $\mathcal B(\widehat A^L)$ with probability at most $\mathrm{e}^{Lf_{|\mathcal A|}(p_0)}/|\mathcal A|^L$.
If $\log|\mathcal A|-f_{|\mathcal A|}(p_0)=0$, no rate $R\geq0$ satisfies~\cref{eq:directcoding-rate}, and there is nothing to prove. Otherwise, choose $J\coloneqq\bigl\lceil\mathrm{e}^{L\varsigma}\bigr\rceil$ with $0<\varsigma<\log|\mathcal A|-f_{|\mathcal A|}(p_0)$, and average~\cref{eq:directcoding-randomcode} over $q\in[J]$ and sum it over the finitely many $s\in\mathcal S_{M,K}$. Then $\E\bigl[\sum_{s\in\mathcal S_{M,K}}P_e(\mathbf C,s)\bigr]$ is at most $|\mathcal S_{M,K}|$ times a quantity that tends to zero as $L\to\infty$. For each $L$, choose a codebook $\mathbf c$ with $\sum_{s\in\mathcal S_{M,K}}P_e(\mathbf c,s)\leq\E\bigl[\sum_{s\in\mathcal S_{M,K}}P_e(\mathbf C,s)\bigr]$. Then
\begin{equation}
 \max_{s\in\mathcal S_{M,K}}P_e(\mathbf c,s)\leq\sum_{s\in\mathcal S_{M,K}}P_e(\mathbf c,s)\leq\E\left[\sum_{s\in\mathcal S_{M,K}}P_e(\mathbf C,s)\right],
\end{equation}
which tends to zero as $L\to\infty$. Its length is $n=\ell L$ and its rate is at least $\varsigma/\ell$. For a length that is not a multiple of $\ell$, append at most $\ell-1$ fixed input symbols and let the decoder ignore the extra outputs. This changes the rate by a vanishing amount. Since $\varsigma\in(0,\log|\mathcal A|-f_{|\mathcal A|}(p_0))$ is arbitrary, every rate in~\cref{eq:directcoding-rate} is achievable. This completes the proof of \cref{lem:directcoding}.
\end{IEEEproof}

\begin{IEEEproof}[Proof of \cref{thm:lower}]
Let $\tau\coloneqq\bigl\lceil8(M+1)^{2/3}\bigr\rceil$ satisfy $\tau\leq(M+1)/4$. Apply \cref{lem:directcoding} with $\mathcal A=[K]$, $\ell=2\tau$, $x(c)=(c,\ldots,c)$, $\psi$ the decoder~\cref{eq:inner-decoder}, $p_\ast=9/128$ by~\cref{eq:inner-uniform-error}, and $p_0=1/12$. The hypothesis~\cref{eq:directcoding-hyp} holds because, given the past, a block that repeats $c$ is a $2\tau$-use constant-color transmission from the current composition, to which~\cref{eq:inner-uniform-error} applies. Since $p_\ast<p_0\leq1-1/K$, the lemma shows that every rate below $(\log K-f_K(1/12))/(2\tau)$ is achievable, which is~\cref{eq:lower-exact}. The numerator is positive because $f_K$ is increasing on $[0,1-1/K]$ (its derivative is $\log((1-u)(K-1)/u)$) and $f_K(1-1/K)=\log K$.

For~\cref{eq:lower-asym}, note first that
\begin{equation}\label{eq:positiveconstant}
 \log K-H_{\mathrm b}\left(\frac1{12}\right)-\frac1{12}\log(K-1)\overset{\text{(a)}}{\geq}\frac{11}{12}\log K-H_{\mathrm b}\left(\frac1{12}\right)\overset{\text{(b)}}{\geq}\gamma\log K,
\end{equation}
where (a) follows from $\log(K-1)\leq\log K$, and (b) follows from $H_{\mathrm b}(1/12)\leq(\log K/\log2)\,H_{\mathrm b}(1/12)$ for $K\geq2$, with $\gamma$ as in~\cref{eq:gamma}. The constant $\gamma$ is positive because $H_{\mathrm b}(u)\leq-u\log u+u$ for $0\leq u<1$ (from $-\log(1-u)\leq u/(1-u)$), so that $H_{\mathrm b}(1/12)\leq(\log12+1)/12<(11/12)\log2$, the last inequality being equivalent to $12\mathrm{e}<2^{11}$. Moreover,
\begin{equation}\label{eq:durationbound}
 \tau\leq8(M+1)^{2/3}+1\leq9(M+1)^{2/3}.
\end{equation}
Substituting~\cref{eq:positiveconstant} and~\cref{eq:durationbound} into~\cref{eq:lower-exact} gives~\cref{eq:lower-asym}, completing the proof of \cref{thm:lower}.
\end{IEEEproof}

\subsection{Merging colors}\label{subsec:merging}
\begin{IEEEproof}[Proof of \cref{prop:merging}]
Let $2\leq k\leq K$. Partition $[K]$ into $k$ nonempty groups $G_1,\ldots,G_k$, for instance $G_j=\{j\}$ for $j<k$ and $G_k=\{k,\ldots,K\}$, choose $c_j\in G_j$, and let $\chi(i)$ be the index of the group that contains color $i$. Run a code for $k$ colors on the channel with $K$ colors as follows: input $x\in[k]$ is sent as color $c_x$, and every output $y$ is replaced by $\chi(y)$, for the decoder and, with feedback, for the encoder. For $z\in\mathcal S_{M,K}$, write $\bar z_j\coloneqq\sum_{i\in G_j}z_i$, so that $\bar z\in\mathcal S_{M,k}$, and let $\bar S_t$ and $\bar s$ be obtained from $S_t$ and $s$ in the same way. Since $c_x\in G_j$ exactly when $x=j$,
\begin{equation}\label{eq:merging-law}
 \PP\bigl\{\chi(Y_t)=j\bigm|S_{t-1}=z,\,X_t=c_x\bigr\}=\sum_{i\in G_j}\frac{z_i+\ind_{\{i\}}(c_x)}{M+1}=\frac{\bar z_j+\ind_{\{j\}}(x)}{M+1},
\end{equation}
and $\bar S_t=\bar S_{t-1}+\ee_x-\ee_{\chi(Y_t)}$, where $\ee_x$ and $\ee_{\chi(Y_t)}$ are standard basis vectors of $\mathbb R^k$. The encoder's inputs depend on the message and the merged past only. Hence, by induction on $t$, the pair consisting of the message and the merged outputs has the same law as the message and the outputs of the channel with $k$ colors started at $\bar s$. The error probability of the code from $s$ therefore equals its error probability on the channel with $k$ colors from $\bar s$, which is at most its worst-case error probability there. Every rate achievable with $k$ colors, with or without feedback, is therefore achievable with $K$ colors. This completes the proof of \cref{prop:merging}.
\end{IEEEproof}

\subsection{A lower bound that grows like \texorpdfstring{$K^{2/3}$}{K\textasciicircum(2/3)}}\label{subsec:kdependence}
\begin{IEEEproof}[Proof of \cref{prop:kdependence}]
Put
\begin{equation}
 r\coloneqq\lfloor K/2\rfloor,\qquad b\coloneqq\left\lceil8\left(\frac{M+1}{r}\right)^{2/3}\right\rceil,\qquad\tau\coloneqq4rb.
\end{equation}
Then
\begin{equation}\label{eq:kdep-tau}
 32(M+1)^{2/3}r^{1/3}\leq\tau\leq36(M+1)^{2/3}r^{1/3}\leq\frac {M+1}4,
\end{equation}
where the upper bounds use $b\leq8((M+1)/r)^{2/3}+1\leq9((M+1)/r)^{2/3}$, which holds because $((M+1)/r)^{2/3}\geq1$, and $(M+1)/r\geq2^{22}>144^3$.

Send a word $u\in[K]^\tau$ twice in the same order, and let $T$ be the statistic~\cref{eq:weights} formed from the $2\tau$ outputs. Let $\mu(u)\in\mathbb R^K$ be the composition of $u$ divided by $\tau$, so that $\mu_i(u)$ is the fraction of the letters of $u$ that equal $i$. We compute the expectation of $T$. The input at uses $j$ and $\tau+j$ is $u_j$, $j=1,\ldots,\tau$. With $u_j$ in place of $c$, the derivation of~\cref{eq:linearstate} gives $\E[S_j]=\rho\,\E[S_{j-1}]+\rho\,\ee_{u_j}$ for $1\leq j\leq\tau$, and $\E[S_j]=\rho\,\E[S_{j-1}]+\rho\,\ee_{u_{j-\tau}}$ for $\tau<j\leq2\tau$. Iterating from $S_0=s$ gives
\begin{align}
 \E[S_\tau]&=\rho^\tau s+F_u,\nonumber\\
 \E[S_{2\tau}]&=\rho^{2\tau}s+\bigl(1+\rho^\tau\bigr)F_u,\label{eq:kdep-statemeans}
\end{align}
where $F_u\coloneqq\sum_{j=1}^{\tau}\rho^{\tau-j+1}\ee_{u_j}$. Each half of the block inserts the composition $\tau\mu(u)$, and each use removes one ball, so the output histograms of the two halves are
\begin{align}
 \sum_{j=1}^{\tau}\ee_{Y_j}&=\tau\mu(u)+s-S_\tau,\nonumber\\
 \sum_{j=\tau+1}^{2\tau}\ee_{Y_j}&=\tau\mu(u)+S_\tau-S_{2\tau}.\label{eq:kdep-histograms}
\end{align}
By~\cref{eq:weights}, $T$ is the first histogram minus $\rho^{-\tau}$ times the second. Therefore,
\begin{align}
 \E[T]
 &\overset{\text{(a)}}{=}\bigl\{\tau\mu(u)+\bigl(1-\rho^\tau\bigr)s-F_u\bigr\}-\rho^{-\tau}\bigl\{\tau\mu(u)+\bigl(\rho^\tau-\rho^{2\tau}\bigr)s-\rho^\tau F_u\bigr\}\nonumber\\
 &=-\tau\bigl(\rho^{-\tau}-1\bigr)\mu(u)\nonumber\\
 &\overset{\text{(b)}}{=}-d\,\mu(u),\label{eq:kdep-mean}
\end{align}
where (a) follows from~\cref{eq:kdep-histograms} and~\cref{eq:kdep-statemeans}, and (b) follows from the definition of $d$ in~\cref{eq:statmean}. Thus, the initial composition cancels as before, the order of the letters cancels because the two halves are identical, and $\E[T]=-d\,\mu(u)$ for every initial composition.

For the fluctuations, replace $\ee_c$ in~\cref{eq:noise} by the basis vector of the input at use $j$. The inputs are fixed, so $S_j-\E[S_j]=\rho(S_{j-1}-\E[S_{j-1}])-N_j$ as before. Hence, \cref{eq:martrepr} holds with the same coefficients $\kappa_t$, and the argument that led to~\cref{eq:statvariance} gives $\E\bigl[\|T+d\,\mu(u)\|_2^2\bigr]\leq18\tau$.

For $\omega\in\{-1,1\}^r$, let $u(\omega)$ be a word whose normalized composition $\mu(\omega)\coloneqq\mu(u(\omega))$ satisfies
\begin{equation}\label{eq:kdep-composition}
 \mu_{2j-1}(\omega)=\frac{2+\omega_j}{4r},\qquad\mu_{2j}(\omega)=\frac{2-\omega_j}{4r},\qquad j=1,\ldots,r,
\end{equation}
with $\mu_K(\omega)=0$ if $K$ is odd. These numbers are integer multiples of $1/\tau$ because $\tau=4rb$, and
\begin{equation}\label{eq:kdep-distance}
 \|\mu(\omega)-\mu(\omega')\|_2^2=\sum_{j=1}^r\frac{2\bigl(\omega_j-\omega'_j\bigr)^2}{16r^2}=\frac{d_{\mathrm H}(\omega,\omega')}{2r^2},
\end{equation}
where $d_{\mathrm H}$ is the Hamming distance. For example, for $K=4$, the sign vectors $(1,1)$ and $(-1,-1)$ give the normalized compositions $(3,1,3,1)/8$ and $(1,3,1,3)/8$, that is, words with $3b,b,3b,b$ and $b,3b,b,3b$ letters of the four colors (\cref{fig:compositions}). A greedy choice gives a set $\Omega\subset\{-1,1\}^r$ with pairwise Hamming distances at least $\lceil r/4\rceil$ and
\begin{equation}\label{eq:kdep-gv}
 |\Omega|\geq\max\bigl\{2,\mathrm{e}^{\eta r}\bigr\},\qquad\eta=\log2-H_{\mathrm b}\left(\frac14\right),
\end{equation}
by Gilbert's greedy argument~\cite[proof of Theorem~1, p.~507]{gilbert1952}: each chosen point excludes a Hamming ball of radius $\lceil r/4\rceil-1<r/4$, which has at most $\mathrm{e}^{rH_{\mathrm b}(1/4)}$ points by the ball bound in the proof of \cref{lem:directcoding} (with $\mathcal A=\{-1,1\}$, $L=r$ and $p_0=1/4$), and two antipodal points are always at distance $r\geq\lceil r/4\rceil$.
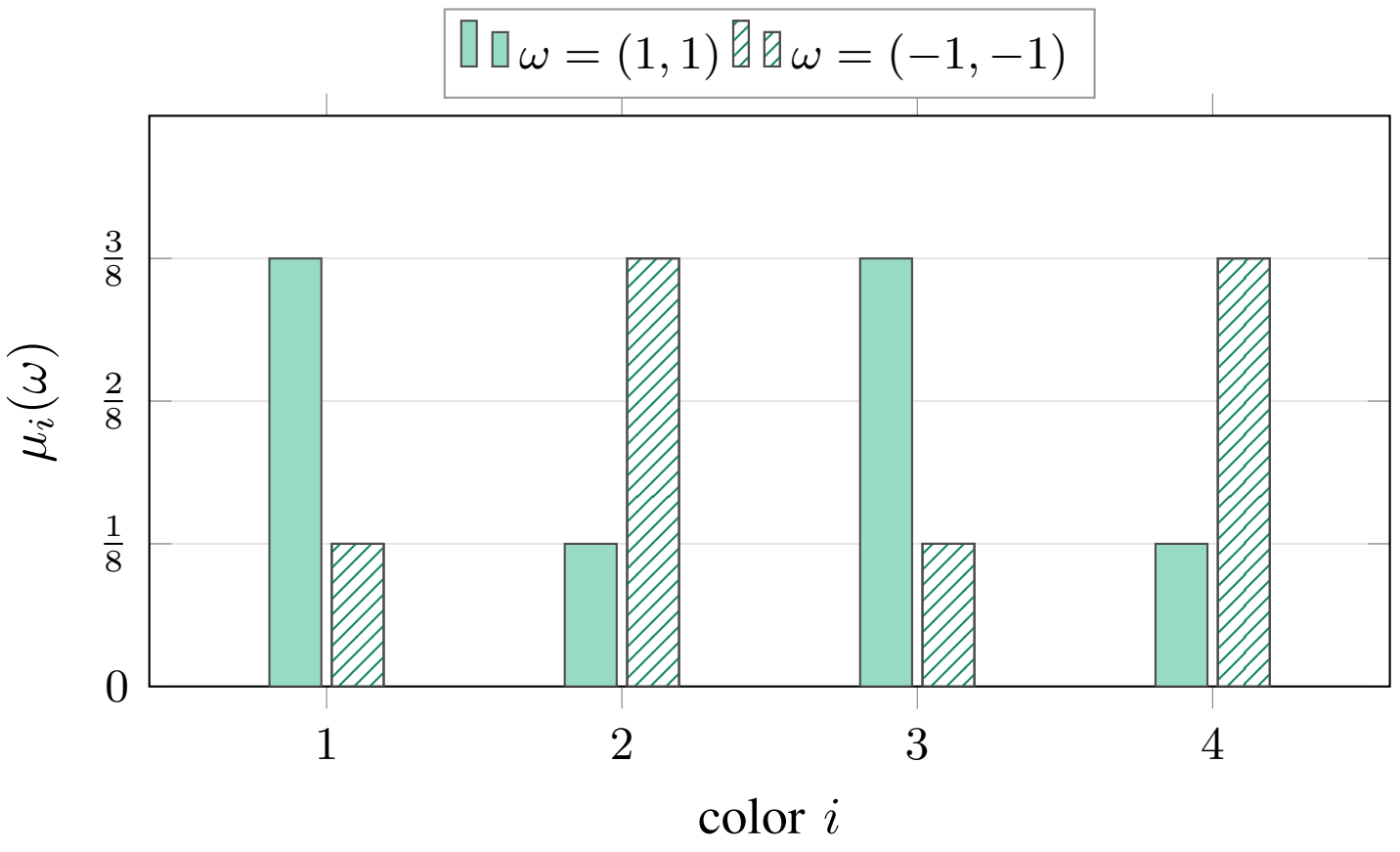
\begin{figure}[t]
\centering
\begin{tikzpicture}
\begin{axis}[
 ybar, bar width=10pt, width=0.55\textwidth, height=0.3\textwidth,
 ymin=0, ymax=0.5, xmin=0.4, xmax=4.6,
 xtick={1,2,3,4}, xlabel={color $i$}, ylabel={$\mu_i(\omega)$},
 ytick={0,0.125,0.25,0.375}, yticklabels={$0$,$\frac18$,$\frac28$,$\frac38$},
 legend style={at={(0.5,1.03)}, anchor=south, legend columns=2, font=\footnotesize, draw=black!40},
 tick label style={font=\footnotesize}, label style={font=\small},
 ymajorgrids, grid style={black!10}]
\addplot[draw=black!70, fill=kbound!45] coordinates {(1,0.375) (2,0.125) (3,0.375) (4,0.125)};
\addlegendentry{$\omega=(1,1)$}
\addplot[draw=black!70, fill=white, postaction={pattern=north east lines, pattern color=kbound!80!black}] coordinates {(1,0.125) (2,0.375) (3,0.125) (4,0.375)};
\addlegendentry{$\omega=(-1,-1)$}
\end{axis}
\end{tikzpicture}
\caption{The normalized compositions $\mu(\omega)$ of~\cref{eq:kdep-composition} for $K=4$ and the sign vectors $\omega=(1,1)$ and $\omega=(-1,-1)$: words with $3b,b,3b,b$ and $b,3b,b,3b$ letters of the four colors, in any fixed order. Each word is sent twice in the same order, and the statistic~\cref{eq:weights} then has the expectation $-d\,\mu(\omega)$ for every initial composition, by~\cref{eq:kdep-mean}. By~\cref{eq:kdep-distance}, $\|\mu(\omega)-\mu(\omega')\|_2^2=d_{\mathrm H}(\omega,\omega')/(2r^2)=1/4$ for these two words.}
\label{fig:compositions}
\end{figure}

Let $\psi$ map the $2\tau$ outputs of a block to a label $\omega'\in\Omega$ that minimizes $\|T+d\,\mu(\omega')\|_2$, with ties broken in a fixed order of $\Omega$. By~\cref{eq:kdep-distance}, distinct $\omega,\omega'\in\Omega$ have $\|d\,\mu(\omega)-d\,\mu(\omega')\|_2^2\geq d^2/(8r)$, so, as in~\cref{eq:inner-error-implies-deviation}, when $u(\omega)$ is sent twice, an error implies $\|T+d\,\mu(\omega)\|_2^2\geq d^2/(32r)$. Hence
\begin{equation}\label{eq:kdep-error}
 \PP\bigl\{\psi(Y^{2\tau})\neq\omega\bigr\}\overset{\text{(a)}}{\leq}\frac{32r\cdot18\tau}{d^2}\overset{\text{(b)}}{\leq}\frac{576\,r(M+1)^2}{\tau^3}\overset{\text{(c)}}{\leq}\frac{576}{32^3}=\frac9{512},
\end{equation}
uniformly over the initial composition, where (a) follows from Markov's inequality, (b) follows from $d\geq\tau^2/(M+1)$ by~\cref{eq:inner-separation-lower}, and (c) follows from~\cref{eq:kdep-tau}. Apply \cref{lem:directcoding} with $\mathcal A=\Omega$, $\ell=2\tau$, $x(\omega)$ the word $u(\omega)$ sent twice, $p_\ast=9/512$ and $p_0=1/12$. Every rate below $(\log|\Omega|-f_{|\Omega|}(1/12))/(2\tau)$ is achievable. As in~\cref{eq:positiveconstant}, this threshold is at least $\gamma\log|\Omega|/(2\tau)$, and, by~\cref{eq:kdep-tau} and~\cref{eq:kdep-gv},
\begin{equation}\label{eq:kdep-rate}
 \frac{\gamma\log|\Omega|}{2\tau}\geq\frac{\gamma\eta r}{72\,(M+1)^{2/3}r^{1/3}}=\frac{\gamma\eta}{72}\left(\frac{r}{M+1}\right)^{2/3}.
\end{equation}
This proves~\cref{eq:kdep} and completes the proof of \cref{prop:kdependence}.
\end{IEEEproof}

\subsection{The order}\label{subsec:order}
\begin{IEEEproof}[Proof of \cref{cor:order}]
The upper bound is~\cref{eq:converse-capacity}. For the lower bound, let
\begin{equation}
 r_1\coloneqq\min\bigl\{\lfloor K/2\rfloor,\lfloor(M+1)/2^{22}\rfloor\bigr\},\qquad k\coloneqq2r_1.
\end{equation} Then, $r_1\geq1$, because $K\geq2$ and $M+1\geq2^{22}$, and $2\leq k\leq K$. Since $\lfloor k/2\rfloor=r_1$ and $M+1\geq2^{22}r_1$, \cref{prop:kdependence} applies to $k$ colors, and, with \cref{prop:merging},
\begin{equation}\label{eq:order-lower}
 \CC\geq C_{M,k}\geq\frac{\gamma\eta}{72}\left(\frac{r_1}{M+1}\right)^{2/3}.
\end{equation}
Finally, $r_1\geq K/2^{23}$. Indeed, if $r_1=\lfloor K/2\rfloor$, then $r_1\geq K/3$. Otherwise, $r_1=\lfloor (M+1)/2^{22}\rfloor\geq (M+1)/2^{23}\geq K/2^{23}$, since $(M+1)/2^{22}\geq1$ and $K\leq M+1$. Substituting into~\cref{eq:order-lower} gives $\CC\geq c_0(K/(M+1))^{2/3}$ with $c_0=2^{-46/3}\gamma\eta/72$, as desired.
\end{IEEEproof}
\section{Discussion}\label{sec:discussion}
\subsection{Random, adversarial, and cooperative removal}
The same bag supports different removal mechanisms. When an adversary chooses the removed ball and the sender does not observe the outputs, the zero-error capacity with two colors and a known initial composition is exactly $(\log2)/(M+1)$ nats per channel use~\cite{ahlswede}. When the message sender chooses the removed balls and a helper supplies the inputs independently of the message, the capacity is the logarithm of the largest eigenvalue of a matrix built from the state transitions~\cite{kobayashi}. Remark~2 in~\cite{kobayashi} suggests that this capacity approaches $\log K$ as $M$ grows. Piret~\cite{piret1989} also studied a permuting mailbox in which both the inserting user and the removing user convey messages. With the initial content known to all users, he constructed rate pairs for a binary alphabet and a store of one symbol that lie beyond time sharing between the two single-user schemes. With a uniformly random removal, \cref{cor:order} gives the intermediate order $(K/M)^{2/3}$. Although the operational
criteria differ, this comparison illustrates how the
removal mechanism affects the capacity scaling.

\subsection{An integrator observed in noise}
Let $K=2$, and let $V_t$ be the initial number of balls of color~1 plus the number of inputs of color~1 up to use $t$. Let $\Pi_t$ be the conditional probability that $Y_t=1$ given the initial state, the inputs up to use $t$ and the outputs before use $t$. Then~\cref{eq:kernel} gives $\Pi_t=\bigl(V_t-\sum_{j=1}^{t-1}\ind_{\{Y_j=1\}}\bigr)/(M+1)$, and hence
\begin{equation}\label{eq:integrator}
 \ind_{\{Y_t=1\}}+\frac1{M+1}\sum_{j=1}^{t-1}\ind_{\{Y_j=1\}}=\frac{V_t}{M+1}+\Delta_t,
\end{equation}
where $\Delta_t\coloneqq\ind_{\{Y_t=1\}}-\Pi_t$ has conditional expectation zero and conditional variance $\Pi_t(1-\Pi_t)\leq1/4$ given the same variables as $\Pi_t$. The receiver therefore observes the integrated input through the gain $1/(M+1)$ in martingale-difference noise. This holds for every encoder, with or without feedback.

Replacing $\Delta_t$ by independent Gaussian noise and the binary input by a power-constrained input, with a fixed positive noise variance and a fixed positive average-input-power constraint, both independent of $M$, gives, after differencing, a Gaussian channel with first-order moving-average noise whose spectrum vanishes at zero frequency. For this surrogate, water-filling gives a nonfeedback capacity
of order $(M+1)^{-2/3}$, and Kim's feedback capacity
formula~\cite{kim2006} gives the same order with feedback.
The bound of Cover and Pombra~\cite{coverpombra}, which they credit to Pinsker and Ebert, limits
the feedback gain to a factor of two.
The surrogate ignores the amplitude constraint of the input, the state dependence of the noise and the boundary of the state space, so it gives no bound for the grabbag channel. It explains the exponent heuristically, and \cref{cor:order} confirms the order.

\subsection{The number of colors}
\begin{figure}[t]
\centering
\begin{tikzpicture}
\begin{loglogaxis}[
 width=0.82\textwidth, height=0.46\textwidth,
 xmin=1.5, xmax=3e6, ymin=5e-4, ymax=1e6,
 xlabel={$K$ (number of colors)},
 ylabel={$(M+1)^{2/3}\times{}$bound (nats per use)},
 xtick={1e1,1e2,1e3,1e4,1e5,1e6},
 ytick={1e-3,1e-1,1e1,1e3,1e5},
 legend pos=north west, legend cell align=left,
 legend style={font=\footnotesize,draw=black!40},
 tick label style={font=\footnotesize}, label style={font=\small},
 grid=major, grid style={black!10}]
\addplot[conv, samples at={2,3,4,5,6,7,8,9,10,11,12,13,14,15,16,17,18,19,20,24,28,32,40,48,56,64,80,96,112,128,160,192,224,256,320,384,448,512,640,768,896,1024,1280,1536,1792,2048,2560,3072,3584,4096,5120,6144,7168,8192,10240,12288,14336,16384,20480,24576,28672,31641,31642,32768,40960,49152,57344,65536,81920,98304,114688,131072,163840,196608,229376,262144,327680,393216,458752,524288,655360,786432,917504,1048576,1310720,1572864,1835008,2097152}]
 {(4+95/24*(1+2*x/2^42)^(1/3))*x^(2/3)};
\addlegendentry{converse $4\lambda_{K,M}+B_{K,M}$ (\cref{thm:converse})}
\addplot[lowb, samples at={2,3,4,5,6,7,8,9,10,11,12,13,14,15,16,17,18,19,20,24,28,32,40,48,56,64,80,96,112,128,160,192,224,256,320,384,448,512,640,768,896,1024,1280,1536,1792,2048,2560,3072,3584,4096,5120,6144,7168,8192,10240,12288,14336,16384,20480,24576,28672,31641,31642,32768,40960,49152,57344,65536,81920,98304,114688,131072,163840,196608,229376,262144,327680,393216,458752,524288,655360,786432,917504,1048576,1310720,1572864,1835008,2097152}]
 {2^28/(2*2^31)*(ln(x)-(-(1/12)*ln(1/12)-(11/12)*ln(11/12))-ln(x-1)/12)};
\addlegendentry{bound~\cref{eq:lower-exact} (\cref{thm:lower})}
\addplot[kdepb, samples at={2,3,4,5,6,7,8,9,10,11,12,13,14,15,16,17,18,19,20,24,28,32,40,48,56,64,80,96,112,128,160,192,224,256,320,384,448,512,640,768,896,1024,1280,1536,1792,2048,2560,3072,3584,4096,5120,6144,7168,8192,10240,12288,14336,16384,20480,24576,28672,31641,31642,32768,40960,49152,57344,65536,81920,98304,114688,131072,163840,196608,229376,262144,327680,393216,458752,524288,655360,786432,917504,1048576,1310720,1572864,1835008,2097152}]
 {(11/12-(-(1/12)*ln(1/12)-(11/12)*ln(11/12))/ln(2))*(ln(2)-(-(1/4)*ln(1/4)-(3/4)*ln(3/4)))/72*floor(x/2)^(2/3)};
\addlegendentry{bound~\cref{eq:kdep} (\cref{prop:kdependence})}
\addplot[only marks, mark=*, mark size=1.2pt, lowerbound, forget plot, samples at={2,3,4,5,6,7,8,9,10,11,12,13,14,15,16,17,18,19,20}]
 {2^28/(2*2^31)*(ln(x)-(-(1/12)*ln(1/12)-(11/12)*ln(11/12))-ln(x-1)/12)};
\addplot[only marks, mark=o, mark size=1.2pt, kbound, forget plot, samples at={2,3,4,5,6,7,8,9,10,11,12,13,14,15,16,17,18,19,20}]
 {(11/12-(-(1/12)*ln(1/12)-(11/12)*ln(11/12))/ln(2))*(ln(2)-(-(1/4)*ln(1/4)-(3/4)*ln(3/4)))/72*floor(x/2)^(2/3)};
\draw[densely dotted, black!60] (axis cs:31642,5e-4) -- (axis cs:31642,1e6);
\node[font=\footnotesize, anchor=north west] at (axis cs:3.4e4,2e5) {$K=31642$};
\node[font=\footnotesize, anchor=south east] at (axis cs:2.5e6,7e-4) {$M+1=2^{42}$};
\end{loglogaxis}
\end{tikzpicture}
\caption{The bounds of \cref{thm:converse,thm:lower} and \cref{prop:kdependence}, multiplied by $(M+1)^{2/3}$, against the number of colors $K$ on logarithmic axes, for $M+1=2^{42}$ and $2\leq K\leq2^{21}$. In this range all three apply, with $\tau=2^{31}$ in \cref{thm:lower}. The converse and the bound~\cref{eq:kdep} both grow like $K^{2/3}$, as \cref{cor:order} states, and the bound~\cref{eq:lower-exact} grows like $\log K$. Of the two lower bounds,~\cref{eq:kdep} is the larger from $K=31642$ on (the dotted line), and~\cref{eq:lower-exact} for every smaller $K$. The lines join the values at integer $K$, and the two lower bounds are marked for $K\leq20$. The curves are proved bounds, not estimates of the capacity.}
\label{fig:kdep}
\end{figure}
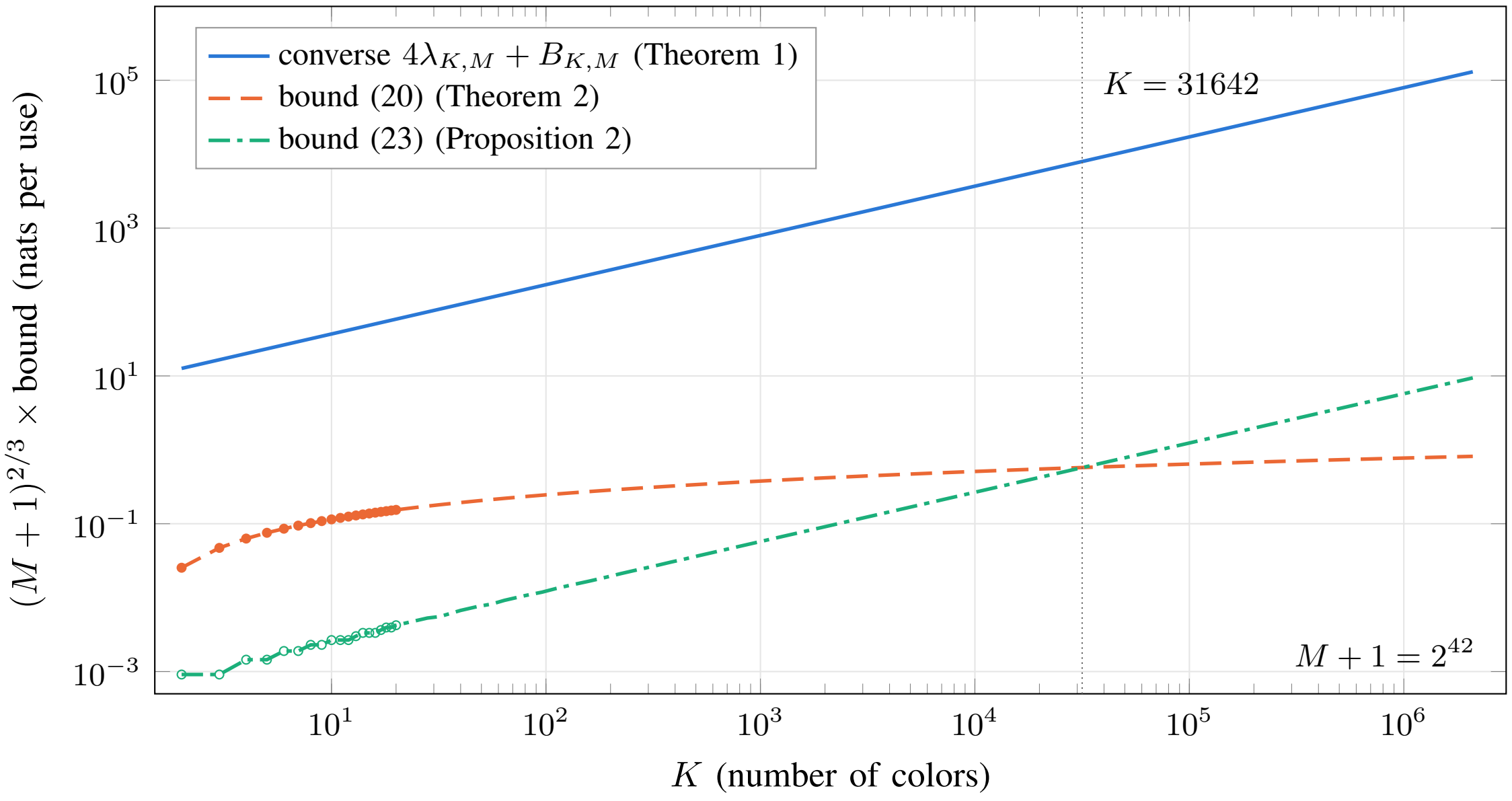

A block of draws resembles a noisy permutation channel acting on the colors inserted during the block. For a strictly positive channel of rank $r$, the capacity of a noisy permutation channel is $(r-1)/2$ in units of the logarithm of the blocklength~\cite{tangpolyanskiy}, matching the general lower bound of Makur~\cite{makur}. This analogy suggested that the number $K-1$ of free coordinates of a composition, rather than $\log K$, governs the dependence on $K$ for large $M$.
\Cref{cor:order} shows that the dependence is $K^{2/3}$ when $K\leq M+1$. Where both lower bounds of \cref{sec:achievability} apply, the bound~\cref{eq:lower-exact}, which grows like $\log K$, is the larger one only for small $K$, because the constant of \cref{prop:kdependence} is small. \Cref{fig:kdep} compares them with \cref{thm:converse} for $M+1=2^{42}$, where both apply for $2\leq K\leq2^{21}$.

\subsection{Related channels}
Several channels carry information through a store, and they differ from the grabbag channel in where their memory comes from. In the single-server queue studied by Anantharam and Verd\'u~\cite{bitsqueues}, the information is carried by the timing of the packets. The capacity is lowest, $\mathrm{e}^{-1}$ nats per average service time, when the service time is exponential, and feedback does not increase it. Sahasranand and Tchamkerten~\cite{queuebounded} study the effect of feedback on queues with bounded service times. The grabbag channel carries its information in the contents of the store instead.

In the channel modeled on contagion of Alajaji and Fuja~\cite{contagion} and in the queue-based channel of Zhong, Alajaji, and Takahara~\cite{qbc}, an urn or a finite queue of balls generates an additive noise process. These channels have memory, but the inputs do not change the contents of the urn or the queue. In the grabbag channel, every input enters the store, so the inputs themselves drive the memory.

For each $M$, the grabbag channel is a strongly connected unifilar finite-state channel (\cref{app:fsc}). Hence, the single-letter upper bound of Sabag, Permuter, and Pfister~\cite{sabagpermuterpfister} and the duality upper bound~\cite{dualityfb} apply to its feedback capacity with an initial state known to both terminals, and therefore to $\CF$. For a fixed $Q$-graph, Sabag, Huleihel, and Permuter~\cite{sabaggraph} formulate the upper bound as a convex optimization over stationary joint distributions. Their lower-bound problem yields graph-based encoders from admissible BCJR-invariant input distributions. Tsur, Aharoni, Goldfeld, and Permuter~\cite{tsurdiscrete} use neural estimation of directed information to optimize input processes for discrete channels with memory. They also use the learned input process to construct $Q$-graphs for bounds on the feedback capacity of unifilar finite-state channels. Aharoni, Sabag, and Permuter~\cite{isinglarge} use reinforcement learning to identify $Q$-graphs that lead to analytic feedback-capacity bounds for Ising channels. In their model, the state is the preceding input. Graph-based encoders provide a possible route to sharper bounds for small stores with a known initial state. How these bounds behave as $M$ grows is not known to us. Feedback does not increase the capacity of the POST($\alpha$) channels of Permuter, Asnani, and Weissman~\cite{postpaw}, whose state is the previous output, nor that of the approximately memoryless surjective POST channels of~\cite{post}. The parameter $\alpha$ of POST($\alpha$) is unrelated to the law $\alpha$ of the state in \cref{sec:converse}. With the initial state known to both terminals, feedback increases the capacity of the grabbag channel for $M=1$ and $K=2$ (\cref{tab:prior}). By \cref{cor:order}, feedback can increase the capacity by at most a constant factor when $K\leq M+1$ and $M$ is large. The size of that factor is open.

\subsection{Variants of the model}\label{subsec:model-variants}
For an integer $b\geq2$, without feedback and up to a shift of the input by one use, a channel that draws a ball from a store of $b$ balls before inserting the input is the grabbag channel with $M=b-1$. Indeed, in that channel the ball drawn at use $t$ is one of the $b-1$ balls kept from earlier uses or the input of use $t-1$. Hence, its nonfeedback capacity scales in $b$ as the nonfeedback capacity of the grabbag channel scales in $M+1$. With feedback, the shift changes which outputs are available when an input is chosen, so the two models need a separate comparison. The model assumes uniform draws and exact observation of the drawn color. Non-uniform drawing probabilities have been studied for a single retained ball under the name chemical channel~\cite{chemical2007}. With the initial state known to both terminals, Wu and Anastasopoulos~\cite{wuanastasopoulos} prove that the feedback capacity of the binary chemical channel is constant on an interval of drawing probabilities. This constant equals the feedback capacity of the uniform trapdoor channel. Sabag, Huleihel, and Permuter~\cite{sabaggraph} also study the non-uniform one-ball trapdoor channel with $Q$-graphs. Whether the order $(K/M)^{2/3}$ persists when each ball is drawn with probability proportional to a weight that is bounded above and below by positive constants independent of $M$ and $K$ is open. The same question arises for noisy observations of the colors and for several insertions per use.

\section{Conclusion}\label{sec:conclusion}
For the insert-then-draw grabbag channel with $M$ retained balls and $K$ colors, both the nonfeedback capacity and the capacity with noiseless strictly causal output feedback are of order $(K/M)^{2/3}$, with absolute constants, when $K\leq M+1$ and $M$ is large. The converse is one bound for feedback codes. For each initial state, it tracks a potential of the posterior law of the state given the earlier outputs, defined through noisy cube roots of the color counts, and a budget on the true state. The lower bound combines a weighted histogram statistic, whose expectation does not depend on the unknown initial composition, with an outer code, and words with prescribed compositions give the power $K^{2/3}$.

Several questions remain open. First, the constants in~\cref{eq:orderK} differ by a factor of about $4.4\times10^8$; for two colors, the sharper bounds in~\cref{eq:twocolor-gap} reduce this ratio to about $613$. Whether $M^{2/3}\CC$ and $M^{2/3}\CF$ converge as $M$ grows, for a fixed $K$, calls for a limit theorem for the capacities on the time scale $M^{2/3}$. The Gaussian surrogate discussed in \cref{sec:discussion}
suggests that such limits may exist, but does not itself
provide bounds on the grabbag channel's capacities. Second, the feedback gain $\CF/\CC$ is bounded in the range of \cref{cor:order}, but its size is open. Third, a sharp characterization of the joint dependence on
$M$ and $K$ when $K>M+1$ remains open. There, the converse of \cref{thm:converse} grows linearly in $K/M$, and for each fixed $M$ it is weaker than the trivial bound $\log K$ when $K$ is large.

Further questions arise from the model. Near a composition of one color, the bag behaves like a timing channel in which a ball of the other color stays for a geometric time with expectation $M+1$. This suggests studying the trade-off between the rate and the number of inserted balls of a signaling color. With feedback delayed by two uses and the initial state known to both terminals, the trapdoor channel loses most of its feedback gain~\cite{delayed}, and it is natural to ask how long a delay the grabbag channel tolerates as $M$ grows. 
The outer-code argument in the proof of \cref{lem:directcoding} already gives exponentially small
error probabilities on the scale $n/M^{2/3}$ at suitable
rates, with a prefactor depending on the number of initial
states. Sharper bounds that control this dependence uniformly
in $M$ and $K$ would clarify the blocklength required at a
given storage size.
 \Cref{prop:accumulation} shows that an additive term of order $M^{1/3}$ is unavoidable in a bound that holds for every length. If the drawn ball is returned instead of removed, the bag grows by one ball per use, and how fast the information about the message can grow with the blocklength in that model is a separate question. Finally, it is open whether both capacities are nonincreasing in $M$.
\section*{Acknowledgment}

The grabbag channel model was introduced by Toby Berger as part of his development of biological information theory. Zhen Zhang and Toby Berger first characterized its asymptotic capacity for binary inputs. Their result was presented at several seminars (see, e.g.,~\cite{bergerseminar}), but the proof remained distributed across a collection of unpublished notes.

In 2009, Berger shared these notes with Jun Chen in the hope that they could be developed into a manuscript. Regrettably, that project remained unfinished at the time of Berger's passing in 2022. The present paper brings this long-delayed undertaking to completion and extends the analysis to nonbinary inputs and to communication with feedback. GPT-6 Astra assisted the authors in organizing and developing the proofs and preparing the manuscript, with access provided through the ChatGPT for Academic Researchers program. Claude Opus 5.5 (Anthropic) provided additional assistance in strengthening the results, verifying the proofs, and refining the manuscript. The authors take full responsibility for the arguments and conclusions presented here.

Zhen Zhang graciously declined coauthorship and encouraged the authors to publish this work. The authors are deeply grateful for his generosity and support. They wish to make clear that the central ideas and the principal intellectual contributions underlying this paper belong to Zhang and Berger. It is a privilege to help bring their work into the published record. The authors offer this paper in grateful remembrance of Berger and in recognition of the insight and generosity with which he shared his ideas.

\appendices
\crefalias{section}{appendix}
\section{Indecomposability and the Finite-State-Channel Formulas}\label{app:fsc}
This appendix records two structural facts that connect the capacities of \cref{def:capacity} with the standard finite-state-channel formulas. The main results do not use them.

\subsection{Indecomposability and the nonfeedback information formula}
For probability laws $P$ and $P'$ on $\mathcal S_{M,K}$, put
\begin{equation}
 \|P-P'\|_{\mathrm{TV}}\coloneqq\frac12\sum_{z\in\mathcal S_{M,K}}|P(z)-P'(z)|.
\end{equation}
For an input string $x^n\in[K]^n$ and an initial state $s$, let $P_{n}(\cdot\mid x^n,s)$ be the law of $S_n$ when the inputs are $x^n$ and $S_0=s$. A finite-state channel is \emph{indecomposable} if
\begin{equation}\label{eq:indecomposable-definition}
 \lim_{n\to\infty}\max_{x^n\in[K]^n}\max_{s,s'\in\mathcal S_{M,K}}\bigl\|P_n(\cdot\mid x^n,s)-P_n(\cdot\mid x^n,s')\bigr\|_{\mathrm{TV}}=0.
\end{equation}
For a finite state space, this is equivalent to the uniform convergence of the individual state probabilities required in~\cite[Eq.~(4.6.26)]{gallager}.

\begin{lemma}[Indecomposability]\label{lem:indecomposable}
For every $n\geq1$, every $x^n\in[K]^n$ and all $s,s'\in\mathcal S_{M,K}$,
\begin{equation}\label{eq:coupling}
 \bigl\|P_n(\cdot\mid x^n,s)-P_n(\cdot\mid x^n,s')\bigr\|_{\mathrm{TV}}\leq M\left(1-\frac1{M+1}\right)^n\leq M\mathrm{e}^{-n/(M+1)}.
\end{equation}
In particular, the grabbag channel is indecomposable.
\end{lemma}

\begin{IEEEproof}
Represent the bag by $M$ slots, numbered $1,\ldots,M$, each holding one ball. At use $t$, let $\Xi_t$ be uniform on $\{0,1,\ldots,M\}$, independently of everything before. If $\Xi_t=0$, output the inserted ball and leave the slots unchanged; if $\Xi_t=k\geq1$, output the ball in slot $k$ and put the inserted ball into that slot. If the slots hold the composition $z$ and the input is $x$, the output is color $i$ with probability $(z_i+\ind_{\{i\}}(x))/(M+1)$, and the composition of the slots becomes $z+\ee_x-\ee_i$; this is the law~\cref{eq:kernel}. Now fix $x^n$, $s$ and $s'$, fill the slots of two copies with the compositions $s$ and $s'$, and run both copies with the same inputs and the same indices $\Xi_1,\ldots,\Xi_n$. Whenever a slot is selected, both copies put the same input ball into it, and the slot agrees in the two copies from then on. On the event $\mathcal E_n$ that every slot is selected at least once by use $n$, the two compositions agree at time $n$. Hence
\begin{equation}\label{eq:coupling-event}
 \bigl\|P_n(\cdot\mid x^n,s)-P_n(\cdot\mid x^n,s')\bigr\|_{\mathrm{TV}}\overset{\text{(a)}}{\leq}\PP\bigl(\mathcal E_n^\complement\bigr)\overset{\text{(b)}}{\leq}M\left(1-\frac1{M+1}\right)^n,
\end{equation}
where (a) follows from the fact that, if $S_n$ and $S'_n$ denote the compositions of the two copies at time $n$, then
\begin{equation}
 \bigl|\PP\{S_n\in\mathcal B\}-\PP\{S'_n\in\mathcal B\}\bigr|\leq\PP\{S_n\neq S'_n\}\leq\PP\bigl(\mathcal E_n^\complement\bigr)
\end{equation}
for every set $\mathcal B\subset\mathcal S_{M,K}$, and (b) follows from the union bound over the $M$ slots, each of which is missed by $n$ independent indices with probability $(1-1/(M+1))^n$. The last inequality in~\cref{eq:coupling} follows from $1-x\leq \mathrm{e}^{-x}$. This completes the proof of \cref{lem:indecomposable}.
\end{IEEEproof}

The channel has finite input and output alphabets and a finite state space, and its transition law is stationary. By \cref{lem:indecomposable}, the coding and capacity theorems for indecomposable finite-state channels apply~\cite[Theorems~3--4]{bbt}, \cite[Theorems~4.6.1, 4.6.2, 4.6.4 and 5.9.2]{gallager}. Blackwell, Breiman, and Thomasian state their theorems for channels whose output is a function of the new state; for the grabbag channel they apply to the state $(S_t,Y_t)$, since the coupling above makes $Y_t$ and $S_t$ agree in the two copies on the event $\mathcal E_{t-1}$.

The upper and lower information capacities coincide~\cite[Theorem~4.6.4]{gallager} and equal the operational capacity under~\cref{eq:error}~\cite[Theorem~5.9.2 and the proof of Theorem~4.6.2]{gallager}; although Gallager states Theorem~4.6.2 for the error probability per source letter, the same argument with Fano's inequality, applied to equiprobable messages, bounds the block error probability~\cref{eq:error}. Consequently, for every $s\in\mathcal S_{M,K}$,
\begin{equation}\label{eq:infoformula}
 \CC=\lim_{n\to\infty}\frac1n\max_{P_{X^n}}I\bigl(X^n\wedge Y^n\bigm|S_0=s\bigr),
\end{equation}
where the limit exists and does not depend on $s$, and the maximum is over the laws of the input string, chosen without feedback.

The formula~\cref{eq:infoformula} gives a second proof of the lower bound~\cref{eq:lower-exact}, which does not need a separate outer code.
\begin{IEEEproof}[Second proof of~\cref{eq:lower-exact}]
Assume the hypotheses of \cref{thm:lower}, let $L\geq1$, and let $n\coloneqq2L\tau$. Take block colors $B_1,\ldots,B_L$ independent and uniform on $[K]$, send each $B_\ell$ as the constant input~\cref{eq:inner-constant-input} of a block of length $2\tau$, and let $\mathbf Y_\ell\in[K]^{2\tau}$ be the outputs of block $\ell$, and $\mathbf Y^L\coloneqq(\mathbf Y_1,\ldots,\mathbf Y_L)$. The decision rule of \cref{subsec:inner} recovers $B_\ell$ from $\mathbf Y_\ell$ with error probability at most $1/12$, whatever the composition at the start of the block. Hence, for every $s\in\mathcal S_{M,K}$,
\begin{align}
 I\bigl(X^n\wedge Y^n\bigm|S_0=s\bigr)
 &\overset{\text{(a)}}{\geq}I\bigl(B^L\wedge\mathbf Y^L\bigm|S_0=s\bigr)\nonumber\\
 &\overset{\text{(b)}}{\geq}L\log K-\sum_{\ell=1}^LH\bigl(B_\ell\bigm|\mathbf Y_\ell,S_0=s\bigr)\nonumber\\
 &\overset{\text{(c)}}{\geq}L\Bigl(\log K-H_{\mathrm b}\bigl(\tfrac1{12}\bigr)-\tfrac1{12}\log(K-1)\Bigr),
\end{align}
where (a) follows from the data-processing inequality for the Markov chain $B^L\to X^n\to Y^n$ given $S_0=s$, the blocks $\mathbf Y^L$ being a grouping of $Y^n$; (b) follows from $H(B^L)=L\log K$ and $H(B^L\mid\mathbf Y^L,S_0=s)\leq\sum_{\ell=1}^LH(B_\ell\mid\mathbf Y_\ell,S_0=s)$, which does not require the output blocks to be independent; and (c) follows from Fano's inequality and the monotonicity of $p\mapsto H_{\mathrm b}(p)+p\log(K-1)$ on $[0,(K-1)/K]$. Dividing by $n$ and applying~\cref{eq:infoformula} gives~\cref{eq:lower-exact}, which completes the second proof.
\end{IEEEproof}
The limit in~\cref{eq:infoformula} is taken with $M$ and $K$ fixed. The blocks of length $2\tau$, of order $M^{2/3}$, need not erase the initial composition: the error bound of the decision rule is uniform over it, and the dependence between successive blocks stays in the conditional entropies.

\subsection{Unifilarity, connectivity and the feedback formula}
Dabora and Goldsmith~\cite[Theorem~1 and Corollary~1]{daboragoldsmith} characterize the feedback capacity of indecomposable finite-state channels whose initial state is unknown to both terminals. Their characterization uses an initial segment without feedback. The length of this segment tends to infinity and is a vanishing fraction of the blocklength. The subsequent encoder ignores the input and output history of that segment.

The next state of the grabbag channel is a deterministic function of the state, the input and the output, $S_t=S_{t-1}+\ee_{X_t}-\ee_{Y_t}$; that is, the channel is \emph{unifilar}. It is also \emph{strongly connected}. Let $s\neq s'$ be states in $\mathcal S_{M,K}$. There are a color $a$ with $s_a>s'_a$ and a color $c$ with $s_c<s'_c$. The input $c$ followed by the output $a$ has probability $s_a/(M+1)>0$ and leads to $s-\ee_a+\ee_c$, and this step lowers $\frac12\sum_{i=1}^K|s_i-s'_i|$ by one. Hence, $s'$ is reached from $s$ in at most $M$ steps of positive probability. Under inputs drawn independently and uniformly from $[K]$, the same paths have positive probability, so the grabbag channel is connected in the sense of~\cite[Definition~2]{trapdoorfb}. By~\cite[Theorem~3]{trapdoorfb}, the feedback capacity with an initial state $s$ known to both terminals is
\begin{equation}\label{eq:fbformula}
 C^{\mathrm{fb,known}}_{M,K}(s)=\lim_{n\to\infty}\frac1n\max_{\{P(x_t\mid s_{t-1},y^{t-1})\}_{t=1}^n}\sum_{t=1}^nI\bigl((X_t,S_{t-1})\wedge Y_t\bigm|Y^{t-1},S_0=s\bigr),
\end{equation}
and $\CF\leq C^{\mathrm{fb,known}}_{M,K}(s)$ for every $s$, because a code that meets the criterion~\cref{eq:error} can be used when the initial state is revealed to both terminals. With $s$ known, the encoder can track the state from its
own inputs and the feedback. The decoder generally cannot
determine the state exactly, because it does not know
the input sequence. Shemuel, Sabag, and Permuter~\cite{shemuelsabagpermuter} study strongly connected finite-state channels with feedback and causal state information at the encoder. Their model makes the initial state available to the encoder. It does not reveal that state to the decoder.
 \Cref{thm:converse} bounds $C^{\mathrm{fb,known}}_{M,K}(s)$ directly and does not use~\cref{eq:fbformula}.

A causal stochastic input law in~\cref{eq:fbformula} can be realized by a random deterministic strategy $Q$, chosen independently of the draws of the channel. With $s$ known, the input $X_t$ and the state $S_{t-1}$ are then functions of $Q$ and $Y^{t-1}$, and $Y_t$ depends on $(Q,Y^{t-1})$ only through them. Hence, each term of the sum in~\cref{eq:fbformula} equals $I(Q\wedge Y_t\mid Y^{t-1},S_0=s)$, and the sum equals $I(Q\wedge Y^n\mid S_0=s)$ by the chain rule. This is in general not $I(X^n\wedge Y^n\mid S_0=s)$.

\section{Proofs of the Converse Lemmas}
\label{app:converse-lemmas}

This appendix proves \cref{lem:oneuse,lem:budget}.
We first record the density representation of the potential
and establish two auxiliary estimates: a shift bound for
Laplace-smoothed random vectors and an error bound for the
cube-root estimator. Throughout, logarithms are natural, and
$Z$ and $g$ are as defined in
\cref{eq:laplace,eq:cuberootmap}. Whenever the auxiliary noise
$Z$ is introduced, it is independent of the channel variables
under consideration.

\subsection{Properties of the potential}
\label{app:potential-properties}

Let $\nu$ be a finitely supported probability law on
$\mathbb Z_{\geq0}^K$, and let $R\sim\nu$ be independent of $Z$.
For $z\in\mathbb Z_{\geq0}^K$, write
$p_z(v)\coloneqq p_Z(v-g(z))$ for the density of $g(z)+Z$.
The random vector $V_\nu=g(R)+Z$ has density
\begin{equation}\label{eq:mixture-density}
 p_\nu(v)
 =\sum_{z\in\mathbb Z_{\geq0}^K:\,\nu(z)>0}\nu(z)p_z(v),
 \qquad v\in\mathbb R^K.
\end{equation}
In particular, $p_\nu$ is strictly positive.

We first verify that its differential entropy is finite.
Since $p_\nu(v)\leq3^K$, the function $-\log p_\nu(v)$ is
bounded below. Fix any $z_0$ with $\nu(z_0)>0$.
The inequality $p_\nu(v)\geq\nu(z_0)p_{z_0}(v)$ gives
\begin{equation}\label{eq:potential-entropy-integrability}
 -\log p_\nu(v)
 \leq
 -\log\nu(z_0)-K\log3+6\|v-g(z_0)\|_1.
\end{equation}
The right side is integrable under $p_\nu$, because $R$
has finite support and $Z$ has a finite first moment.
Consequently, $h(V_\nu)$ is finite. The same argument
applies to every finite mixture of translates of $p_Z$,
including mixtures whose shifts lie in $\mathbb R^K$.

Since $h(g(z)+Z)=h(Z)$ for every $z$~\cite[Theorem~8.6.3]{covertho}, we have
\begin{align}
 \Phi(\nu)
 &=-\int_{\mathbb R^K}p_\nu(v)\log p_\nu(v)\,\mathrm dv+\sum_{z\in\mathbb Z_{\geq0}^K:\,\nu(z)>0}\nu(z)\int_{\mathbb R^K}p_z(v)\log p_z(v)\,\mathrm dv\nonumber\\
 &\overset{\text{(a)}}{=}\sum_{z\in\mathbb Z_{\geq0}^K:\,\nu(z)>0}\nu(z)\int_{\mathbb R^K}p_z(v)\log\frac{p_z(v)}{p_\nu(v)}\,\mathrm dv
 =\sum_{z\in\mathbb Z_{\geq0}^K:\,\nu(z)>0}\nu(z)\,D\bigl(p_z\,\big\Vert\,p_\nu\bigr),\label{eq:potential-divergence}
\end{align}
where (a) follows from~\cref{eq:mixture-density}, which gives
\begin{equation}
 \int_{\mathbb R^K}p_\nu(v)\log p_\nu(v)\,\mathrm dv=\sum_{z\in\mathbb Z_{\geq0}^K:\,\nu(z)>0}\nu(z)\int_{\mathbb R^K}p_z(v)\log p_\nu(v)\,\mathrm dv,
\end{equation}
and $D(\cdot\,\Vert\,\cdot)$ is the relative entropy between densities~\cite[Eq.~(8.46)]{covertho}. Consequently,
\begin{equation}\label{eq:potential-range-proof}
 0\overset{\text{(a)}}{\leq}\Phi(\nu)\overset{\text{(b)}}{\leq}\sum_{z\in\mathbb Z_{\geq0}^K:\,\nu(z)>0}\nu(z)\log\frac1{\nu(z)}=H(R),
\end{equation}
where (a) follows from~\cref{eq:potential-divergence} and the nonnegativity of relative entropy~\cite[Theorem~8.6.1]{covertho}, and (b) follows from~\cref{eq:potential-divergence} and $p_\nu\geq\nu(z)\,p_z$, which gives $\log(p_z/p_\nu)\leq\log(1/\nu(z))$ wherever $p_z>0$. In particular, $\Phi(\nu)=0$ whenever $\nu$ is a point mass.

\subsection{A shift bound}
\label{app:shift}

\begin{lemma}[A shift bound]\label{lem:shift}
Let $A_0$ and $A_1$ be random vectors in $\mathbb R^K$,
defined on one probability space and taking finitely many
values, and let $Z$ be independent of $(A_0,A_1)$.
Then
\begin{equation}\label{eq:shift}
 h(A_1+Z)-h(A_0+Z)
 \leq6\,\E\bigl[\|A_1-A_0\|_1\bigr].
\end{equation}
\end{lemma}

\begin{IEEEproof}
Put $V_0\coloneqq A_0+Z$ and $V_1\coloneqq A_1+Z$, and let $p_{V_0}$ be the density of $V_0$. For all $v,w,a\in\mathbb R^K$,
\begin{equation}\label{eq:laplace-lipschitz}
 p_Z(v-a)=\prod_{i=1}^K3\mathrm{e}^{-6|v_i-a_i|}\overset{\text{(a)}}{\leq}\prod_{i=1}^K3\mathrm{e}^{-6|w_i-a_i|+6|v_i-w_i|}=\mathrm{e}^{6\|v-w\|_1}\,p_Z(w-a),
\end{equation}
where (a) follows from the triangle inequality $|w_i-a_i|\leq|v_i-a_i|+|v_i-w_i|$. Since $p_{V_0}$ is a convex combination of translates $p_Z(\cdot-a)$ over the realizations $a$ of $A_0$ with positive probability, \cref{eq:laplace-lipschitz} gives $p_{V_0}(v)\leq \mathrm{e}^{6\|v-w\|_1}p_{V_0}(w)$, and, exchanging $v$ and $w$,
\begin{equation}\label{eq:logp0-lipschitz}
 \bigl|\log p_{V_0}(v)-\log p_{V_0}(w)\bigr|\leq6\|v-w\|_1,\qquad v,w\in\mathbb R^K.
\end{equation}
Then
\begin{align}
 h(V_1)-h(V_0)
 &\overset{\text{(a)}}{\leq}-\E\bigl[\log p_{V_0}(V_1)\bigr]+\E\bigl[\log p_{V_0}(V_0)\bigr]\nonumber\\
 &=\E\bigl[\log p_{V_0}(V_0)-\log p_{V_0}(V_1)\bigr]\nonumber\\
 &\overset{\text{(b)}}{\leq}6\,\E\bigl[\|V_0-V_1\|_1\bigr]\nonumber\\
 &\overset{\text{(c)}}{=}6\,\E\bigl[\|A_0-A_1\|_1\bigr],\label{eq:shift-chain}
\end{align}
where (a) follows from $h(V_0)=-\E[\log p_{V_0}(V_0)]$ and from the nonnegativity of the relative entropy between the density of $V_1$ and $p_{V_0}$~\cite[Theorem~8.6.1]{covertho}, which gives $h(V_1)\leq-\E[\log p_{V_0}(V_1)]$; (b) follows from~\cref{eq:logp0-lipschitz}; and (c) follows from $V_0-V_1=A_0-A_1$. Since $V_1=A_1+Z$ and $V_0=A_0+Z$, this is~\cref{eq:shift}, which completes the proof of \cref{lem:shift}.
\end{IEEEproof}

\subsection{A cube-root estimator}
\label{app:cuberoot}

For real numbers $r\geq0$ and $A>0$, put
\begin{equation}\label{eq:dpoisson}
 d_{\mathrm P}(r\,\Vert\,A)\coloneqq r\log\frac rA-r+A,
\end{equation}
with $0\log0=0$. This is the relative entropy between the Poisson laws with expectations $r$ and $A$, because
\begin{equation}\label{eq:dpoisson-derivation}
 \sum_{k=0}^\infty\frac{\mathrm{e}^{-r}r^k}{k!}\log\frac{\mathrm{e}^{-r}r^k/k!}{\mathrm{e}^{-A}A^k/k!}
 =\sum_{k=0}^\infty\frac{\mathrm{e}^{-r}r^k}{k!}\left(A-r+k\log\frac rA\right)
 =A-r+r\log\frac rA,
\end{equation}
where the last equality uses $\sum_{k=0}^\infty k\,\mathrm{e}^{-r}r^k/k!=r$ (for $r=0$ only the term $k=0$ is present, and both sides equal $A$). Hence, $d_{\mathrm P}(r\,\Vert\,A)\geq0$~\cite[Theorem~2.6.3]{covertho}.

\begin{lemma}[A cube-root estimator]\label{lem:cuberoot}
Let $r\geq0$, put $\theta\coloneqq(r+2)^{1/3}$, and define
$A\coloneqq\max\{1,\theta+Z_1\}^3$. Then
\begin{equation}\label{eq:cuberoot}
 \E\bigl[d_{\mathrm P}(r\,\Vert\,A)\bigr]
 \leq\frac{95}{24}\,\theta.
\end{equation}
\end{lemma}

\begin{IEEEproof}
Put $\sigma\coloneqq1/6$. For every integer $k\geq1$, the positive part $Z_1^+$ and the negative part $Z_1^-$ of $Z_1$ satisfy
\begin{equation}\label{eq:laplace-moments}
 \E\Bigl[\bigl(Z_1^+\bigr)^k\Bigr]\overset{\text{(a)}}{=}\E\Bigl[\bigl(Z_1^-\bigr)^k\Bigr]\overset{\text{(b)}}{=}\int_0^\infty v^k\,3\mathrm{e}^{-6v}\,\mathrm dv\overset{\text{(c)}}{=}\frac{3\,k!}{6^{k+1}}\overset{\text{(d)}}{=}\frac{k!\,\sigma^k}2,
\end{equation}
where (a) follows from the symmetry of the density $3\mathrm{e}^{-6|v|}$ about $0$; (b) follows from $Z_1^-=\max\{-Z_1,0\}$, the density $3\mathrm{e}^{-6|v|}$ and the substitution $v\mapsto-v$; (c) follows from $k$ integrations by parts; and (d) follows from $\sigma=1/6$.
Put $\Theta\coloneqq\max\{1,\theta+Z_1\}$, so that $A=\Theta^3$, and put $f(x)\coloneqq x^3-1-3\log x$ for $x>0$. By~\cref{eq:dpoisson},
\begin{equation}\label{eq:cuberoot-f}
 d_{\mathrm P}(\theta^3\,\Vert\,A)=\theta^3\log\frac{\theta^3}{\Theta^3}-\theta^3+\Theta^3=\theta^3f\left(\frac\Theta\theta\right).
\end{equation}
Since $r\geq0$, we have $\theta\geq2^{1/3}>1$. We first bound $d_{\mathrm P}(\theta^3\,\Vert\,A)$ on three events.

\emph{The event $\{Z_1\geq0\}$.} Here, $\Theta=\theta+Z_1$. For the real number $x\coloneqq Z_1/\theta\geq0$ (pointwise on this event),
\begin{equation}\label{eq:cuberoot-case1}
 f(1+x)=3x+3x^2+x^3-3\log(1+x)\overset{\text{(a)}}{\leq}3x+3x^2+x^3-3\left(x-\frac{x^2}2\right)=\frac92x^2+x^3,
\end{equation}
where (a) follows from $\log(1+x)\geq x-x^2/2$ for $x\geq0$. By~\cref{eq:cuberoot-f}, $d_{\mathrm P}(\theta^3\,\Vert\,A)\leq\frac92\theta Z_1^2+Z_1^3$ on this event.

\emph{The event $\{-\theta/2\leq Z_1<0\}$.} Here, $\theta/2\leq\Theta\leq\theta$ and $|\Theta-\theta|\leq|Z_1|$, because $\theta+Z_1\leq\Theta\leq\theta$. We have $f(1)=0$ and $f'(1)=0$, and $f''(x)=6x+3x^{-2}$ is decreasing on $[1/2,1]$, so that $f''(x)\leq f''(1/2)=15$ there. By Taylor's theorem, $f(x)\leq\frac{15}2(x-1)^2$ for $x\in[1/2,1]$, and~\cref{eq:cuberoot-f} gives
\begin{equation}\label{eq:cuberoot-case2}
 d_{\mathrm P}(\theta^3\,\Vert\,A)\leq\frac{15}2\,\theta^3\left(\frac\Theta\theta-1\right)^2=\frac{15}2\,\theta\,(\Theta-\theta)^2\leq\frac{15}2\,\theta Z_1^2.
\end{equation}

\emph{The event $\{Z_1<-\theta/2\}$.} Here, $1\leq\Theta\leq\theta$, so that $\Theta/\theta\in[1/\theta,1]$ and $f(\Theta/\theta)\leq-3\log(\Theta/\theta)\leq3\log\theta$. Hence, $d_{\mathrm P}(\theta^3\,\Vert\,A)\leq3\theta^3\log\theta$ on this event. Moreover,
\begin{equation}\label{eq:cuberoot-case3}
 3\theta^3\log\theta\cdot\PP\left\{Z_1<-\frac\theta2\right\}
 \overset{\text{(a)}}{\leq}3\theta^3\log\theta\cdot\frac{16\,\E\bigl[(Z_1^-)^4\bigr]}{\theta^4}
 \overset{\text{(b)}}{\leq}24\,\E\Bigl[\bigl(Z_1^-\bigr)^4\Bigr]\,\theta,
\end{equation}
where (a) follows from Markov's inequality applied to $\bigl(Z_1^-\bigr)^4$, and (b) follows from $\log\theta\leq\theta^2/2$.

Adding the three bounds and using~\cref{eq:laplace-moments}, we obtain
\begin{align}
 \E\bigl[d_{\mathrm P}(\theta^3\,\Vert\,A)\bigr]
 &\leq\frac92\,\theta\,\E\Bigl[\bigl(Z_1^+\bigr)^2\Bigr]+\E\Bigl[\bigl(Z_1^+\bigr)^3\Bigr]+\frac{15}2\,\theta\,\E\Bigl[\bigl(Z_1^-\bigr)^2\Bigr]+24\,\theta\,\E\Bigl[\bigl(Z_1^-\bigr)^4\Bigr]\nonumber\\
 &\overset{\text{(a)}}{=}12\sigma^2\theta+3\sigma^3+288\sigma^4\theta\nonumber\\
 &\overset{\text{(b)}}{\leq}\bigl(12\sigma^2+3\sigma^3+288\sigma^4\bigr)\theta=\frac{41}{72}\,\theta,\label{eq:cuberoot-cubesum}
\end{align}
where (a) follows from~\cref{eq:laplace-moments}, and (b) follows from $\theta\geq1$.

It remains to pass from $\theta^3$ to $r=\theta^3-2$. Since $\partial d_{\mathrm P}(x\,\Vert\,A)/\partial x=\log(x/A)$,
\begin{align}
 d_{\mathrm P}(r\,\Vert\,A)-d_{\mathrm P}(\theta^3\,\Vert\,A)
 &\overset{\text{(a)}}{=}\int_r^{\theta^3}\log\frac Ax\,\mathrm dx\nonumber\\
 &\overset{\text{(b)}}{=}2\log\frac A{\theta^3}+\int_r^{\theta^3}\log\frac{\theta^3}x\,\mathrm dx\nonumber\\
 &\overset{\text{(c)}}{\leq}2\left(\frac A{\theta^3}-1\right)+2=\frac{2A}{\theta^3},\label{eq:cuberoot-shift}
\end{align}
where (a) follows from the fundamental theorem of calculus and $\partial d_{\mathrm P}(x\,\Vert\,A)/\partial x=\log(x/A)$; (b) follows from $\log(A/x)=\log(A/\theta^3)+\log(\theta^3/x)$ and $\theta^3-r=2$; and (c) follows from $\log y\leq y-1$ and from
\begin{equation}\label{eq:cuberoot-integral}
 \int_r^{\theta^3}\log\frac{\theta^3}x\,\mathrm dx\overset{\text{(a)}}{=}\theta^3\int_{1-\varepsilon}^1\log\frac1y\,\mathrm dy\overset{\text{(b)}}{=}\theta^3\bigl[\varepsilon+(1-\varepsilon)\log(1-\varepsilon)\bigr]\overset{\text{(c)}}{\leq}\theta^3\varepsilon=2,\qquad\varepsilon\coloneqq\frac2{\theta^3}\in(0,1].
\end{equation}
In~\cref{eq:cuberoot-integral}, (a) follows from the substitution $x=\theta^3y$ and $r=\theta^3(1-\varepsilon)$; (b) follows from the fact that $y-y\log y$ is an antiderivative of $\log(1/y)$; and (c) follows from $(1-\varepsilon)\log(1-\varepsilon)\leq0$. Here, $\varepsilon\leq1$ because $\theta^3=r+2\geq2$. For $r=0$, the integrals in~\cref{eq:cuberoot-shift} and~\cref{eq:cuberoot-integral} are improper at $x=0$, and these identities hold by continuity, since $x\log x\to0$ as $x\downarrow0$.
Since $\Theta\leq\theta+|Z_1|$ and $\theta\geq1$,
\begin{equation}\label{eq:cuberoot-moment}
 \frac{\E[A]}{\theta^3}\leq\E\Biggl[\left(1+\frac{|Z_1|}\theta\right)^3\Biggr]\leq\E\bigl[(1+|Z_1|)^3\bigr]\overset{\text{(a)}}{=}1+3\sigma+6\sigma^2+6\sigma^3=\frac{61}{36},
\end{equation}
where (a) follows from~\cref{eq:laplace-moments}, which gives $\E\bigl[|Z_1|^k\bigr]=k!\,\sigma^k$. Combining~\cref{eq:cuberoot-cubesum}, \cref{eq:cuberoot-shift} and~\cref{eq:cuberoot-moment},
\begin{equation}\label{eq:cuberoot-final}
 \E\bigl[d_{\mathrm P}(r\,\Vert\,A)\bigr]\leq\frac{41}{72}\,\theta+\frac{61}{18}\overset{\text{(a)}}{\leq}\left(\frac{41}{72}+\frac{244}{72}\right)\theta=\frac{95}{24}\,\theta,
\end{equation}
where (a) follows from $\theta\geq1$. This completes the proof of \cref{lem:cuberoot}.
\end{IEEEproof}

\subsection{Proof of the one-use information bound}
\label{app:oneuse-proof}

\begin{IEEEproof}[Proof of \cref{lem:oneuse}]
Use the notation preceding \cref{lem:oneuse}.
In addition, let $\beta$ be the law of $U$, and, for
$q_i>0$, let $\beta^i$ be the conditional law of $U$
given $Y=i$.
Introduce $Z$ independently of $(S,X,Y)$ and put
$V\coloneqq g(U)+Z$.
The density of $V$ is $p_\beta$, and its conditional
density given $Y=i$ is $p_{\beta^i}$.
The density $p_\beta$ is positive because $p_Z$ is positive.
For $v\in\mathbb R^K$, Bayes' rule gives the conditional law of $(U,Y)$
given $V=v$: each pair $(u,i)$ has probability
$\PP\{U=u,Y=i\}\,p_u(v)/p_\beta(v)$.
Conditional probabilities and expectations given $V=v$ are taken
under this law.
For $v\in\mathbb R^K$, define the probability vector $\bar p(v)$ on $[K]$ by
\begin{equation}\label{eq:oneuse-pbar}
 \bar p_i(v)\coloneqq\frac1{p_\beta(v)}\sum_{u\in\mathcal S_{M+1,K}}\beta(u)\,\frac{u_i}{M+1}\,p_u(v),\qquad i\in[K].
\end{equation}
Since $\PP\{Y=i\mid U=u\}=u_i/(M+1)$, the right side of~\cref{eq:oneuse-pbar} is the conditional probability of $\{Y=i\}$ given $V=v$, so that
\begin{equation}\label{eq:oneuse-bayes}
 \E\left[\frac{U}{M+1}\,\middle|\,V=v\right]=\bar p(v),\qquad
 q_i\,p_{\beta^i}(v)=p_\beta(v)\,\bar p_i(v),\qquad i\in[K]\text{ with }q_i>0,\ v\in\mathbb R^K;
\end{equation}
if $q_i=0$, then $u_i=0$ on the support of $\beta$, and $\bar p_i\equiv0$.

We first decompose $I(U\wedge Y)$. For every $v$ and every probability vector $\hat p$ on $[K]$,
\begin{equation}\label{eq:oneuse-compensation}
 \E\left[D\left(\frac{U}{M+1}\,\middle\Vert\,\hat p\right)\;\middle|\;V=v\right]=\E\left[D\left(\frac{U}{M+1}\,\middle\Vert\,\bar p(v)\right)\;\middle|\;V=v\right]+D\bigl(\bar p(v)\,\big\Vert\,\hat p\bigr),
\end{equation}
with both sides in $[0,\infty]$. If $\hat p_i=0<\bar p_i(v)$ for some $i$, both sides are infinite, since then $u_i>0$ for some $u$ with $\beta(u)>0$, and $p_u(v)>0$. Otherwise, for every $u$ with $\beta(u)>0$ and every $i$ with $u_i>0$, we have $\bar p_i(v)>0$ and $\hat p_i>0$. For these $u$ and $i$, write $\log(u_i/((M+1)\hat p_i))=\log(u_i/((M+1)\bar p_i(v)))+\log(\bar p_i(v)/\hat p_i)$, average over $i$ with the weights $u_i/(M+1)$ and then over $U$ given $V=v$, and use the first identity in~\cref{eq:oneuse-bayes}. The coordinates with $\bar p_i(v)=0$ contribute zero to both sides. This is the identity of~\cite[Eq.~(10)]{lapidothmoser} for the conditional law of $U$ given $V=v$ and the channel that maps $u$ to the law $u/(M+1)$. Since $\PP\{Y=\cdot\mid U\}=U/(M+1)$ and $\PP\{Y=\cdot\}=q$, the definition of mutual information gives $I(U\wedge Y)=\E[D(U/(M+1)\,\Vert\,q)]$. Applying~\cref{eq:oneuse-compensation} with $\hat p=q$ and averaging over $V$, we obtain
\begin{equation}\label{eq:oneuse-split}
 I(U\wedge Y)=\E\Bigl[D\bigl(\bar p(V)\,\big\Vert\,q\bigr)\Bigr]+\E\left[D\left(\frac{U}{M+1}\,\middle\Vert\,\bar p(V)\right)\right].
\end{equation}
We bound the two terms in turn.

\emph{The first term in~\cref{eq:oneuse-split}.} We have
\begin{align}
 \E\Bigl[D\bigl(\bar p(V)\,\big\Vert\,q\bigr)\Bigr]
 &=\int_{\mathbb R^K}p_\beta(v)\sum_{i=1}^K\bar p_i(v)\log\frac{\bar p_i(v)}{q_i}\,\mathrm dv\nonumber\\
 &\overset{\text{(a)}}{=}\sum_{i\in[K]:\,q_i>0}q_i\int_{\mathbb R^K}p_{\beta^i}(v)\log\frac{p_{\beta^i}(v)}{p_\beta(v)}\,\mathrm dv\nonumber\\
 &\overset{\text{(b)}}{=}-\int_{\mathbb R^K}p_\beta(v)\log p_\beta(v)\,\mathrm dv+\sum_{i\in[K]:\,q_i>0}q_i\int_{\mathbb R^K}p_{\beta^i}(v)\log p_{\beta^i}(v)\,\mathrm dv\nonumber\\
 &\overset{\text{(c)}}{=}\Phi(\beta)-\sum_{i\in[K]:\,q_i>0}q_i\,\Phi(\beta^i)\nonumber\\
 &=\Phi(\alpha)-\sum_{i\in[K]:\,q_i>0}q_i\,\Phi(\alpha^i)+\bigl[\Phi(\beta)-\Phi(\alpha)\bigr]+\sum_{i\in[K]:\,q_i>0}q_i\bigl[\Phi(\alpha^i)-\Phi(\beta^i)\bigr],\label{eq:oneuse-VY}
\end{align}
where (a) follows from the second identity in~\cref{eq:oneuse-bayes}, which gives $p_\beta\bar p_i=q_ip_{\beta^i}$ and $\bar p_i/q_i=p_{\beta^i}/p_\beta$; (b) follows from $\sum_{i\in[K]:\,q_i>0}q_ip_{\beta^i}=p_\beta$; and (c) follows from the definition~\cref{eq:potential}, the finite entropies being added and subtracted.

The first bracket in~\cref{eq:oneuse-VY} is the cost of the insertion. By~\cref{eq:potential}, it equals $h(g(S+\ee_X)+Z)-h(g(S)+Z)$. Apply \cref{lem:shift} with $A_0=g(S)$ and $A_1=g(S+\ee_X)$. Only the coordinate $X$ of $g$ changes, from $(S_X+2)^{1/3}$ to $(S_X+3)^{1/3}$, so
\begin{equation}\label{eq:oneuse-insertion}
 \Phi(\beta)-\Phi(\alpha)\leq6\,\E\bigl[(S_X+3)^{1/3}-(S_X+2)^{1/3}\bigr]\overset{\text{(a)}}{\leq}6\,\E\left[\frac13(S_X+2)^{-2/3}\right]=2a(\pi),
\end{equation}
where (a) follows from the inequality $(x+1)^{1/3}-x^{1/3}\leq\frac13x^{-2/3}$ for $x>0$, which holds because the derivative of $x\mapsto x^{1/3}$ is decreasing.

The last sum in~\cref{eq:oneuse-VY} is the cost of the removal. Fix $i$ with $q_i>0$. Given $Y=i$, we have $U_i\geq1$, the law of $U$ is $\beta^i$, the law of $U-\ee_i$ is $\alpha^i$, and $Z$ is independent of $U$. Apply \cref{lem:shift} under the conditional law given $Y=i$, with $A_0=g(U)$ and $A_1=g(U-\ee_i)$. Only the coordinate $i$ changes, by $\delta(U_i)$, where
\begin{equation}\label{eq:oneuse-delta}
 \delta(r)\coloneqq(r+2)^{1/3}-(r+1)^{1/3}\leq\frac13(r+1)^{-2/3},\qquad r\geq0,
\end{equation}
by the same inequality. Hence, $\Phi(\alpha^i)-\Phi(\beta^i)\leq6\,\E[\delta(U_i)\mid Y=i]$, and
\begin{align}
 \sum_{i\in[K]:\,q_i>0}q_i\,\E\bigl[\delta(U_i)\bigm|Y=i\bigr]
 &\overset{\text{(a)}}{=}\E\left[\sum_{i=1}^K\frac{U_i}{M+1}\,\delta(U_i)\right]\nonumber\\
 &\overset{\text{(b)}}{\leq}\frac1{3(M+1)}\,\E\left[\sum_{i\in[K]:\,U_i\geq1}U_i(U_i+1)^{-2/3}\right]\nonumber\\
 &\overset{\text{(c)}}{\leq}\frac1{3(M+1)}\,\E\left[\sum_{i\in[K]:\,U_i\geq1}U_i^{1/3}\right]\nonumber\\
 &\overset{\text{(d)}}{\leq}\frac13\left(\frac{\min\{K,M+1\}}{M+1}\right)^{2/3}=\frac13\lambda_{K,M},\label{eq:oneuse-removal}
\end{align}
where (a) follows from
\[
 q_i\,\E\bigl[\delta(U_i)\bigm|Y=i\bigr]=\E\bigl[\ind_{\{Y=i\}}\,\delta(U_i)\bigr]=\E\bigl[(U_i/(M+1))\,\delta(U_i)\bigr],
\]
which uses $\PP\{Y=i\mid U\}=U_i/(M+1)$; (b) follows from~\cref{eq:oneuse-delta}; (c) follows from $(U_i+1)^{-2/3}\leq U_i^{-2/3}$; and (d) follows from the fact that at most $\min\{K,M+1\}$ coordinates of $U$ are positive and they sum to $M+1$, so that Jensen's inequality for the concave function $x\mapsto x^{1/3}$~\cite[Theorem~2.6.2]{covertho} bounds the sum of their cube roots by $\min\{K,M+1\}^{2/3}(M+1)^{1/3}$. Combining~\cref{eq:oneuse-VY}, \cref{eq:oneuse-insertion} and~\cref{eq:oneuse-removal},
\begin{equation}\label{eq:oneuse-VYbound}
 \E\Bigl[D\bigl(\bar p(V)\,\big\Vert\,q\bigr)\Bigr]\leq\Phi(\alpha)-\sum_{i\in[K]:\,q_i>0}q_i\,\Phi(\alpha^i)+2a(\pi)+2\lambda_{K,M}.
\end{equation}

\emph{The second term in~\cref{eq:oneuse-split}.} For every measurable map $v\mapsto\hat p(v)$ into the probability vectors on $[K]$,
\begin{equation}\label{eq:oneuse-predictor}
 \E\left[D\left(\frac{U}{M+1}\,\middle\Vert\,\bar p(V)\right)\right]\overset{\text{(a)}}{\leq}\E\left[D\left(\frac{U}{M+1}\,\middle\Vert\,\hat p(V)\right)\right],
\end{equation}
where (a) follows from~\cref{eq:oneuse-compensation} with $\hat p=\hat p(v)$ and the nonnegativity of relative entropy~\cite[Theorem~2.6.3]{covertho}.

Choose $A_i\coloneqq\max\{1,V_i\}^3$ for $i\in[K]$, $A_\Sigma\coloneqq\sum_{i=1}^KA_i$ and $\hat p_i(V)\coloneqq A_i/A_\Sigma$. For a composition $u\in\mathcal S_{M+1,K}$,
\begin{align}
 D\left(\frac{u}{M+1}\,\middle\Vert\,\hat p(V)\right)
 &=\sum_{i=1}^K\frac{u_i}{M+1}\log\frac{u_iA_\Sigma}{(M+1)A_i}\nonumber\\
 &\overset{\text{(a)}}{=}\frac1{M+1}\sum_{i=1}^Ku_i\log\frac{u_i}{A_i}+\log\frac{A_\Sigma}{M+1}\nonumber\\
 &\overset{\text{(b)}}{=}\frac1{M+1}\sum_{i=1}^Kd_{\mathrm P}(u_i\,\Vert\,A_i)-\frac1{M+1}\,d_{\mathrm P}(M+1\,\Vert\,A_\Sigma)\nonumber\\
 &\overset{\text{(c)}}{\leq}\frac1{M+1}\sum_{i=1}^Kd_{\mathrm P}(u_i\,\Vert\,A_i),\label{eq:oneuse-poisson}
\end{align}
where (a) and (b) use $\sum_{i=1}^Ku_i=M+1$ and $\sum_{i=1}^KA_i=A_\Sigma$ (for (b), expand both sides with~\cref{eq:dpoisson}), and (c) follows from $d_{\mathrm P}(M+1\,\Vert\,A_\Sigma)\geq0$. Given $U=u$, we have $V_i=(u_i+2)^{1/3}+Z_i$ for every $i$, so \cref{lem:cuberoot} applies to each coordinate, including the coordinates with $u_i=0$. Therefore,
\begin{align}
 \E\left[D\left(\frac{U}{M+1}\,\middle\Vert\,\hat p(V)\right)\;\middle|\;U=u\right]
 &\overset{\text{(a)}}{\leq}\frac{95}{24(M+1)}\sum_{i=1}^K(u_i+2)^{1/3}\nonumber\\
 &\overset{\text{(b)}}{\leq}\frac{95}{24(M+1)}K^{2/3}(M+1+2K)^{1/3}=B_{K,M},\label{eq:oneuse-estimator}
\end{align}
where (a) follows from~\cref{eq:oneuse-poisson} and \cref{lem:cuberoot}, and (b) follows from Jensen's inequality for the concave function $x\mapsto x^{1/3}$~\cite[Theorem~2.6.2]{covertho} and $\sum_{i=1}^K(u_i+2)=M+1+2K$. Averaging~\cref{eq:oneuse-estimator} over $U$ and using~\cref{eq:oneuse-predictor} gives $\E[D(U/(M+1)\,\Vert\,\bar p(V))]\leq B_{K,M}$. Substituting this and~\cref{eq:oneuse-VYbound} into~\cref{eq:oneuse-split} proves~\cref{eq:oneuse}, completing the proof of \cref{lem:oneuse}.
\end{IEEEproof}

\subsection{Proof of the state budget}
\label{app:budget-proof}

\begin{IEEEproof}[Proof of \cref{lem:budget}]
Fix $t$ and condition on $(Q,Y^{t-1})$. This fixes $z=S_{t-1}$, $x=X_t$ and $u=z+\ee_x$. The insertion raises $F(z)$ to
\begin{equation}\label{eq:budget-insertion}
 \sum_{i=1}^K(u_i+1)^{1/3}=F(z)+(z_x+2)^{1/3}-(z_x+1)^{1/3}\overset{\text{(a)}}{\geq}F(z)+\frac13(z_x+2)^{-2/3},
\end{equation}
where (a) follows from the fact that the increment of a concave function over an interval of length one is at least its derivative at the right end. The draw gives $F(S_t)=\sum_{i=1}^K(u_i+1)^{1/3}-\Bigl[\bigl(u_{Y_t}+1\bigr)^{1/3}-u_{Y_t}^{1/3}\Bigr]$, and the expected decrease satisfies
\begin{equation}\label{eq:budget-removal}
 \sum_{i=1}^K\frac{u_i}{M+1}\bigl[(u_i+1)^{1/3}-u_i^{1/3}\bigr]\overset{\text{(a)}}{\leq}\frac1{3(M+1)}\sum_{i\in[K]:\,u_i\geq1}u_i^{1/3}\overset{\text{(b)}}{\leq}\frac13\lambda_{K,M},
\end{equation}
where (a) follows from $(x+1)^{1/3}-x^{1/3}\leq\frac13x^{-2/3}$ for $x\geq1$ (the terms with $u_i=0$ vanish), and (b) is step (d) of~\cref{eq:oneuse-removal}. Combining~\cref{eq:budget-insertion} and~\cref{eq:budget-removal},
\begin{equation}\label{eq:budget-step}
 \E\bigl[F(S_t)-F(S_{t-1})\bigm|Q,Y^{t-1}\bigr]\geq\frac13\bigl(S_{t-1,X_t}+2\bigr)^{-2/3}-\frac13\lambda_{K,M}.
\end{equation}
Taking expectations in~\cref{eq:budget-step} and summing over $t=1,\ldots,n$ give the first inequality in~\cref{eq:budget}, since $S_0=s$.

For the second inequality, note that $F$ extends to a concave function on the simplex $\bigl\{z\in\mathbb R_{\geq0}^K\bigm|\sum_{i=1}^Kz_i=M\bigr\}$. Its minimum over the simplex is attained at a vertex $M\ee_j$, where $F=(M+1)^{1/3}+K-1$. By symmetry and concavity, its maximum is attained at the center, where $F=K(M/K+1)^{1/3}=K^{2/3}(M+K)^{1/3}$. The values of $F$ on $\mathcal S_{M,K}$ lie between these two numbers, so $\E[F(S_n)]-F(s)\leq\varrho_{K,M}$. This gives the second inequality in~\cref{eq:budget} and completes the proof of \cref{lem:budget}.
\end{IEEEproof}

\end{document}